\documentclass [12pt] {book}
\usepackage {epsfig}

\usepackage {url}
\usepackage {alltt}

\usepackage [pdftex,
bookmarksopen=true,
backref=false,
colorlinks=true,
linkcolor=blue,
urlcolor=blue,
citecolor=blue,
pdftitle={Multi-Scale Molecular Dynamics Simulations},
pdfauthor={Frédéric Boussinot},
] {hyperref}

\newcommand{\temp}[1] {{{#1}\:K}}

\newcommand{\source}[1] {{}}

\newenvironment{abstract}[1]{%
  \begin{center}\normalfont\textbf{Abstract}\end{center}
  \begin{quotation} #1 \end{quotation}
}{%
  \vspace{1cm}
  }

\newenvironment{myfigure}
{\begin{figure}[!htb]
\begin{center}}
{\end{center}
\end{figure}}

\newcommand {\image} [1] {#1}

\newcommand{\alkane}[2] {{C_\mathit {#1}H_\mathit {#2}}}

\newcommand {\verlet} {{\it Velocity-Verlet}}

\newcommand {\position} [1]{{\textbf r}_{#1}}
\newcommand {\speed} [1]{{\textbf v}_{#1}}
\newcommand {\acceleration} [1]{{\textbf a}_{#1}}

\newcommand {\mvector} [1] {\overrightarrow {#1}}
\newcommand {\normal} [1] {norm ({#1})}

\newcommand {\AllAtom} {{\it AA}}
\newcommand {\CG} {{CG}}
\newcommand {\UA} {{UA}}
\newcommand {\MD} {{MD}}

\newcommand{\code} [1] {{\tt #1}}

\begin{document}

\title {\bf \Huge {\textbf {Multi-Scale Molecular
      Dynamics Simulations}}\\
}

\author {{\bf Frédéric Boussinot}\\
frederic.boussinot@gmail.com}

\date{October 2024}

\maketitle
\begin {abstract}
  
  {\small
    In molecular dynamics (MD), systems are molecules made up of
    atoms, and the aim is to determine their evolution over time.  MD
    is based on a numerical resolution algorithm, whose role is to
    apply the forces generated by the various components, according to
    the equations of Newtonian physics. Molecular Dynamics is
    currently mainly used in materials science and molecular biology.

  In this document, we limit ourselves to {\it alkanes} which are
  non-cyclic carbon-hydrogenated chains. In the basic ``All-atom''
  (AA) scale, all the atoms are directly simulated. In the
  ``United-atom'' (UA) scale, one considers grains that are composed
  of a carbon atom with the hydrogen atoms attached to it. Grains in
  the ``Coarse-grained'' (CG) scale are composed of two consecutive UA
  grains. In the multi-scale approach, one tries to use as much as
  possible the UA and CG scales which can be more efficiently
  simulated than the AA scale.

  In this document, we mainly put the focus on three topics.

  First, we describe an MD system, implemented in the Java
  programming language, according to the Synchronous Reactive
  Programming approach in which there exists a notion of a global
  logical time.  This system is used to simulate molecules and also to
  build the potentials functions at the UA and CG scales.

  Second, two methods to derive UA and CG potentials from AA
  potentials are proposed and analysed. Basically, both methods rely
  on strong geometrical links with the AA scale. We use these links
  with AA to determine the forms and values of the UA and CG
  potentials.  In the first method (called ``inverse-Boltzmann''), one
  considers data produced during several AA scale molecule
  simulations, and one processes these data using a statistical
  approach. In the second method (``minimisation method''), one
  applies a constrained-minimisation technique to AA molecules. The
  most satisfactory method clearly appears to be the
  minimisation-based one.  The UA potentials we have determined have
  standard forms: they only differ from AA potentials by parameter
  values. On the opposite, CG potentials are non-standard
  functions. We show how to implement them with functions defined ``by
  cases''.

  Finally, we consider ``reconstructions'' which are means to
  dynamically change molecule scales during simulations.  In
  particular, we consider automatic reconstructions based on the
  proximity of molecules.

}
  \end{abstract}


{\small \tableofcontents}


\chapter {Introduction}\label{chapter:introduction}

Numerical simulation in physics consists in modeling systems as computer programs and in running them to study the system properties. In molecular dynamics ({\MD}) \cite {MolecularDynamics}, systems are molecules made up of atoms, and the aim is to determine how they evolve over time. The first work on atomic-scale simulation dates back to the 50s \cite{1953JChPh..21.1087M}, at a time when computing resources were extremely limited. The first simulated systems consisted of independent atoms (not grouped into molecules) subjected to perfectly elastic shocks (perfect gas). Then, inter-atomic (van Der Waals) forces were introduced, to obtain simulations closer to reality. Intra-molecular interactions were then taken into account, to obtain true molecular models. {\MD} is currently mainly used in materials science and molecular biology.

\section {Molecular Dynamics}
The basis of {\MD} is classical (Newtownian) physics, with the fundamental equation :

\begin {equation}
\mvector {F} = m\mvector {\mathbf a} 
\end {equation}
where $\mvector {F}$ is the force experienced by a particle of mass $m$ and $\mvector {\mathbf a}$ its acceleration (the second derivative of the variation of its position, with respect to time).

The elementary components used to model molecules are as follows:

\begin {itemize}
\item[$\bullet$]
Atoms, with 6 degrees of freedom (coordinates $x, y, z$ and velocities $sx, sy, sz$).
 
\item[$\bullet$]
Bonds that link two atoms (said to be {\it bonded}) within a molecule; the bond that links two atoms $a, b$ tends to keep the distance $ab$ constant.

\item[$\bullet$]
Valence angles defined by two atoms $a$ and $c$ bonded to the same third atom $b$ within a molecule. The valence angle tends to keep the $\widehat{abc}$ angle at a fixed value.
 
\item[$\bullet$]
  Torsion angles, also called {\it dihedrals}, which link four atoms $a, b, c, d$ within a molecule; $a$ is linked to $b$, $b$ to $c$ and $c$ to $d$; the torsion angle tends to make the two $abc$ and $bcd$ planes coincide.
  
\item[$\bullet$]
  Van Der Waals interactions, which concern two atoms, not necessarily belonging to the same molecule; these interactions depend on the kind of atoms considered.
\end {itemize}

Molecular models can also take into account electrostatic interactions (Coulomb's law); we will not consider this aspect here.

Molecules are made up of linked atoms. An intra-molecular distance is defined to delimit the molecule atoms for which one considers that van Der Waals interactions between them are not to be treated directly, but are taken intp account by the bonds and angles the atoms are involved in.

{\MD} is based on a numerical resolution algorithm, whose role is to apply the forces generated by the various components, according to the equations of Newtonian physics. The resolution algorithm applied to a molecule has a parameter which is the $\Delta t$ simulation time-step. This is of the order of the femto-second ($10^{-15} s$), for molecules at AA scale (see below).

A resolution method often used in {\MD} systems is the $\verlet$ resolution named after its designer who proposed it in 1967 \cite{Verlet}. It is based on a two-stages resolution: in the first stage, velocity is calculated for a half-time-step of $\Delta t / 2$
and position is calculated according to this velocity; in the second stage, the velocity is calculated for the full time-step $\Delta t$. This method has two main advantages: (1) it is very stable (simulations can be long, in a sense that will be made precise later); (2) it is energy-preserving. Energy preservation (i.e. the fact that the resolution neither adds nor removes energy during the simulation) is a fundamental criterion for physics simulations. It corresponds to a deep symmetry of classical physics (independence from the direction of time flow).

Intra-molecular forces (of bonds, valence, dihedrals) and inter-molecular forces (van Der Waals) are conservative: the work done between two points is independent of the path followed. They can therefore be defined as derivatives of scalar fields, called {\it potentials}. We then have:

\begin {equation}
\mvector {F}(\mathbf r) = -\mvector {\nabla} U(\mathbf r)
\end {equation}
where $\mathbf r$ denotes the coordinates of the point to which the force $\mvector {F}(\mathbf r)$ applies and $U$ is the potential from which the force is derived.

The potentials of the various components are grouped together as {\it force-fields}. Several force-fields exist and are used in different contexts. The OPLS force-field \cite{DFTJ97} is often used in the context of liquid simulations \cite{MolecularDynamics}; it is this force-field that we will consider in this document.

The definition of a force-field may require calculations on a quantum scale; we will not deal with this aspect here.

All {\MD} systems comprise the following elements:
\begin {itemize}
\item[$\bullet$] Means of defining atoms and molecules.
\item[$\bullet$] Implementation of the various potentials.
\item[$\bullet$] Implementation of a resolution method.
\item[$\bullet$] Means of defining and performing simulations.
\end {itemize}
In addition, we generally find in {\MD} systems:
\begin {itemize}
\item[$\bullet$] Means of performing simulations in the context of a thermostat (temperature control) or of a barostat (pressure control).
 
\item[$\bullet$] Means to simulate unbounded quantities of atoms (periodic conditions).
\end {itemize}
We will not consider these last two aspects here, as we are only concerned with the fundamentals of {\MD}.

Several dozen {\MD} systems are available, some of which have been developed in the academic world and are freely available. These include DL\_POLY \cite{DLPOLY}, GROMACS \cite {GROMACS} and CHARMM \cite {CHARMFF10}. These systems are written in FORTRAN or C/C++ and are interfaced with 3D visualisation tools, using translation tools between appropriate formats.  The work presented in this document was carried out during the development of a {\MD}\cite{IJMPC-RPSP} system that has been used to carry out all the simulations described in the sequel.

\section {Multi-Scale Simulations}\label{section:multi-echelle}

{\MD} simulations at the all-atom level ({\AllAtom}) are a very powerful tool for analysing molecular systems, but they have two main limitations associated with the number of atoms that can be simulated and to the time-step of the resolution scheme used.

As far as the size of the systems that can be simulated is concerned, the main restriction lies in the size of the memory that has to be used. A few thousand atoms can reasonably be processed by a standard machine. This number can be significantly increased by using networks of distributed machines running in parallel and sharing a memory distributed across the network. In this way, clusters of distributed machines are expected to be able to simulate systems of up to a million atoms.

The time-step of the simulations is imposed by physics: it is a fraction of the period of the shortest vibration occurring in the system. For example, for hydrocarbon molecules described on the {\AllAtom} scale, the time-step depends on the vibration of the CH bond, which is $10^{-14}$ s ($\lambda = 2860 cm^{-1}$ in the infrared spectrum). Usually, the time-step is at most one tenth of this vibration, i.e. $10^{-15}$ s (1 femto-second). With such a time-step, a million steps are needed to predict the dynamics of the system for just one nano-second. Thus, {\AllAtom} models are the most realistic ones at the chemical level, but the time that is simulated can rarely exceed a few nano-seconds. Many simulations, dealing with diffusion phenomena for example, require much longer time-scales. The limitation concerning the simulation time-step appears to be the main limitation, compared with that concerning the size of the system being simulated.

There are two approaches to bypass the time-step limitation, thereby increasing the simulated time. The first approach is called {\it Hyper-molecular dynamics} or {\it Accelerated-molecular dynamics} \cite{Voter97, MF04}. It consists of a modification of the force-field, obtained by reducing the potential energy barrier between two states corresponding to two rare events, so as to increase the probability of a transition between these states.

The second approach, which is widely used and which is the one adopted in this document, consists in reducing the complexity of the molecular system by grouping certain atoms together to form ``grains''. This general approach is called {\it coarse-grained}. The simplest reduction consists in grouping a carbon atom and the hydrogen atoms attached to it into a single grain; this is known as the ``Unified Atom‘’ model ({\UA}).

In what follows, we will also consider grains formed by two bonded carbon atoms, together with the hydrogen atoms that are linked to them; it is to this reduction that we will refer as {\CG} in the following.

Reducing the complexity of molecular systems allows for longer simulated times. This is due to several reasons: 
\begin {itemize}
  
\item[$\bullet$] The time-step of the resolution method can be increased with respect to the {\AllAtom} scale since the shortest period of vibration is also increased (the mass of grains is greater than that of AA atoms).

\item[$\bullet$] The number of degrees of freedom and the number of intra-molecular components (bonds and angles) is reduced compared with their number at {\AllAtom} scale.
In particular, the number of inter-molecular interactions (van Der Waals forces) between grains is reduced compared to the number of interations between atoms. This is important because the number of inter-molecular interactions can be very high and therefore costly to simulate.
  
\end {itemize}

With all these points taken into account, the simulated time can be increased by more than two orders of magnitude by moving from the {\AllAtom} scale to the {\CG} scale. This increase is so significant that most {\MD} simulations are in fact carried out on {\UA} or {\CG} scales.

However, the choice to use the {\UA} or {\CG} scale requires the force field {\AllAtom} to be transported to this scale. This transport is fairly easy from {\AllAtom} to {\UA} \cite {DFTJ97, JMS84, AAProteins98, CHARMFF10}. In fact, the integration of hydrogen atoms into {\UA} grains only requires a change in the parameters of the torsion angles potentials and of van Der Waals interactions. The new parameters can be quite easily deduced analytically from those of {\AllAtom}.

For transport from {\AllAtom} to {\CG}, the parameters are generally deduced from a statistical analysis in dense materials to reproduce certain fundamental functions of these materials (density, energy, distribution) valid on the {\AllAtom} scale. More precisely, the {\CG} force field is constructed from an inverse-Boltzmann treatment in the references \cite {MDKP12,RPMP03,RMMP02}.

An important point is that this type of transfer assumes that the form of the potentials in {\CG} is the same as in {\AllAtom}. This is a point that will be contested in the rest of the text, by proposing a different approach, justifying an analytical construction of {\CG} and allowing {\it reconstructions} that are dynamic changes of scale between AA, UA and CG.

\section {Dynamic Creations and Destructions}\label{section:dynamicity}

Not being able to simulate systems in which numbers of
atoms and bonds may vary during simulations, is a constraint that we may wish to overcome in at least two cases:

\begin{itemize}
\item[$\bullet$] To simulate a chemical reaction, in which a new bond between two atoms may appear during the simulation (for example, between an oxygen atom and an iron atom, when they become too close).

\item[$\bullet$] In the context of a change of scale of a molecule {\it during the simulation}. The change of scale can be implemented by the dynamical addition of the new version of the molecule {\it simultaneously} with the destruction of the old
  version. This is particularly important in the context of reconstructions.
  \end{itemize}

  It therefore seems interesting that a {\MD} system should offer mechanisms
  to destroy or create simulation components dynamically, i.e. at runtime.
  It is a system with such a capability, based on a particular programming style called Synchronous Reactive Programming (RP), on which this document is based.
   
\section {Plan of the Text}

In Chap.\ref{chapter:alkanes}, alkanes, which are linear hydrocarbon molecules, are presented. Alkanes will be used throughout the document. The components of these molecules (bonds, angles) are described and their potentials defining their potential energy are presented in the form of curves (these curves correspond to the OPLS\cite{OPLS} force field, which is that of the AA scale in this document).

Chap.\ref{chapter:forcefield} describes with the help of vector algebra how the
intra-molecular and inter-molecular forces are applied to atoms at the AA scale.

The implementation is considered in Chap.\ref{chapter:implementation} starting with an introduction to reactive programming, in which programs are defined in relation to a notion of time. The Java library SugarCubes\cite{SC} is briefly described, along with the
{\verlet} time-resolution method.

AA simulations are considered in Chap.\ref{chapter:simulationsAA} from the point of view of stability and determinism. The general framework of MD is deterministic chaos.

In Chap.\ref{chapter:echelles} we introduce the two scales of description
{\UA} and {\CG} for which we shall seek to determine potential. 

The inverse-Boltzmann method for determining a {\UA} potential is
presented in Chap.\ref{chapter:boltzmannUA}. The case of {\CG} is
considered in Chap.\ref{chapter:boltzmannCG}.
The inverse-Boltzmann method proves to be unsatisfactory in several
aspects, which justifies the consideration of an alternative ``minimisation'' approach.
This alternative minimisation method is defined for the UA scale in
Chap.\ref{chapter:minimisationUA}.  The minimisation method is based on
a geometric link between the scales {\UA} and {\AllAtom},
and on minimisation of the potential energy at the
{\AllAtom} scale.  The case of {\CG} is considered in
Chap.\ref{chapter:minimisationCG}. The minimisation method proves to be
much more satisfactory than the inverse-Boltzmann method.

UA scale simulations are considered in
Chap.\ref{chapter:simulationsUA}. Those at the CG scale are considered in Chap.\ref{chapter:simulationsCG}.

The reconstructions between the three scales AA, UA, and CG are described in
Chap.\ref{chapter:reconstructions}.

Finally, Chap.\ref{chapter:conclusion} concludes the document.


\chapter {Alkanes}\label{chapter:alkanes}
{\it Alkanes} are linear chains of carbon atoms to which hydrogen
atoms are attached.  Alkanes are designated by formulae of the form
$\alkane{n}{2n+2}$ where $n$ is the number of carbon atoms.  Fig.\ref
{figure:carbonchain} shows an alkane molecule $\alkane{6}{14}$,
composed of 6 carbon atoms and 14 hydrogen atoms.

\begin{myfigure}
\includegraphics[width=3cm] {\image 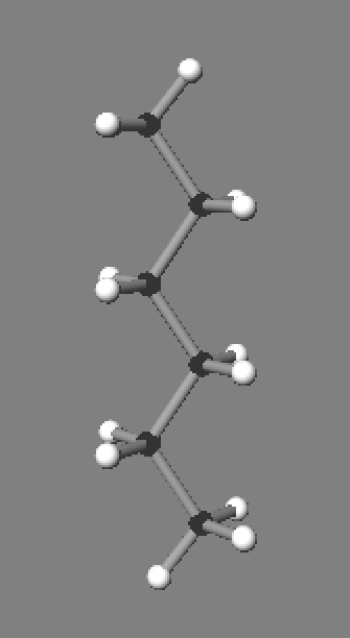}
\caption {\small Hydro-carbon chain $\alkane{6}{14}$ (6 carbon
  atoms, 14 hydrogen atoms).}
\label {figure:carbonchain}
\source {aa/ImageAAAp.java frag = false make image}
\end{myfigure}

In {\MD}, molecules are structured into components that determine
their structure and the forces that apply to their atoms.  Each of
these components is associated with a {\it potential} which is a
function describing the energy of the component.

Potentials are usually grouped together in {\it force-fields}.  The
inter-molecular forces exerted between molecules are generally also
included in force-fields.  In what follows, at the AA scale, we will
consider only one particular force-field, called OPLS, which forms the
basis of the {\MD} system DL\_POLY\cite{DLPOLY}.

In OPLS, the main components of molecules are bonds connecting two
atoms, valence angles connecting three atoms, and torsion angles
connecting four atoms.  For example, in OPLS, the molecule
$\alkane{6}{14}$ has 19 bonds, 36 valence angles and 45 torsion angles
(also called {\it dihedrals}).

At both ends of alkanes there are three hydrogen atoms, while only two
are linked to the other carbon atoms. To simplify, we shall often
consider {\it fragments} of alkanes which are of the form
$\alkane{n}{2n}$, i.e. all carbons atoms without exception have two
hydrogen bonds.

Fig.\ref{figure:fragment} shows the fragment $\alkane{6}{12}$.

\begin{myfigure}
\includegraphics[width=3cm] {\image 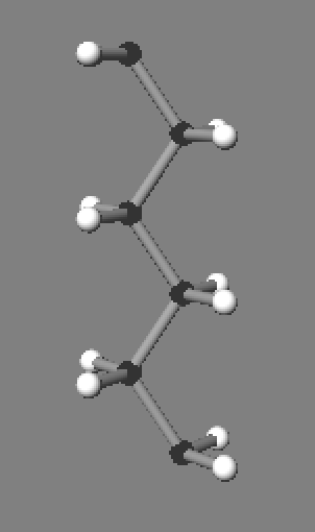}  
\caption {\small Fragment $\alkane{6}{12}$ (6 carbon atoms, 12
  hydrogen atoms).}
\label {figure:fragment}
\source {aa/ImageAAAp.java frag = true make image}
\end{myfigure}
The fragment $\alkane{6}{12}$ has 17 bonds, 30 valence angles
and 39 torsion angles.

We will now describe in more detail the main constituents of the OPLS
force-field.

\section {Bonds}

A {\it bond} models a sharing of electrons between two atoms,
generating a force between them.  In OPLS, the potentials of bonds are
harmonic: a {\it harmonic bond potential} is a scalar field $\cal U$
defining the binding (potential) energy between two atoms at a
distance $r$ as being:

\begin{equation}
  {\cal U}(r) = k (r-r_0)^2
\label{equation:aa-bond}
\end{equation}
where $k$ is the bond strength and $r_0$ is the equilibrium distance
(distance at which no force is exerted on the two atoms).

Fig.\ref {figure:opls-bond} shows the parabolic curves of the OPLS
bonding potentials between two carbons atoms (CC) and between a carbon
atom and a hydrogen atom (CH).

\begin{myfigure}
\includegraphics[width=13cm] {\image 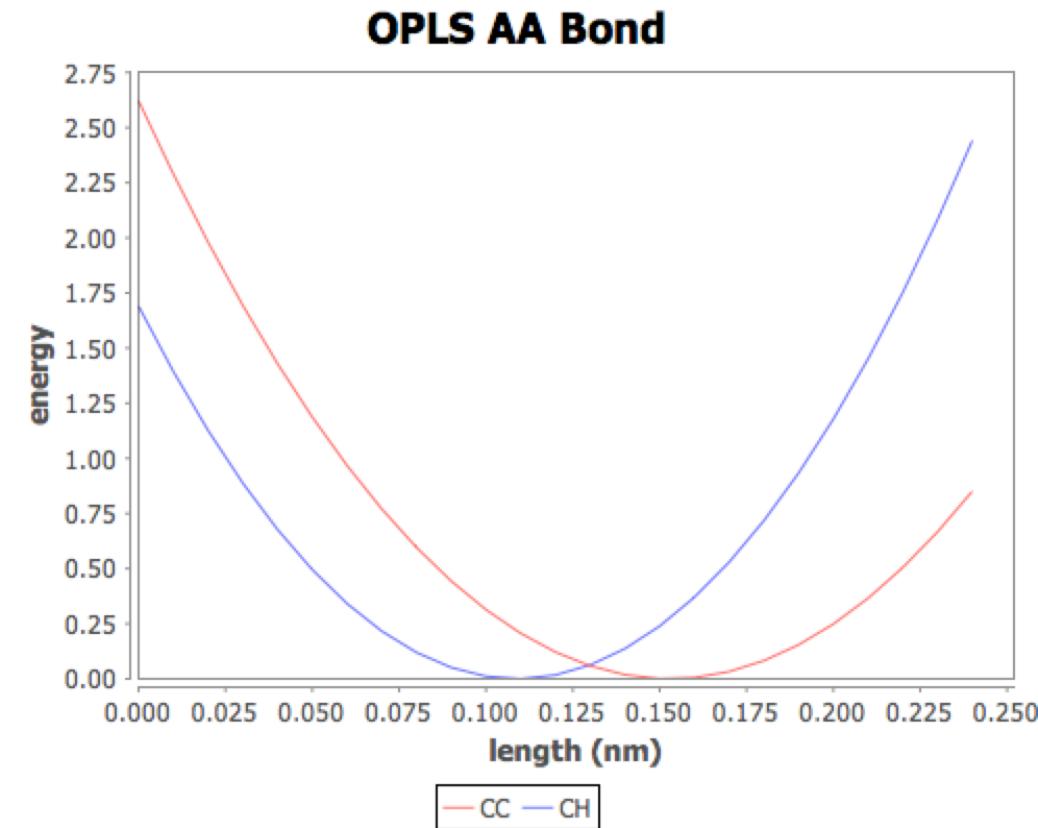}
\caption {\small OPLS potentials ({\AllAtom} scale); bond between two
  carbon atoms: CC; bond between a carbon atom and a hydrogen atom: CH.}
\label {figure:opls-bond}
\source {curves/VisualiseAABondApp.java AAonly = true make aabond}
\end{myfigure}

Distances and energies are given in the internal units of the {\MD}
system (see Chap.\ref{chapter:simulationsAA} for their definitions).

\section {Valence Angles}
Valence angles tend to maintain constant the angle $\theta$ 
between three linked atoms. In OPLS, the valence angle potentials
are harmonic: a {\it harmonic valence potential} is a 
scalar field $\cal U$ which defines the potential energy of an angle
by:
 \begin{equation}
   {\cal U}(\theta) = k (\theta-\theta_0)^2
\label{equation:aa-valence}   
\end{equation}
where $k$ is the strength of the valence angle and $\theta_0$ is the
angle of equilibrium (the one for which no force is exerted on the
three atoms by the valence angle).

Fig.\ref {figure:opls-valence} shows the parabolic curves of the OPLS
valence potentials between three carbon atoms (CCC), between one
carbon atom and two hydrogen atoms (HCH), and between two carbons
atoms and one hydrogen atom (CCH).

\begin{myfigure}
\includegraphics[width=13cm] {\image 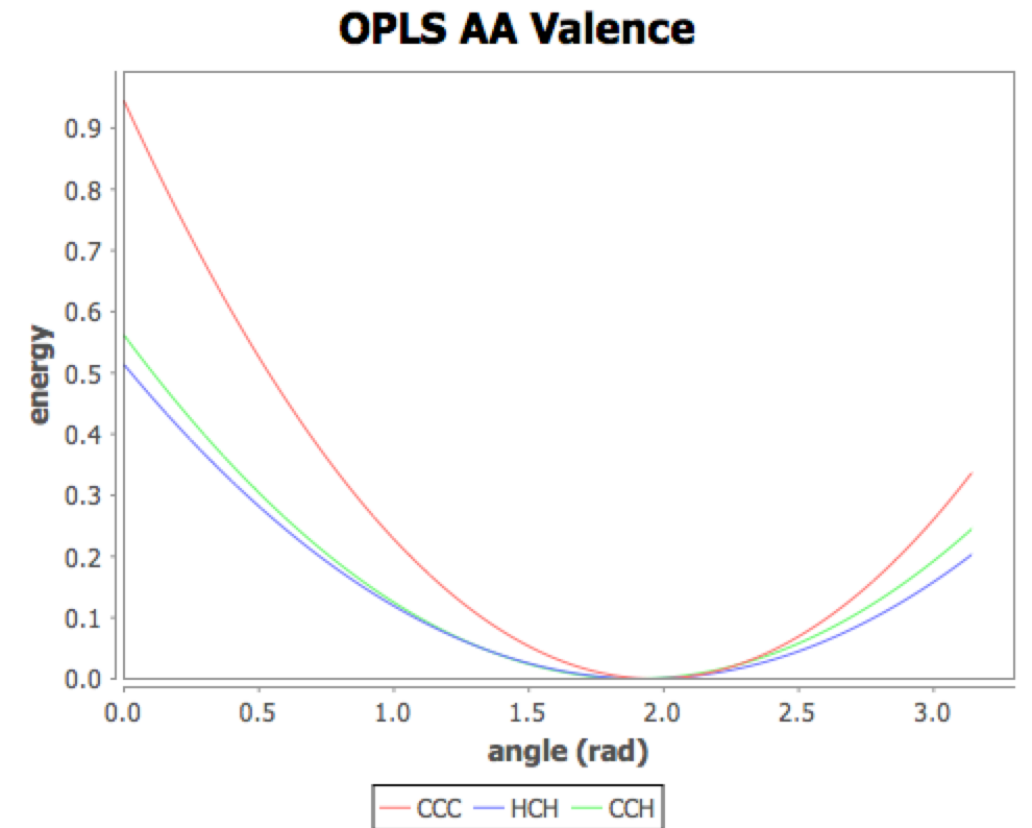}
\caption {\small OPLS potentials of valence angles; between three
  carbons: CCC; between one carbon and two hydrogens: HCH; between one
  hydrogen and two carbons: CCH.}
\label {figure:opls-valence}
\source {curves/VisualiseAAValenceApp.java AAonly = true make aavalence}
\end{myfigure}

\section {Torsion Angles}
A torsion angle (also called {\it dihedral}) tends to keep constant
the angle formed between two planes determined by four linked atoms.
In OPLS, the potentials of the torsion angles have a ``triple cosine''
form, which means that the potential $\cal U$ of a torsion angle
$\theta$ is given by~:

\begin {equation} 
{\cal U}(\theta) = 0.5 [
   A_1 (1+cos (  \theta))
+ A_2 (1 - cos (2\theta)) 
+ A_3 (1 + cos (3\theta))
]
\label{equation:aa-torsion}
\end {equation}

Fig.\ref {figure:opls-torsion} shows the curves of the torsion angle
potentials between four carbons (CCCC), between two carbon atoms and
two hydrogen atoms (HCCH) and between three carbon atoms and one
hydrogen atom (CCCH).  Note that for alkanes, in all torsion angles
the two central atoms are carbon atoms.  Furthermore, in OPLS no force
is exerted on the four atoms when they belong to the same plane ($\pi$
torsion angle).

\begin{myfigure}
\includegraphics[width=13cm] {\image 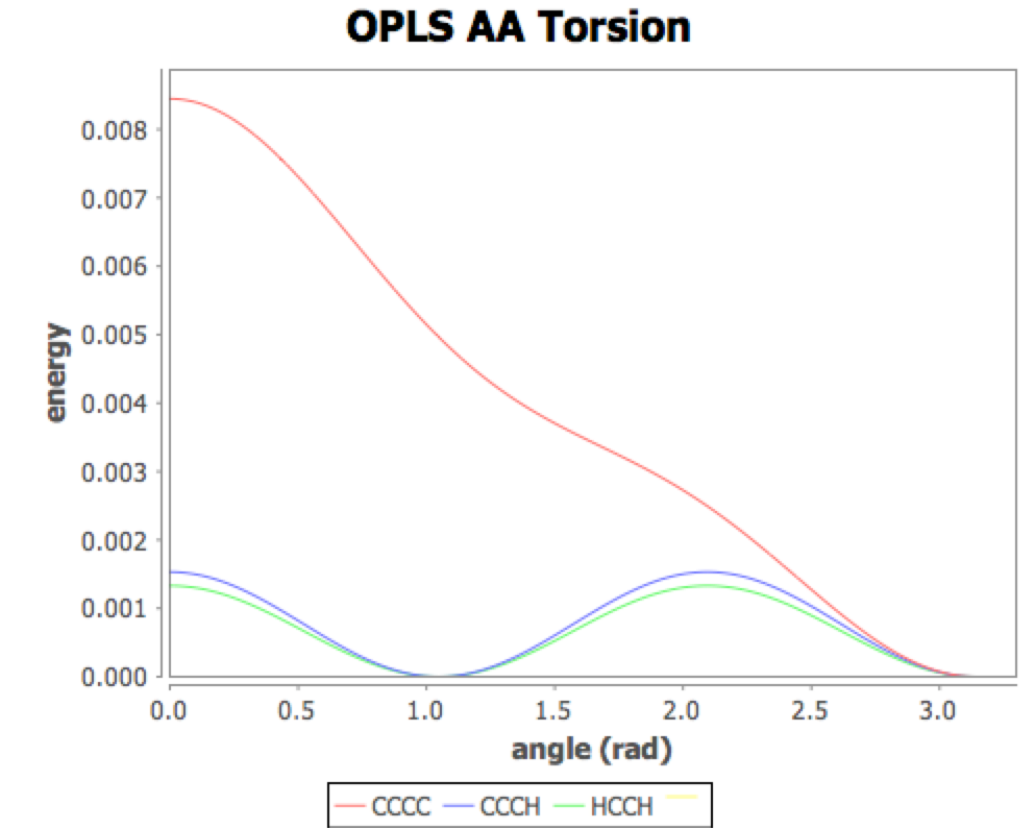}
\caption {\small OPLS torsion angle potentials ({\AllAtom} scale);
  between four carbon atoms: CCCC; between three carbon atoms and one
  hydrogen atom: CCCH; between two carbon atoms and two hydrogen
  atoms: HCCH.}
\label {figure:opls-torsion}
\source {curves/VisualiseAADihedralApp.java AAonly = true make aadihedral}
\end{myfigure}

\section {Inter-Molecular Forces}
The van Der Waals forces exerted between two atoms are extremely
repulsive at short distances and weakly attractive at long distances.
In OPLS, van Der Waals forces are described by {\it 6-12
  Lennard-Jones} potentials.  A 6-12 Lennard-Jones potential is
defined by two parameters $\sigma$ and $\epsilon$; $\sigma$ is the
distance at which the potential is zero and $\epsilon$ is the depth of
the potential (the maximum of the attractive energy). The potential
energy ${\cal U}(r)$ between two atoms at a distance $r$ is defined
by:

\begin {equation} \label {equation:lj-1}
{\cal U}(r) = 4 \epsilon 
[
  {( \frac {\sigma} {r} )}^{12} -
  {( \frac {\sigma} {r} )}^{6} 
]
\end {equation}

Fig.\ref {figure:opls-lj} shows the curves of the OPLS 6-12
Lennard-Jones potentials between two carbon atoms (CC), between a
carbon atom and a hydrogen atom (CH). and between two hydrogen atoms
(HH).

\begin{myfigure}
\includegraphics[width=13cm] {\image 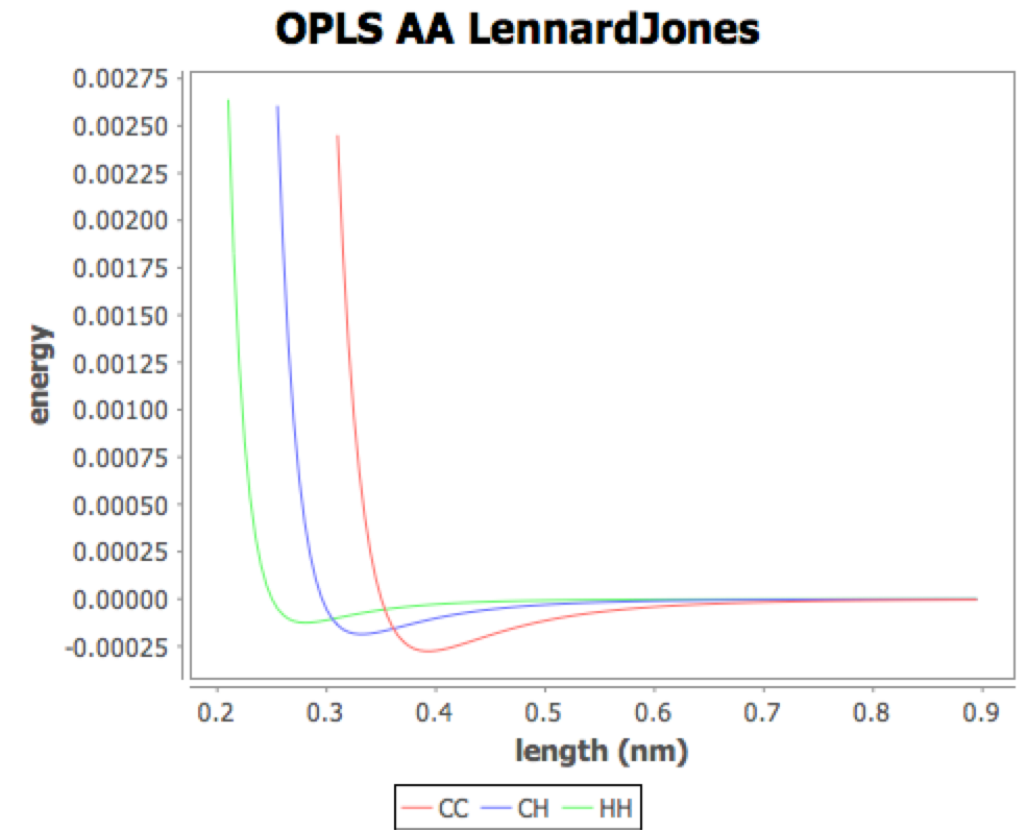}
\caption {\small Lennard-Jones OPLS potentials (AA scale) between two
  carbon atoms: CC; between one carbon atom and one hydrogen atom: CH;
  between two hydrogen atoms: HH.
}
\label {figure:opls-lj}
\source {curves/VisualiseAALjApp.java AAonly = true make aalj}
\end{myfigure}

From now on, the 6-12 Lennard-Jones functions will simply be called
``Lennard-Jones functions''.


\chapter {Forces at AA Scale}\label {chapter:forcefield}
We now describe the forces that apply at the  {\AllAtom} scale\footnote {This
  chapter is directly taken from \cite{monasse:hal-00880202} (also
  available in \cite{monasse:arXiv:1401.1181}).}.

The definition of the forces that apply to atoms must be be very
precise, otherwise some energy may be introduced or lost when
simulating closed molecular systems.

One uses the following notations of vector algebra:

\begin {itemize}

\item if $a$ and $b$ are two atoms, we note $\mvector {ab}$ the vector
  with origin $a$ and end $b$; the distance between the two atoms is
  noted $|ab|$.

\item The null vector is noted $0$.

\item The length of vector $\mvector u$ is noted $| \mvector u |$. One
  thus has: $| \mvector {ab} | = |ab|$.

\item Multiplication of $\mvector u$ by the scalar $n$ is noted
  $n . \mvector {u}$, or more simply $n \mvector {u}$.

\item The vectorial product of $\mvector u$ and $\mvector v$ is noted
  $\mvector {u} \times \mvector {v}$.

\item The scalar product of $\mvector u$ and $\mvector v$ is noted
  $\mvector {u} \bullet \mvector {v}$.

\item We write $\mvector u \bot \mvector v$ when $\mvector {u}$ and
  $\mvector {v}$ are orthogonal
  ($\mvector {u} \bullet \mvector {v} = 0$).

\item We note $\normal {\mvector {u}}$ the normalized vector from
  $\mvector u$ (same direction, but length equal to 1) defined by
  $\normal {\mvector {u}} = (1/| \mvector u |).\mvector u$.

\item If $a$, $b$ and $c$ are atoms, we note $\widehat {abc}$ the
  angle formed by $a$, $b$ and $c$.

\end {itemize}

\section {Bonds \label {section:liaisons}}
A bond models a sharing of electrons between two atoms which produces
a force between them. This force is the derivative of the bond
potential defined between the two atoms. Fig. \ref {figure:liaison}
shows a (attractive) force produced between two linked atoms $a$ and
$b$.

\begin{myfigure} 
\includegraphics[width=13cm] {\image 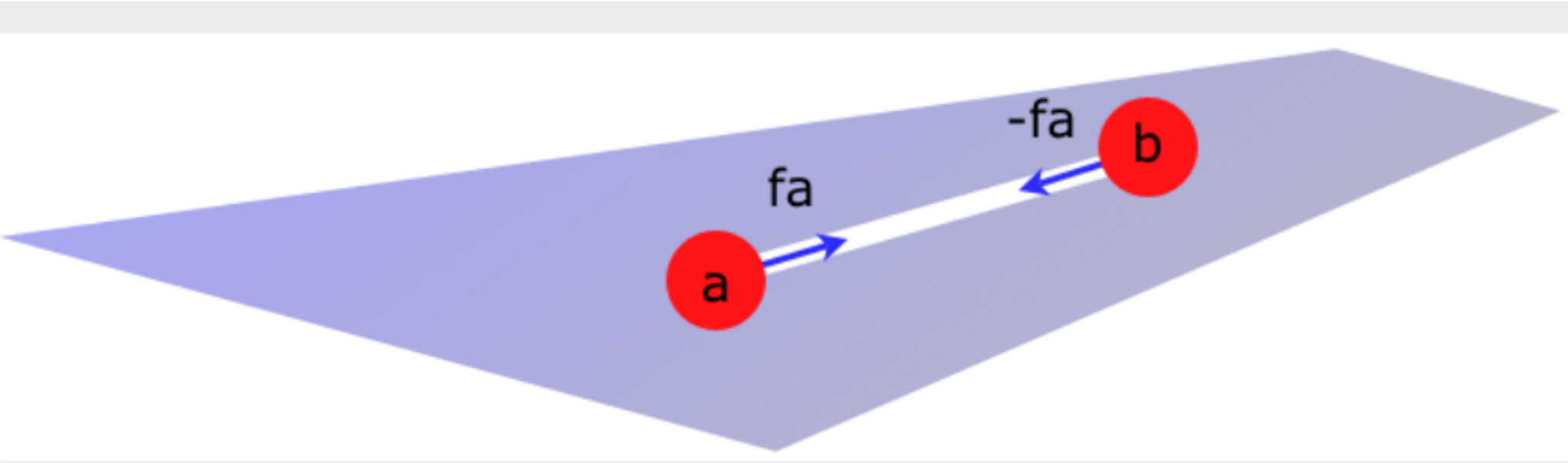}
\caption {\small Attractive forces between two bonded atoms.}
\label {figure:liaison}
\end{myfigure}

A {\it harmonic bond potential} is a scalar field $\cal
U$ which defines the potential energy of two atoms placed at distance
$r$ as:
\begin{equation}
{\cal U}(r) = k (r-r_0)^2
\end{equation}
where $k$ is the strength of the bond and $r_0$ is the equilibrium
distance (the distance at which the force between the two atoms is
null). We thus have:
\begin{equation}
{\frac {\partial {\cal U} (r)} {\partial {r}}} = 2k (r - r_0)
\end{equation}

\noindent
The partial derivative of $\cal U$ according to the position $r_a$ of
$a$ is:
\begin{equation}
{\frac {\partial {\cal U} (r)} {\partial {r_a}}} = {\frac {\partial
    {\cal U} (r)} {\partial r}} . {\frac {\partial {r}} {\partial {r_a}}} . 
\end{equation}

\noindent
But:
\begin{equation}
{\frac {\partial {r}} {\partial {r_a}}} = 1
\end{equation}

\noindent
We thus have:
\begin{equation}
{\frac {\partial {\cal U} (r)} {\partial r_a}}= 2k (r - r_0)
\end{equation}

Let $a$ and $b$ be two atoms, and
$\mvector {u} = \normal {\mvector {ba}}$ be the normalization of
vector $\mvector {ba}$. The force produced on atom $a$ is:
\begin{equation} \label {eq:bond:fa} 
\mvector {f_a} = - {\frac {\partial {\cal U} (r)} {\partial r_a}} .\mvector {u} = -2k(r-r_0) .\mvector {u} 
\end{equation}
and the one on $b$ is the opposite, according to the action/reaction principle: 
\begin{equation} \label {eq:bond:fb} 
\mvector {f_b} = -\mvector {f_a}
\end{equation} 
Therefore, if $r > r_0$, the force on $a$ is a vector whose direction
is opposite to $\mvector {u}$ and tends to bring $a$ and $b$ closer
(attractive force), while it tends to bring them apart (repulsive
force) when $r < r_0$.

According to the definition of $\mvector {f_a}$ and $\mvector {f_b}$,
the sum of the forces applied to $a$ and $b$ is null (i.e. equilibrium
of forces):
\begin{equation} \label {eq:bond:null-sum} 
\mvector {f_a} + \mvector {f_b} = 0
\end{equation} 
Note that no torque (moment of forces) is produced as the two forces are colinear.

\section {Valence Angles\label {section:valence}}
Valence angles tend to maintain at a fixed value the angle between three atoms 
$a$, $b$ and $c$ such that $a$ is linked to $b$ and $b$ to $c$,
as shown on Fig. \ref  {figure:valence}.

\begin{myfigure}
\includegraphics[width=13cm] {\image 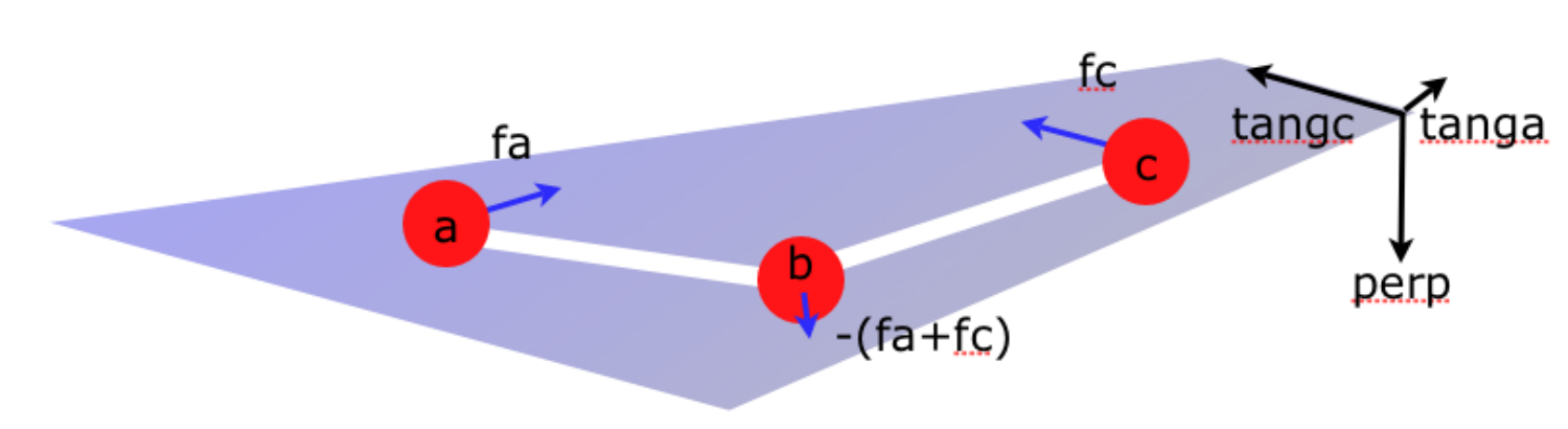}
\caption {\small Valence Angle}
\label {figure:valence}
\end{myfigure}

The forces applied to the three atoms all belong to the plane $abc$
defined by the points $a$, $b$, $c$.

The potential  $\cal U$ of a valence angle is harmonic and verifies
equation Eq.\ref{equation:aa-valence}.

The partial derivative of $\cal U$ according to the angle $\theta$ is thus:
\begin{equation}
{\frac {\partial {\cal U} (\theta)} {\partial \theta}}  = 2k (\theta-\theta_0)
\end{equation}

The partial derivative of $\cal U$ according to the position $r_a$ of $a$ is:
\begin{equation}
{\frac {\partial {\cal U} (\theta)} {\partial r_a}}= 
{\frac {\partial {\cal U} (\theta)} {\partial \theta}} . 
{\frac {\partial \theta} {\partial r_a}}
\end{equation}
that is:
\begin{equation}
{\frac {\partial {\cal U} (\theta)} {\partial r_a}} = 2k
(\theta-\theta_0) . 
{\frac {\partial \theta} {\partial r_a}}
\end{equation}
As $a$ describes a circle with radius $|ab|$, centered on $b$, we
have\footnote {
The length of an arc of circle is equal to the product
  of the radius by the angle (in radians) corresponding to the arc of
  circle.
}:
\begin{equation}
{\frac {\partial \theta} {\partial r_a}} = \frac{1}{|ab|}
\end{equation}

Let $\mvector {p_a}$ be the normalized vector in the plane $abc$,
orthogonal  to $\mvector {ba}$~:
\begin{equation}
\mvector {p_a} = \normal {\mvector
    {ba} \times (\mvector {ba} \times \mvector {bc})}
\end{equation}
The force applied on $a$ is then:
\begin{equation} \label {eq:valence:fa}
\mvector {f_a} = - {\frac {\partial {\cal U} (\theta)} {\partial
    {r_a}}} .\mvector {p_a} = -2k (\theta-\theta_0)/|ab| .\mvector {p_a}
\end{equation}
In the same way, the force applied on $c$ is:
\begin{equation} \label {eq:valence:fc}
\mvector {f_c} = -2k (\theta-\theta_0)/|bc|.\mvector {p_c}
\end{equation}
where $\mvector {p_c}$ is the normalized vector in plane $abc$,
orthogonal to $\mvector {cb}$~:
\begin{equation}
\mvector {p_c} = \normal {\mvector
    {cb} \times (\mvector {ba} \times \mvector {bc})}
\end{equation}

\noindent
The sum of the forces should be null:
\begin{equation}\label {eq:valence:null-sum}
\mvector {f_a} + \mvector {f_b} + \mvector {f_c}  = 0
\end{equation}
Thus, the force applied to $b$ is:
\begin{equation} \label {eq:valence:fb}
\mvector {f_b} = - \mvector {f_a} - \mvector {f_c}
\end{equation}
Moreover, the two momenta exerted on 
$b$ by $\mvector {f_a}$ and
$\mvector {f_c}$ are opposite because~:
\begin{equation}
\mvector {ab} \times \mvector {f_a}= - \mvector {cb} \times \mvector {f_d}
\end{equation}

As a consequence, no rotation around $b$ can result from the
application of the two forces $\mvector {f_a} $ and $\mvector {f_c}$.

\section {Torsion Angles \label {section:dihedrals}}
A torsion angle $\theta$ defined by four atoms $a,b,c,d$ is shown on
Fig. \ref {figure:torsion}.

\begin{myfigure}
\includegraphics[width=13cm] {\image 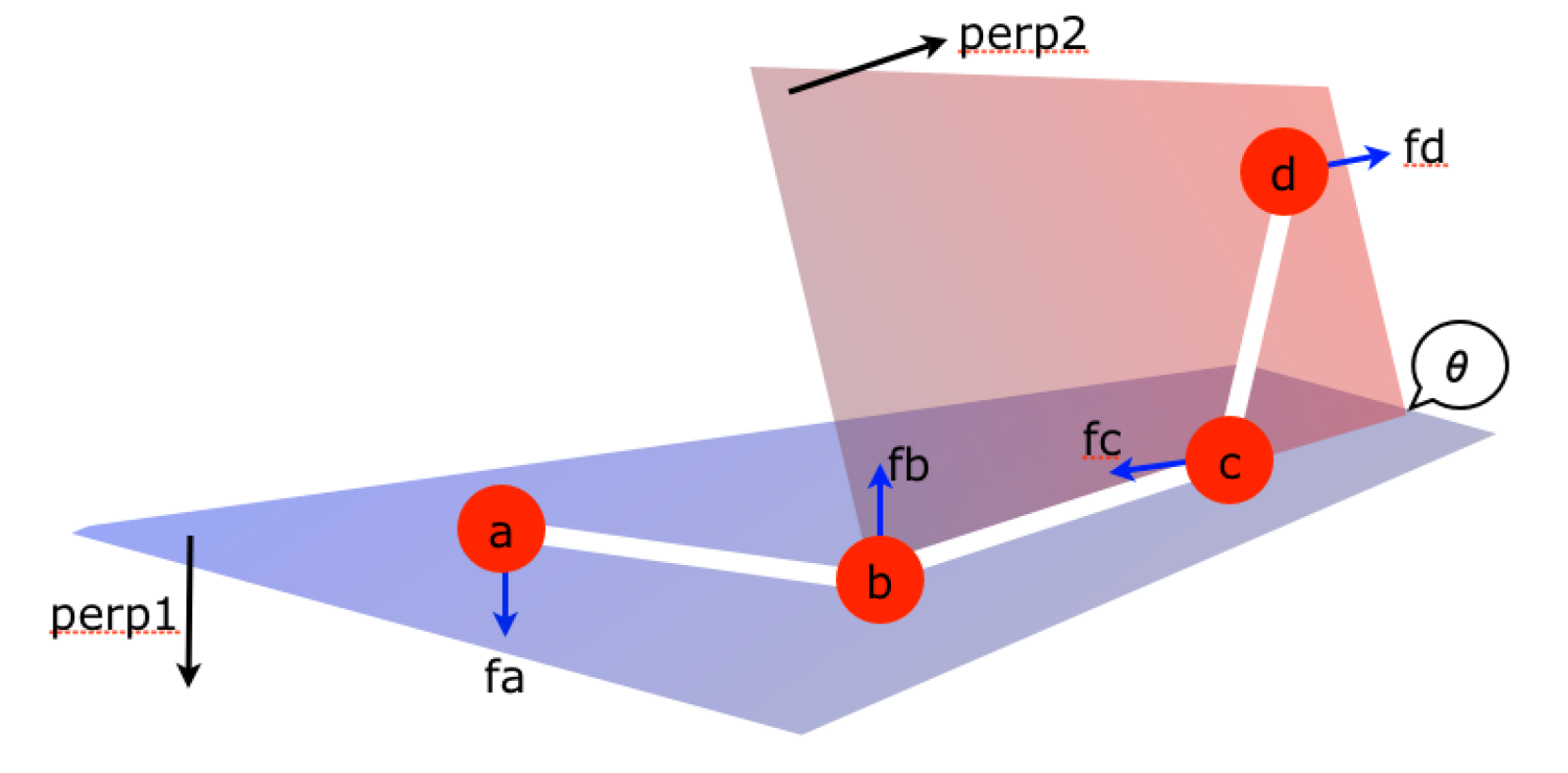}
\caption {\small Torsion angle $\theta$}
\label {figure:torsion}
\end{myfigure}

Potentials of torsion angles have a {\it ``triple-cosine''} form verifying
Eq.\ref{equation:aa-torsion}.

The partial derivative of the torsion angle potential according to the
position $r_a$ of $a$ is:
\begin{equation}
{\frac {\partial {\cal U} (\theta)} {\partial r_a}} = 
{\frac {\partial {\cal U} (\theta)} {\partial \theta}} . 
{\frac {\partial \theta} {\partial r_a}} 
\end{equation}
The partial derivative of the potential according to the angle $\theta$ is:
\begin {eqnarray} \label{equation:dihf}
\frac {\partial {\cal U}(\theta)} {\partial \theta} 
&=& 0.5 (
     - A_1 sin (  \theta)
     + 2A_2 sin (2\theta) 
     - 3A_3 sin (3\theta)
)\\
&=& -0.5 (
     A_1 sin (  \theta)
     - 2A_2 sin (2\theta) 
     + 3A_3 sin (3\theta)
)
\end {eqnarray}

\subsection* {Forces on a and d}
Let us call $\theta_1$ the angle $\widehat {abc}$.  Atom $a$ turns
around direction $bc$, on a circle of radius $|ab|sin (\theta_1)$. The
partial derivative of $\theta$ according to the position of $a$ is:
\begin{equation}
{\frac {\partial \theta} {\partial r_a}} = \frac{1} {|ab|sin (\theta_1)}
\end {equation}

\noindent
We thus have:
\begin{equation}
{\frac {\partial {\cal U} (\theta)} {\partial r_a}} = 
\frac {-0.5} {|ab|sin (\theta_1)}
(
     A_1 sin (  \theta)
- 2A_2 sin (2\theta) 
+ 3A_3 sin (3\theta)) 
\end{equation}

\noindent
Similarly, for atom $d$, where $\theta_2$ is the angle $\widehat {bcd}$:
\begin{equation}
{\frac {\partial {\cal U} (\theta)} {\partial r_d}} = 
\frac {-0.5} {|cd|sin (\theta_2)}
(
     A_1 sin (  \theta)
- 2A_2 sin (2\theta) 
+ 3A_3 sin (3\theta))
\end{equation}

\noindent
Let $\mvector {p_1}$ the normalized vector orthogonal to the plane
$abc$, and $\mvector {p_2}$ the normalized vector orthogonal to the
plane $bcd$ (the angle between $\mvector {p_1}$ and $\mvector {p_2}$
is $\theta$):
\begin{eqnarray}
\mvector {p_1} = \normal {\mvector {ba} \times \mvector {bc}}\\
\mvector {p_2} = \normal {\mvector {cd} \times \mvector {cb}}
\end{eqnarray}

\noindent
The force applied on $a$ is:
\begin{equation} \label {eq:torsion:fa}
\mvector {f_a} = \frac {0.5} {|ab|sin (\theta_1)} (
     A_1 sin (  \theta)
- 2A_2 sin (2\theta) 
+ 3A_3 sin (3\theta)).\mvector {p_1}
\end{equation}

\noindent
In the same way, the force applied on $d$ is:
\begin{equation} \label {eq:torsion:fd}
\mvector {f_d} = \frac {0.5} {|cd|sin (\theta_2)} (
     A_1 sin (  \theta)
- 2A_2 sin (2\theta) 
+ 3A_3 sin (3\theta)).\mvector {p_2}
\end{equation}

\subsection* {Forces on b and c}
We now have to determine the forces $\mvector {f_b}$ and
$\mvector {f_c}$ to be applied on $b$ and $c$.  The equilibrium
conditions imply two constraints: (A) the sum of the forces has to be
null:
\begin{equation}\label{eq:dih:null}
\mvector {f_a} + \mvector {f_b} + \mvector {f_c} + \mvector {f_d} = 0
\end{equation}
and (B) the sum of torques also has to be
null\footnote {
It is not possible to simply define $\mvector {f_b} = -\mvector {f_a}$
and $\mvector {f_c}
= - \mvector {f_d}$, as the sum of torques would be non-null,
thus leading to an increase of potential energy.}.
Calling $o$ the center of bond $bc$, this means:
\begin{equation}\label {eq:dih:null-sum-torque}
\mvector {oa} \times \mvector {f_a} +
 \mvector {od} \times \mvector {f_d} +
 \mvector {ob} \times \mvector {f_b} +
 \mvector {oc} \times \mvector {f_c}  
= 0
\end{equation}

\noindent
From (\ref {eq:dih:null-sum-torque}) it results:
\begin{equation}
( \mvector {ob} + \mvector {ba} ) \times \mvector {f_a} +
( \mvector {oc}  + \mvector {cd} ) \times \mvector {f_d} +
 \mvector {ob} \times \mvector {f_b} +
 \mvector {oc} \times \mvector {f_c}  
= 0
\end{equation}

\noindent
and:
\begin{equation}
( -\mvector {oc} + \mvector {ba} ) \times \mvector {f_a}+
( \mvector {oc}  + \mvector {cd} ) \times \mvector {f_d} -
 \mvector {oc} \times \mvector {f_b} +
 \mvector {oc} \times \mvector {f_c}  
= 0
\end{equation}

\noindent
which implies:
\begin{equation}\label {eq:dih:sum1}
 \mvector {oc} \times ( -\mvector {f_a} + \mvector {f_d} - \mvector
{f_b} + \mvector {f_c}  )
+ \mvector {ba} \times \mvector {f_a} +  \mvector {cd}  \times
\mvector {f_d} 
= 0
\end{equation}

\noindent
From (\ref{eq:dih:null}) it results:
\begin{equation}\label {eq:dih:sum2}
-\mvector {f_a} + \mvector {f_d} - \mvector {f_b} + \mvector {f_c}  =
2 ( \mvector {f_d} + \mvector {f_c})
\end{equation}

\noindent
Substituting (\ref {eq:dih:sum2}) in (\ref {eq:dih:sum1}), one gets:
\begin{equation}
 \mvector {oc} \times ( 2 ( \mvector {f_d} + \mvector {f_c}) )
+ \mvector {ba} \times \mvector {f_a} +  \mvector {cd}  \times
\mvector {f_d} 
= 0
\end{equation}
thus:
\begin{equation}
 2\mvector {oc} \times \mvector {f_d} + 2 \mvector {oc} \times \mvector {f_c}
+ \mvector {ba} \times \mvector {f_a} +  \mvector {cd}  \times \mvector {f_d} 
= 0
\end{equation}
which implies:
\begin{equation}
 2 \mvector {oc} \times \mvector {f_c}  =
-2\mvector {oc} \times \mvector {f_d} -  \mvector {cd} \times \mvector {f_d} -  
    \mvector {ba} \times \mvector {f_a} 
\end{equation}
and finally we get the condition that the torque from $\mvector {f_c}$
should verify in order (\ref{eq:dih:null-sum-torque}) to be true:
\begin{equation} \label {eq:cpl}
\mvector{oc} \times \mvector{f_c} =
- (\mvector {oc} \times \mvector {f_d} + 0.5  \mvector {cd} \times \mvector {f_d} +
    0.5 \mvector {ba} \times \mvector {f_a})
\end{equation}
Let us state:
\begin{equation}
\mvector {t_c} = - (\mvector {oc} \times \mvector {f_d} + 0.5  \mvector {cd} \times \mvector {f_d} +
    0.5 \mvector {ba} \times \mvector {f_a})
\end{equation}

\noindent
Equation $\mvector {oc} \times \mvector {x} = \mvector {t_c}$ has an
infinity of solutions in $\mvector x$, all having the same component
perpendicular to $\mvector {oc}$. We thus simply choose as solution
the force perpendicular to $\mvector {oc}$ defined by:
\begin{equation}\label {eq:torsion:fc} 
\mvector {f_c} = (1/ |oc|^2) \mvector {t_c} \times \mvector
{oc} 
\end{equation}

\noindent
Equation (\ref {eq:cpl}) is verified because:
\begin{equation}
\mvector {oc} \times \mvector {f_c} =  (1/ |oc|^2) \mvector {oc} \times (\mvector {t_c} \times \mvector {oc})
\end{equation}
thus\footnote {
if $u \bot v$, then $u \times (v \times u) = |u|^2 v$.
}~:
\begin{equation} 
\mvector {oc} \times \mvector {f_c} = 
 (1/ |oc|^2) |oc|^2 \mvector {t_c} = \mvector {t_c}
\end{equation}

The value of $\mvector {f_b}$ is finally deduced from equation
(\ref{eq:dih:null}) stating the equilibrium of forces:
\begin{equation} \label {eq:torsion:fb}
\mvector {f_b}= - \mvector {f_a}  - \mvector {f_c} - \mvector {f_d}
\end{equation}

We have thus determined four forces $\mvector {f_a},\mvector {f_b},\mvector
{f_c},\mvector {f_d}$ whose sum is null (\ref {eq:dih:null}) and whose sum of torques is
also null (\ref {eq:dih:null-sum-torque}).

\newcommand {\mdiff} {\mvector {\delta}}

\section {Inter-Molecular Forces\label {section:ljpotentials}}
Inter-molecular potentials are Lennard-Jones potentials of the form defined by 
equation \ref {equation:lj-1}.

Letting $A = \sigma^{12}$ and $B = \sigma^{6}$, this equation
becomes:
\begin {equation} \label {equation:lj-2}
{\cal U}(r) = 
4 \epsilon 
(
  \frac {A} {{r}^{12}} -
  \frac {B} {{r}^{6}}
)
\end {equation}

The partial derivative of $\cal U$ according to distance is thus:
\begin {eqnarray}
{\frac {\partial {\cal U} (r)} {\partial r}} &=&
4 \epsilon
(  -12  \frac {A} {{r}^{13}} +   
   6    \frac {B} {{r}^{7}}
) \\
&=& 24 \epsilon
(  - 2  \frac {A} {{r}^{13}} +  
        \frac {B} {{r}^{7}}
) \\
&=& \frac {24 \epsilon}  {r}
(  -2  \frac {A} {{r}^{12}} +   
         \frac {B} {{r}^{6}}
) \\
&=& -\frac{24 \epsilon}  {r}
(  2  {(\frac {\sigma} {r})}^{12} -   
       {(\frac {\sigma} {r})}^{6}
) 
\end{eqnarray}

Let $a$ and $b$ be two atoms. The force on $a$ is:
\begin{equation} \label {eq:lj:fa}
\mvector {f_a} = \frac {24 \epsilon}  {r}
(  2  {(\frac {\sigma} {r})}^{12} -   
       {(\frac {\sigma} {r})}^{6}
).\mvector {u} 
\end{equation}
where $\mvector {u}$  is the normalization of $\mvector {ba}$. 

The force on $b$ should be the opposite of the force on $a$:
\begin{equation} \label {eq:lj:fb}\label{eq:lj:null-sum}
\mvector {f_b} = -\mvector {f_a}
\end{equation}
The sum of the forces applied to $a$ and $b$ is thus null.
As for bonds, no torque is produced because the two forces are colinear.

\section* {Resume \label {section:resume}}
The forces defined in the previous sections are summed up in the
following table:

\begin {center}
\begin {tabular} {c|c|l}
Bond & \ref {eq:bond:fa}      &  $\mvector {f_a} = -2k(r-r_0).\mvector {u} $  \\
 $ab$   & \ref {eq:bond:fb}      &  $\mvector {f_b} =  -\mvector {f_a}$   \\ \hline
Valence & \ref {eq:valence:fa}  &  $\mvector {f_a} = -2k (\theta-\theta_0)/|ab|.\mvector {p_a}$    \\
 $abc$  &  \ref {eq:valence:fb} & $\mvector {f_b} = -( \mvector {f_a} + \mvector {f_c} )$ \\ 
                         &  \ref {eq:valence:fc} & $\mvector {f_c} = -2k (\theta-\theta_0)/|bc|.\mvector {p_c}$ \\\hline
Torsion & \ref {eq:torsion:fa} & $\mvector {f_a} = \frac {0.5} {|ab|sin (\theta_1)} (
     A_1 sin (  \theta)
- 2A_2 sin (2\theta) 
+ 3A_3 sin (3\theta)).\mvector {p_1}$ \\
 $abcd$           & \ref {eq:torsion:fb}& $\mvector {f_b}= - \mvector {f_a}  - \mvector {f_c} - \mvector {f_d}$ \\ 
                        & \ref {eq:torsion:fc}& $\mvector {f_c} = (1/|oc|^2) \mvector {cpl} \times \mvector {oc}$\\
                         & \ref {eq:torsion:fd}& $\mvector {f_d} = \frac {0.5} {|cd|sin (\theta_2)} (
     A_1 sin (  \theta)
- 2A_2 sin (2\theta) 
+ 3A_3 sin (3\theta)).\mvector {p_2}$ \\ \hline
L-J             & \ref {eq:lj:fa} & $\mvector {f_a} = \frac {24 \epsilon}  {r}
(  2  {(\frac {\sigma} {r})}^{12} -   
       {(\frac {\sigma} {r})}^{6}
).\mvector {u}$\\
$ab$    & \ref {eq:lj:fb} & $\mvector {f_b} = -\mvector {f_a}$
 \end {tabular}
\end {center}

\begin {itemize}

\item[{\bf Bond}] In Eq. \ref {eq:bond:fa}, $k$ is the bond strength
  constant, $r$ is the distance between atoms $a$ and $b$, and $r_0$
  is the equilibrium distance, for which energy is null.  Vector
  $\mvector u$ is defined by $\mvector u = \normal {\mvector {ba}}$.

\item[{\bf Valence}] In \ref {eq:valence:fa} and \ref {eq:valence:fc},
  $k$ is the angle strength constant, $\theta$ is the angle
  $\widehat {abc}$, and $\theta_0$ is the equilibrium angle, for which
  energy is null.
  In \ref {eq:valence:fa}, $\mvector {p_a}$ is  defined by
  $\mvector {p_a} = \normal {\mvector {ba} \times (\mvector {ba}
    \times \mvector {bc})}$.
  In \ref {eq:valence:fc}, $\mvector {p_c}$  is defined by
  $\mvector {p_c} = \normal {\mvector {cb} \times (\mvector {ba}
    \times \mvector {bc})}$.

\item[{\bf Torsion}] In \ref {eq:torsion:fa} and \ref {eq:torsion:fd},
  $\theta$ is the torsion angle, $\theta_1$ is the angle
  $\widehat {abc}$, $\theta_2$ is the angle $\widehat {bcd}$ and
  $A_1$, $A_2$ and $A_3$ are the parameters which define the
  ``three-cosine'' form of the torsion angle. Vector $\mvector {p_1}$
  is defined by
  $\mvector {p_1} = \normal {\mvector {ba} \times \mvector {bc}}$ and
  $\mvector {p_2} = \normal {\mvector {cd} \times \mvector {cb}}$. In
  \ref {eq:torsion:fc}, $o$ is the middle of $bc$ and $\mvector {t_c}$
  is defined by
  $\mvector {t_c}= - (\mvector {oc} \times \mvector {f_d} + 0.5
  \mvector {cd} \times \mvector {f_d} + 0.5 \mvector {ba} \times
  \mvector {f_a})$.

\item[{\bf LJ}] In \ref {eq:lj:fa}, $\sigma$ is the distance at which
  the potential is null and $\epsilon$ is the depth of the potential
  (minimum of energy). As for bonds, one has
  $\mvector u = \normal {\mvector {ba}}$.

\end {itemize}

In each case (bond, valence, torsion, LJ interaction), the sum of the
forces that are applied to atoms is always null (Eq. (\ref
{eq:bond:null-sum}), (\ref {eq:valence:null-sum}),
(\ref{eq:dih:null}), (\ref {eq:lj:null-sum})).  Moreover, no torque is
induced by application of these forces: no torque is produced by bonds
and LJ interactions, as the produced forces are colinear; we have
verified that no torque is produced by valence angles; for torsion
angles, we have chosen the forces in such a way that the sum of the
forces and the global sum of torques are always null (\ref
{eq:dih:null-sum-torque}).  This means that no energy is ever added by
the application of the forces during the simulation process.

It should be noted that torsion angles are the only components that
bring about changes in the 3D geometry of the molecules.  All the
other components are producing forces that remain systematically in a
same plane.

In conclusion, in this chapter we have precisely defined the forces
that apply on atoms in {\MD} simulations. The definitions are given in
a purely vectorial formalism (with no use of a specific coordinate
system). We have shown that the sum of the forces and the sum of the
torques are always null, which means that the energy of (isolated)
molecular systems is preserved while the forces are applied.


\chapter {Implementation}\label{chapter:implementation}
This chapter considers the question of implementing {\MD} by
describing a system implemented in the Java programming language. The
library JavaFX is used for 3D visualisation\footnote{Previous version
  of the system was using Java3D for visualisation.}.  This system is
presented in \cite{IJMPC-RPSP}, the main elements of which are
summerised here.

The implementation is a prototype that does not take into account a
number of functions generally offered by {\MD} systems, such as
temperature control (thermostat) or pressure control (barostat), or
the possibility to define molecular systems using periodic conditions
(crystals).

In fact, the system we are going to consider only implements the core
of {\MD}, in other words Newtonian mechanics, and the only molecular
systems that will be considered are linear chains of carbon and
hydrogen atoms (alkanes) introduced in Chap.\ref{chapter:alkanes}.

The main objective is to provide an implementation of multi-scale
molecular systems (cf Sec.\ref{section:multi-echelle}), allowing
changes in the scale of description during the course of the
simulation (Sec.\ref{section:dynamicity}).

In IT terms, this implementation must be able to handle molecular
systems that are not defined once and for all, not frozen from the
start. On the contrary, the implementation must allow for {\it
  modular} definitions in which molecules can be created or removed
during execution.

A central characteristics of the Newtonian physics on which {\MD} is
based is {\it determinism} or, what amounts to the same thing, the
preservation of the energy of isolated systems over time.  It is
obviously imperative that the total energy of an isolated system
remains the same over the course of the simulation, even if changes of
scale take place.

The {\MD} system built and used here is based on a programming
paradigm called ``Synchonous Reactive programming‘’\cite{Boussinotrc91}, which
reconciles modularity and determinism.  The central modularity tool in
reactive programming is the {\it deterministic parallelism} operator,
which is used through a Java library - SugarCubes\cite{SC}.

The rest of this chapter describes reactive programming and the
SugarCubes library that implements it in Java. We then present the
resolution method used to obtain simulations that are stable over time
(the stability of the implementation is discussed in
Chap.\ref{chapter:simulationsAA}).

\section {Reactive Programming} \label {section:reactive-programming}
Sysnchronous Reactive programming (RP) offers a simple programming para\-digm with
clear and precise semantics. The central feature of RP is that it
provides primitives for expressing parallelism directly at the
programming level. In RP, parallelism is a {\it logical} one, to be
clearly distinguished from the execution parallelism linked to the
operating system on which simulations are run.  Logical parallelism is
an extremely powerful modularity means, enabling complex systems to be
broken down and coded into communicating sub-systems whose structure
can evolve dynamically (dynamicity).

The logical parallelism of RP has also a fundamental characteristic:
it is deterministic. Reconciling parallelism, determinism and
dynamicity may seem paradoxical, but RP provides a way of resolving
this paradox in a coherent computing framework.

In the reactive approach, systems are composed of parallel components
sharing the same {\it instants} which thus define a {\it logical
  clock} shared by all the components. The components synchronise at
the end of each instant and thus run at the same rate.  During each
instant, the components can communicate with each other using signals
(called {\it events} in SugarCubes) which are broadcast
instantaneously. These signals are analogous to radio transmissions
where all the receivers listening on the same frequency immediately
receive the same message.  In the reactive approach, dynamicity is
only taken into account at the boundaries of instants.

There exist several variants of RP, extending various general-purpose
programming languages (for example, ReactiveC \cite {Boussinotrc91}
which extends C, and ReactiveML \cite {RML} which extends the ML
language). RP is also strongly related to the synchronous programming
language Esterel \cite{Esterel}, the main difference being that dynamic
program evolution is forbidden in Esterel while allowed in RP.  One of
the variants of RP that extends the Java language is called
SugarCubes \cite{SC}. The {\bf merge} parallelism operator in
SugarCubes is completely deterministic, which means that at each time
a SugarCubes program has a unique output, function of the inputs, and
that the execution trace is unique.

We will now describe SugarCubes in the following section \footnote{We
  only present here the main concepts; a full description is available
  in \cite{SC}.}.

\section{SugarCubes}
The two main SugarCubes classes are the \code{Instruction} class of
reactive instructions defined with reference to instants, and the
\code{Machine} class of reactive machines which execute the reactive
instructions.

A reactive machine executes the reactive instructions for which it is
responsible in a coordinated manner, allowing instructions to
communicate with each other using {\it events} which are
instantaneously broadcast. Instantaneity here means that an event
generated during an instant is received during that very same instant
by all instructions waiting for it.  

One consequence is that the reaction to the absence of an event can
only take place at the next instant.  Thus, there is no case where an
event is seen as present at one instant by one of the instructions
executed by the reactive machine, while it is seen as absent by
another instruction.

Events are automatically reset to absent by the reactive machine at
the start of each new instant.

Values can be associated with events, and if this is the case, the
instructions receive all of them, in the same order (the order in
which they are generated during the current instant).

Reactive machines proceed by instants: execution takes place during
the first instant, then during the second, then during the third, and
so on indefinitely.  
All the instructions present in the
reactive machine are executed at each instant, and the next instant
only takes place when there is nothing left to execute during the
current instant.  In particular, execution in the current instant is
continued if there remain reactive instructions blocked on 
generated events. In this way, we can be sure that at the end of each instant
all reactive instructions have completed their execution for that
instant and have reacted to all the events generated during the
instant.

New reactive instructions can only be introduced into the reactive
machine between two instants. In this way, the determinism of
execution during an instant is not at risk of being disrupted when
new instructions are dynamically introduced in the execution machine.

Reactive instructions are always executed in parallel and
deterministically by the reactive machine. This deterministic dynamic
parallelism is also directly available at the coding level in the form
of a reactive instruction called \code{merge}.  Let us now describe
the main reactive instructions of SugarCubes.

The main SugarCubes reactive instructions are as follows:

\paragraph {Next Instant.} 
The \code{stop} instruction suspends execution for the current instant
of the reactive instruction it appears in. Execution will resume after
the \code{stop} instruction at the next instant.

\paragraph{Sequence.} 
The instruction \code {seq (i1,i2)} behaves as \code{i1} but the
execution switches immediately (i.e. without waiting for the next
instant) to \code{i2} as soon as \code{i1} terminates.

\paragraph{Parallelism.} 
The \code {merge (i1,i2)} instruction executes one instant of the
\code{i1} instruction and one of the \code{i2} instruction. It ends
when both \code{i1} and \code{i2} have themselves terminated.
Execution always starts with \code{i1} then passes to \code{i2} when
\code{i1} ends or is suspended (i.e. waiting for a non-generated event).

\paragraph{Loop.}
The instruction \code {loop (i)} executes \code{i} in a loop: the
execution of \code{i} is immediately restarted by the loop as soon as
it terminates. We assume that it is impossible for \code{i} to start
and terminate during the same instant (otherwise, we would have an
{\it instantaneous loop} which could cycle indefinitely, preventing
the reactive machine from detecting the end of the current instant and
therefore preventing it to start the next instant).

\paragraph{Java Code.} 
The instruction \code{action (act)} runs the \code{execute} method of
the Java object \code{act} (of type \code{JavaAction}) and terminates
immediately. Note that \code{act} can be executed several time during
the same instant if \code{action (act)}  appears in a loop,
provided it is not an instantaneous one.

\paragraph{Event Generation.} 
The instruction \code {generate (event,value)} generates \code{event}
with \code{value} as associated value, and then terminates
immediately.

\paragraph{Waiting for an Event.} 
The instruction \code {await (event)} terminates immediately if
\code{event} is present (i.e. it has been previously generated during
the current instant). Otherwise, the same execution will start again
at the next instant.

\paragraph{Generated Values.} 
The instruction \code {callback (event,call)} executes the
\code{execute} method of the Java object (of type \code{JavaCallback})
for each value of \code{event} generated during the current
instant. To avoid the risk of losing values, execution of the callback
suspends after each value processed, and in all cases it terminates only
at the next instant.

\paragraph{Preemption.} 
The instruction \code {until (event,i)} executes \code{i} and
terminates either because \code{i} terminates, or because \code{event}
is present (i.e. has been generated during the current instant).

The sequence and parallelism instructions are naturally extended to
more than two branches: for example, the instruction \code {seq
  (i1,i2,i3)} puts in sequence the three instructions \code{i1,i2,i3},
and has the meaning of \code {seq (i1,seq (i2,i3))}.

A reactive machine of class \code{Machine} executes its program which
is a reactive instruction.  The new instructions added in the machine
are placed in parallel (\code{merge}) with the existing program.  The
role of the reactive machine is to execute its program, to detect the
end of the current instant, i.e. when all the parallel branches of its
program are either finished or suspended, and when this is the case,
to move on to the next instant.

New instructions cannot be directly added during the current instant
but only when it is finished, before the start of the next instant.

The execution of the program (the reactive instruction that the
machine holds) for a given instant can be broken down into
several successive phases, during which new events 
are generated, causing the instructions waiting for them to react.

For example, consider the following code, assuming that the event
\code{e} has not been previously generated during the current instant:

\begin{tabular}{ll}
 \code{merge} & (\\
    &\code{await e,} \\
              &\code{generate (e,0)}\\
  &)
                
\end{tabular}

Execution begins with the \code{await} instruction (line 2) which
suspends since \code{e} is absent. Execution then proceeds to the
\code{generate} instruction (line 3) which generates \code{e} (with
the value 0) and then terminates. The first phase of execution of the
\code{merge} instruction is complete but the reactive machine detects
that a suspended instruction remains, waiting for \code{e}. 

A new phase then begins and the \code{await} instruction is executed
again and terminates since \code{e} is now present. The instruction
\code{merge} then also terminates because its two branches
are now both terminated.
  
The execution of the program by a reactive machine is totally
deterministic: only one trace of execution is possible. The execution
of SugarCubes programs is sequential and the parallelism of SugarCubes
is purely logical.  The precise (formal) definition of the semantics
of SugarCubes and a comparison with the standard {\it thread-based}
approach to concurrency is available in \cite{fb-jfs:sugar2000}.

\section {Resolution Method \label {section:resolution}}
The resolution method we use is known as ``{\verlet}''. For each atom,
it works in two stages: the position of the atom is determined at the
end of the first stage, and its velocity is determined at the end of
the second stage. The acceleration is computed (by Newton's law) in
the second stage by summing all the forces applied to the atom during
this stage.  At the end of the second stage, velocity and acceleration
are stored in order to be used in the next stage.

We are going to describe how  {\verlet} works in more detail.

Let $\position t$ be the position of an atom at time $t$, $\speed t$
its velocity at time $t$, and $\acceleration t$ its acceleration at
time $t$.  The {\verlet} resolution method is defined by the two
following equations, where $\Delta t$ is a time interval:

\begin{equation}\label {equation:verlet-1}
  \position {t+\Delta t} =
  \position {t} + \speed {t} \Delta t + 0.5 \acceleration {t} \Delta t^2
\end{equation}
\begin{equation}\label {equation:verlet-2}
  \speed {t+\Delta t} = \speed {t} + 0.5 (
  \acceleration {t} +\acceleration {t + \Delta t} ) \Delta t
\end{equation}

The acceleration $\acceleration {t+\Delta t}$ of an atom is obtained
by collecting the intra- and inter-molecular forces exerted on the
atom at its position at time $t+\Delta t$. In order to do this, two
stages are necessary: in the first stage, the atom is positioned and
in the second the forces exerted on it are collected; the algorithm is
as follows:

\begin{enumerate}

\item Calculation of the speed at half the time-step, from the speed
  and acceleration computed at the previous stage:
\begin{equation}
\speed {t+ 0.5\Delta t} = \speed {t} + 0.5  \acceleration {t} \Delta t
\end{equation}

Then, use of its result to calculate the position at the end of the
complete time-step:
\begin{equation}
\position {t+\Delta t} = \position {t} + \speed {t +
  0.5\Delta t} \Delta t
\end{equation}

\item Calculation of the acceleration at the end of the complete
  time-step from the forces applied to the atom, and calculation of
  the velocity at the end of the time-step from the forces applied to
  the atom. Finally, calculation of the velocity at the end of the
  complete time-step, using the velocity previously calculated for
  half the time-step:
\begin{equation}
\speed {t+\Delta t} = \speed
  {t + 0.5 \Delta t} + 0.5 \acceleration {t+\Delta t} \Delta t
\end{equation}

\end{enumerate}

The instants of RP are naturally identified with the stages of the
resolution method. Thus, two consecutive reactive instants are
required to implement each pair of resolution stages corresponding to
one time-step $\Delta t$.

It should be noted that the time-step may vary depending on the
objects simulated. It is this characteristics that makes it possible to
simultaneously simulate molecules at different scales of
description.

\section*{Conclusion}
In RP, programming is based on logical parallelism, instantaneous
events and dynamic creation/destruction of instructions and
events. Moreover, in the SugarCubes version of RP, programs are
completely deterministic ``by construction''.  The {\MD} system
implemented in SugarCubes and used in this text shows that reactive
programming is an interesting tool for implementing essential aspects
of classical physics, particularly the following points:

\begin {itemize}

\item[$\bullet$] The numerical simulations are based on a
  discretisation of time and on a resolution method implementing an
  integration algorithm. The division of time into instants, which is
  the basis of reactive programming, naturally discretises time, and
  allows us to code in a very simple way the ``{\verlet}'' resolution
  method, which is proving to be very effective in the context of
  {\MD}.

\item[$\bullet$] The forces of classical physics are
  instantaneous and are naturally implemented as instantaneously
  broadcast events in RP.

\item[$\bullet$] Newton's laws are deterministic. Their implementation
  is made much easier by using SugarCubes, whose programs are
  deterministic ``by construction''.  Newton's laws are also
  reversible in time. With the help of a time-reversible resolution
  method (as is {\verlet}), the intrinsic determinism of SugarCubes is
  also an advantage for obtaining the time reversibility property of
  classical physics.
 
\item[$\bullet$] Changes in the chemical structure of the simulated
  systems may appear during simulations. The dynamic nature of RP,
  i.e. the possibility of creating/destroying objects during
  execution, allows these changes to be implemented in a natural,
  semantically clear, and deterministic way.
  
\end {itemize}

Logical parallelism and RP broadcast events are a powerful modularity
tool that allows new parallel components to be introduced into a
system without having to adapt the other components. This modularity
is effective both at the coding and execution levels.  In particular,
{\it observers} can be introduced without disruption, at various
stages of coding, in parallel with the program to be tested. This
possibility allows programmers to adopt modular coding strategies, in
line with modern approaches to programming.


\chapter {Simulations at  AA Scale}\label{chapter:simulationsAA}
We will now simulate molecules at the AA scale using the potentials of
OPLS.

The shapes of the components of the OPLS potentials for alkanes were
described in Chap.\ref{chapter:alkanes}. The forces exerted on atoms
have been defined in Chap.\ref{chapter:implementation}. In order to
simulate alkanes at AA scale, we still need to give the values of the
parameters of the OPLS components.

The system units are the following:
\begin{itemize}
\item[$\bullet$] time~: pico-seconde ($ps$, $10^{-12}$ second)~;
\item[$\bullet$] distances~: nano-mètre ($nm$, $10^{-9}$ meter)~;
\item[$\bullet$] masses~: $kg/mol$~;
\item[$\bullet$] angles~: radian ($rad$)~;
\item[$\bullet$] energies~:  $kg/mol . nm^2 / ps^2$~;
\item[$\bullet$] velocities~: $nm/ps$ (also $km/s$).
\end{itemize}

\noindent
OPLS parameter values for alkanes are given in system units, in Fig.\ref{table:opls}.

\begin{myfigure}
  \begin{eqnarray*}
   & k_{CC} = 112.1312~~~ r_{CC} = 0.1529
  \\
   & k_{CH} = 142.256 ~~~ r_{CH} = 0.109
\\  
  & k_{CCC} = 0.2441364 ~~~ \theta_{CCC} = 1.9669860669976096
\\
  & k_{CCH} = 0.1569 ~~~ \theta_{CCH} = 1.8920924724829928
\\  
   & k_{HCH} = 0.138072 ~~~ \theta_{HCH} =  1.9288037832146987
\\
    & A1_{CCCC} = 0.00728016 ~~~  A2_{CCCC} = -6.56888.10^{-4} \\
    &A3_{CCCC} = 0.001167336
\\
  & A1_{CCCH} = 0 ~~~  A2_{CCCH} = 0 ~~~  A3_{CCCH} = 0.001531344
  \\
   & A1_{HCCH} = 0 ~~~  A2_{HCCH} = 0 ~~~  A3_{HCCH} = 0.001330512
  \\
  & \epsilon_{CC} = 2.7614.10^{-4} ~~~ \sigma_{CC} = 0.35
  \\
  & \epsilon_{CH} = 1.86188.10^{-4} ~~~ \sigma_{CH} =  0.2958
  \\
  & \epsilon_{HH} = 1.2552.10^{-4} ~~~ \sigma_{HH} =  0.25
 \end{eqnarray*}
 \caption{\small Values of OPLS parameters for alkanes, given in
   system units.}
\label{table:opls}
\end{myfigure}

In Fig.\ref{table:opls} one has:

\begin{itemize}
\item[$\bullet$] $k_{l}$ is the strength of the bond $l$ and $r_{l}$
  is its equilibrium distance, for $l = CC$ or $l = CH$
  (Eq.\ref{equation:aa-bond});

\item[$\bullet$] $k_{a}$ is the strength of the valence angle $a$ and
  $r_{a}$ is its equilibrium value, for $a = CCC$ or $a = CCH$ or
  $a = HCH$ (Eq.\ref{equation:aa-valence})~;

\item[$\bullet$] $A1_{a}$, $A2_{a}$ and $A3_{a}$ are the parameters of
  the torsion angles, for $a = CCCC$ or $a = CCCH$ or $a = HCCH$
  (Eq.\ref{equation:aa-torsion});

\item[$\bullet$] $\epsilon_{p}$ and $\sigma_{p}$ are the parameters
  for pairs of atoms, for $p = CC$, $p = CH$ or $p = HH$
  (Eq.\ref{equation:lj-1}).
\end{itemize}


\section {Stability}
The {\MD} system makes it possible to run very long simulations,
preserving energy. To illustrate this stability, consider the
simulation of a $\alkane{8}{16}$ molecule which initially has been
given energy by slightly lengthening its CC bonds. The simulation is
carried out with a time-step of $10^{-4}$ $ps$ (i.e. $0.1$
femto-second). Fig.\ref{figure:simul1} shows the molecule being
simulated on the left, and the energies measured every $10^5$ instants
on the right. Kinetic energy is shown in blue, internal energy in
green, total energy in red (inter-molecular energy, in yellow, is
always zero). Stability of the total energy appears clearly.

\begin{myfigure}
\includegraphics[height=7cm] {\image 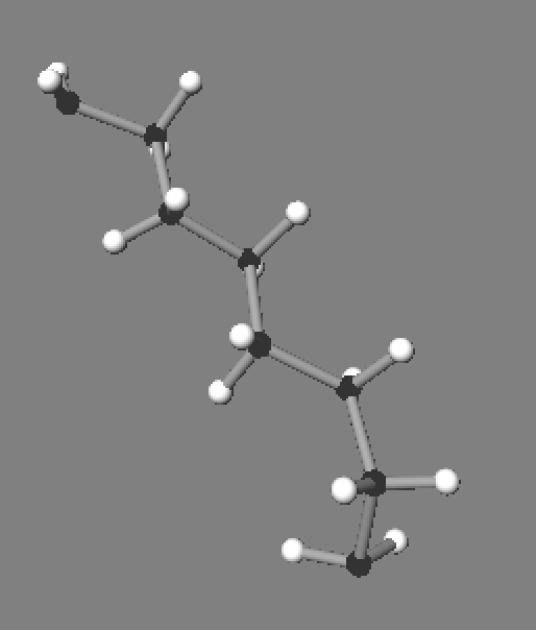} \\
\includegraphics[width=12cm] {\image 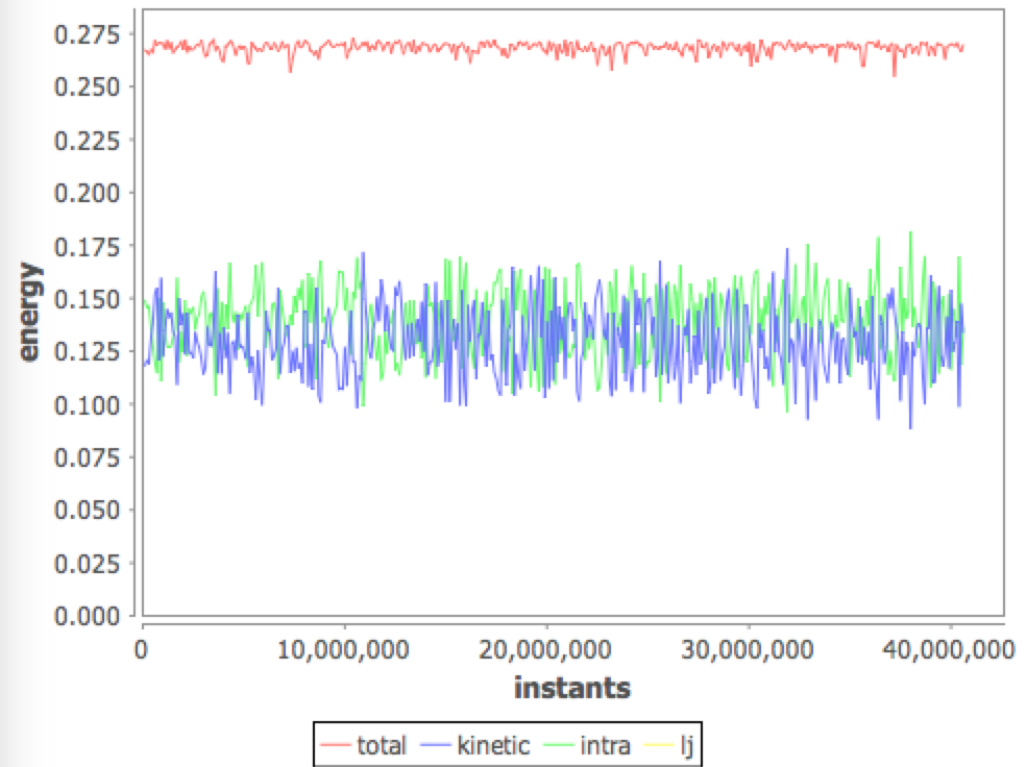}
\caption {\small Top: molecule simulated.
  Bottom: energy measured every $10^5$ instants.}
\label {figure:simul1}
\source {aa/SimpleAAApp.java one=true initialShift = 0.02 tStep=1E-4 make simple}
\end{myfigure}
The $4.10^7$ instants correspond to a simulated time of $2$ $ns$.

In a second simulation, we consider two molecules placed face-to-face
at a distance of $0.9$ $nm$. The two molecules are initially at
equilibrium and the initial energy of the system is therefore the only
(attractive) inter-molecular energy. Fig.\ref{figure:simul2} shows the
two molecules during simulation on the left, and the energies measured
every $10^5$ instants, on the right. We show in red the total energy
which remains close to 0, and in yellow the inter-molecular energy
which remains negative. In blue is the kinetic energy and in green the
intra-molecular energy.

\begin{myfigure}
\includegraphics[height=7cm] {\image 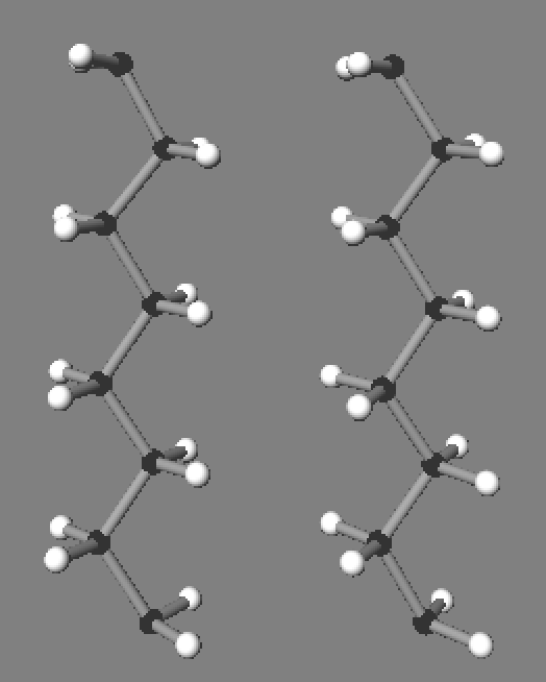} 
\includegraphics[width=12cm] {\image 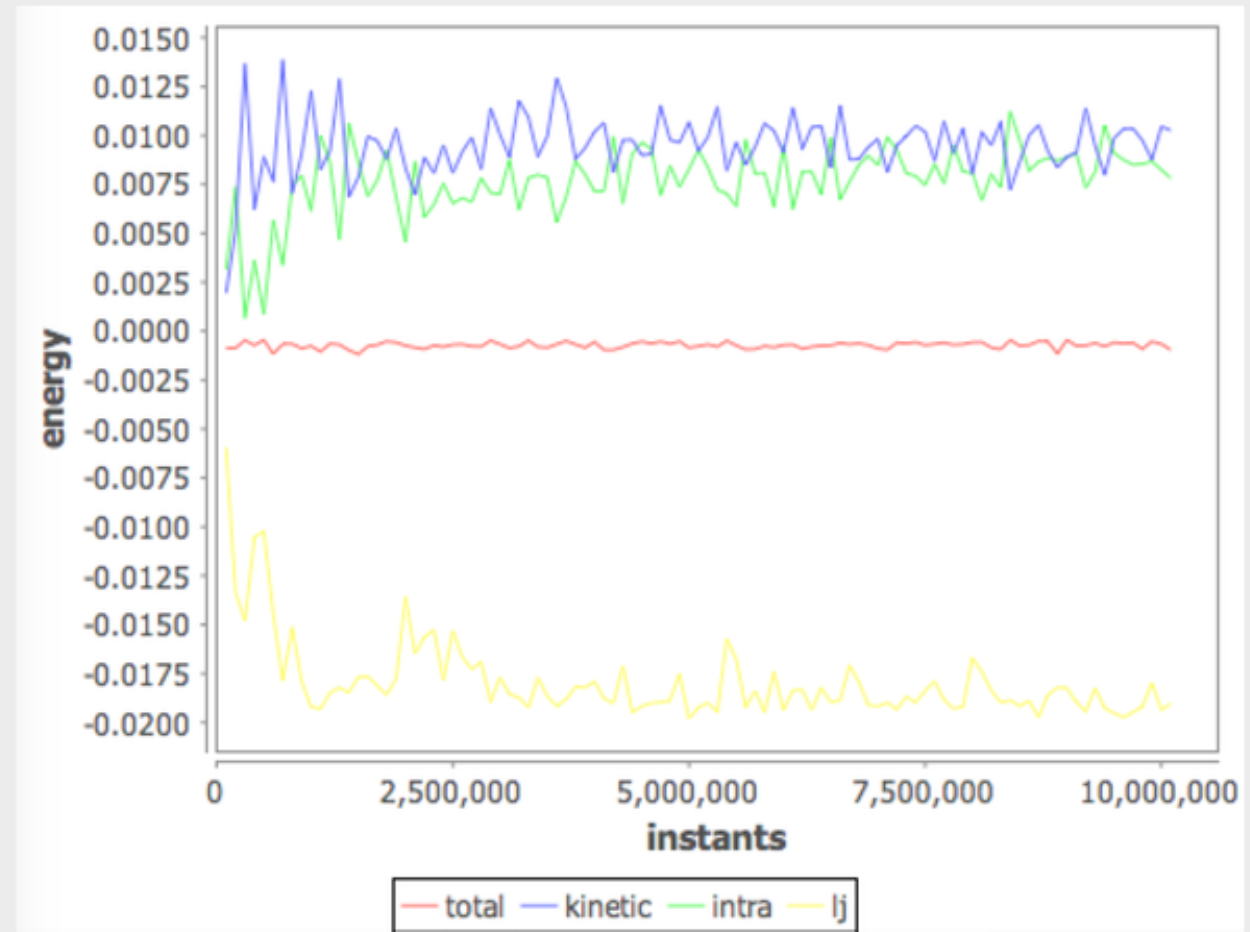}
\caption {\small Top: two molecules initially at equilibrium, during simulation.
  Bottom: energy measured every $10^5$ instants.}
\label {figure:simul2}
\source {aa/SimpleAAApp.java one=false initialShift = 0 tStep=1E-4 make simple}
\end{myfigure}

Let us finally consider a third simulation in which two molecules are
placed face-to-face and given an initial energy, obtained by
translating their first atoms upwards ($y>0$).  Fig.\ref
{figure:simul3} shows the result obtained.

\begin{myfigure}
 \includegraphics[height=4cm] {\image 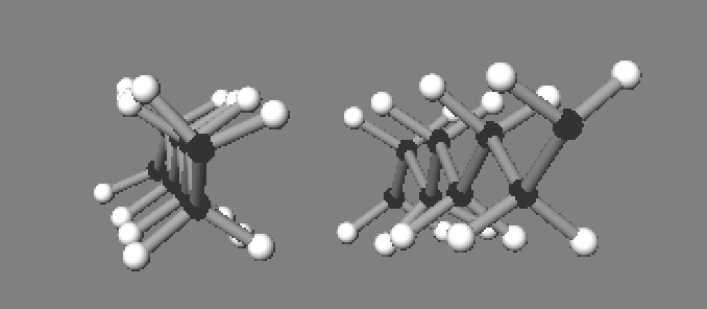}\\
  \includegraphics[width=12cm] {\image 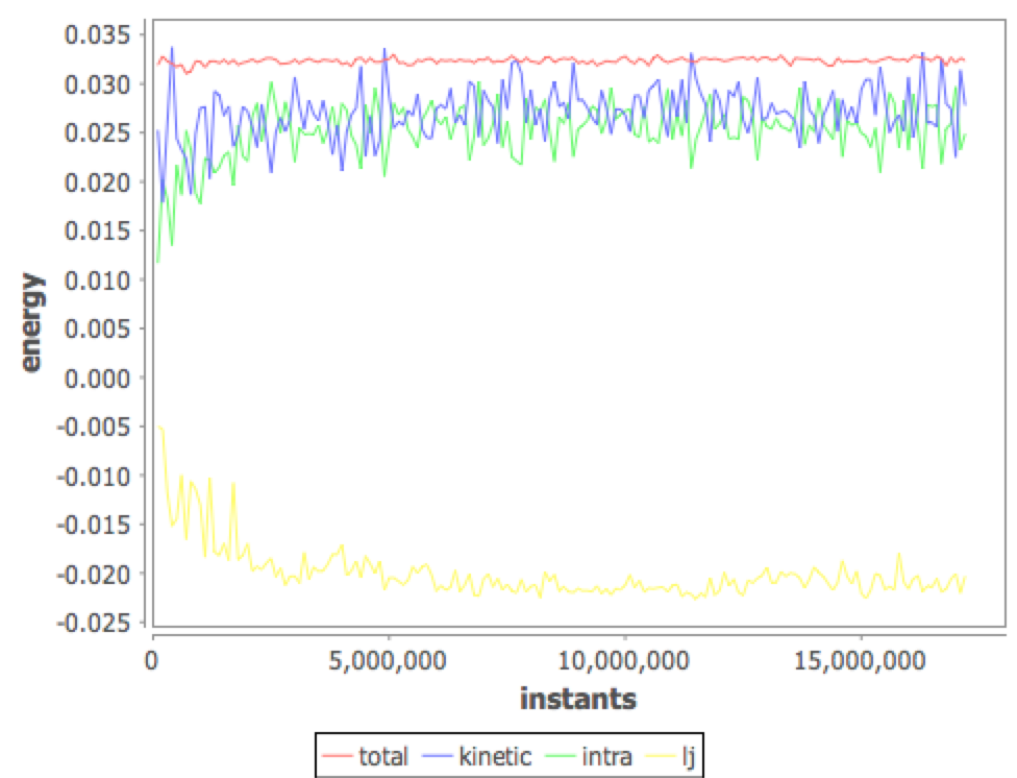}
\caption {\small Top: two molecules, initially not at equilibrium,
  during simulation. Bottom: energies measured every $10^5$ instants.
}
\label {figure:simul3}
\source {aa/SimpleAAApp.java one=false initialShift = 0.005 tStep=1E-4 make simple}
\end{myfigure}

\section {Deterministic Chaos}
The general context of molecular simulations is that of {\it
  deterministic chaos} (see for example \cite{Krivine} for an
enlightening discussion about this notion). Deterministic chaos is
often identified with sensitivity to initial conditions: in systems
exhibiting deterministic chaos, the precision of the initial
conditions is never sufficient to prevent the appearance of chaos
after a certain period of time.

The physics underlying {\MD} is classical Newtonian physics, which is
perfectly deterministic and based on a set of simple laws that are
reversible over time.  The reversibility can be simply illustrated by
choosing a negative time-step; doing so, we obtain simulations that
are similar to simulations with a positive time-step. Reversibility
actually corresponds to preservation of the total energy of isolated
systems, over time.

In the context of MD, deterministic chaos can be demonstrated quite
simply by a pair of simulations that we are going to describe now.


For both simulations, we disconnect the part of the implementation
that processes van Der Waals forces. Thus, only intra-molecular forces
are taken into account in the two simulations.  We place ourselves in an
ortho-normed reference frame where the $x$ axis is horizontal, the $y$
axis is vertical, and the $z$ axis is perpendicular to the two other
axes.

We first simulate two identical molecules ($\alkane{10}{22}$),
initially {\it superimposed}, at a temperature of 1000 $K$. The
superposition of the two molecules is possible because van Der Waals
forces are not taken into account. The initial energy comes from the $y$ translation of the
first carbon of both molecules (with their linked hydrogens).
The initial situation is shown on
Fig.\ref{figure:chaos-superimposed}.

\begin{myfigure}
\includegraphics[width=4.5cm] {\image 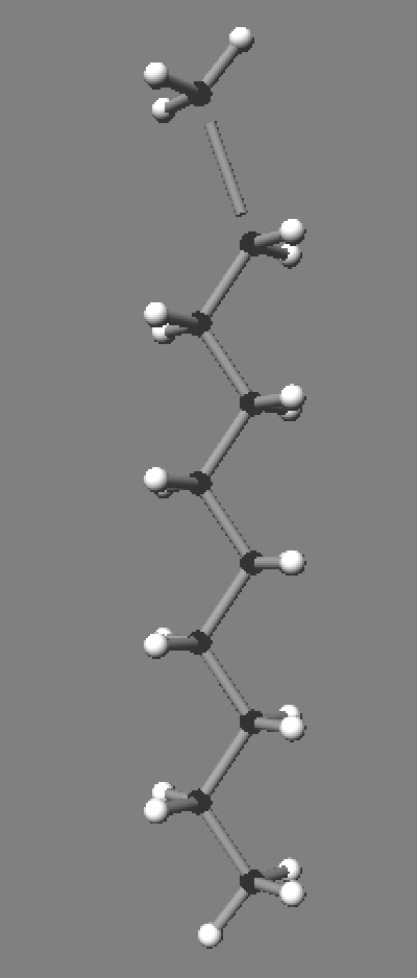} 
\caption {\small Initial situation: superposition of the two
  molecules. Initial energy comes from the $y$ translation of the
  first carbon of both molecules (with their linked hydrogens).}
  \label {figure:chaos-superimposed}
  \end{myfigure}

During the simulation, we observe that the superposition of the two
molecules is maintained over time (tested up to 5 nano-seconds
as shown on Fig.\ref {figure:chaos-final-superimposed}.

\begin{myfigure}
\includegraphics[width=9cm] {\image 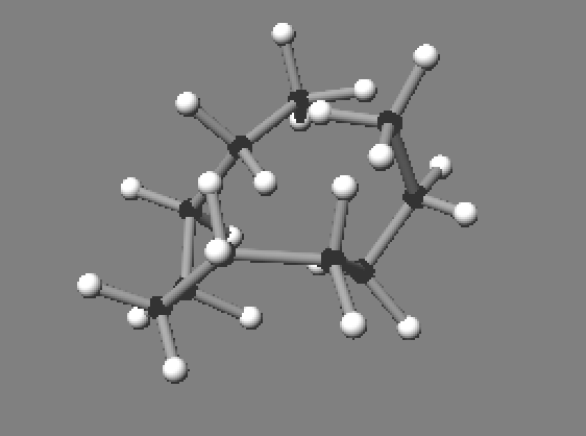} 
\caption {\small Situation after $5$ $ns$: the two molecules are still
  superimposed.}
\label {figure:chaos-final-superimposed}
\end{myfigure}

The preservation of the superposition reflects the complete
determinism of the execution of the two molecules.

In the second simulation, the initial conditions are slightly changed
for one molecule: the $y$ coordinate of the first carbon is translated
by a very small distance of $10^{-17}$ $nm$.  The difference in
placement is of course far too small to be observed on the screen.

However, we can observe quickly (actually,
after $2.7$ pico-seconds) a clear divergence in the positioning of the
two molecules which are no more superimposed, as shown on Fig. \ref{figure:chaos}.

\begin{myfigure}
\includegraphics[height=16cm] {\image 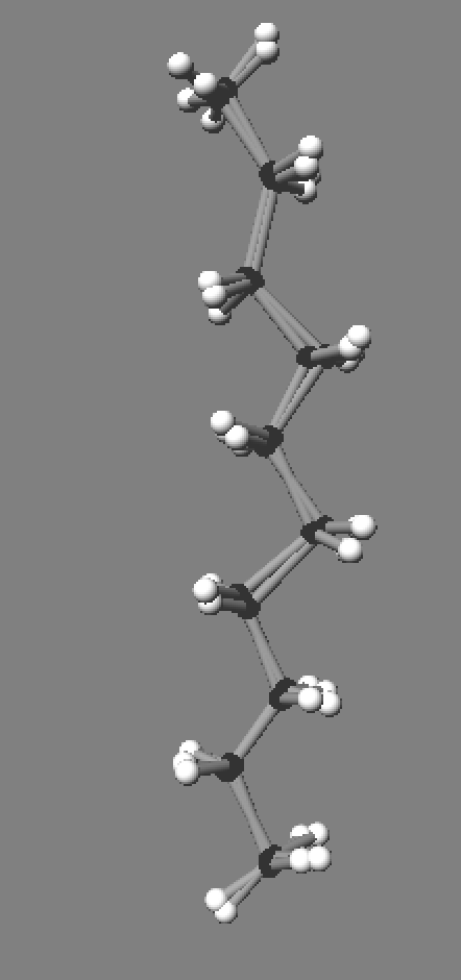}
\caption {\small After adding to the initial $y$ translation of
  only one molecule of a distance of $10^{-17}$ $nm$, one observes a divergence
  at $2.7$ $ps$.}
\label {figure:chaos}
\end{myfigure}

This divergence is the manifestation of the ``sensitivity to initial
conditions'' specific to deterministic chaos.

One important point needs to be emphasised: the total determinism of
the implementation of {\MD} is an essential asset for the development
of simulation programs.  In particular, a surprising result or one
revealing an error {\it can always be reproduced}, which is not the
case with implementations using execution threads or network
communications that are non-deterministic by nature.

\chapter {Multi-Scale Approch}\label{chapter:echelles}
This chapter considers the multi-scale approach by defining two scales
based on AA. The first scale is called UA and consists of ignoring
hydrogen atoms: a UA grain is a carbon atom with the hydrogen atoms
attached to it.  The second scale, called CG, combines two UA
grains into a single CG grain.

\section {{\UA} Scale}
Molecules on the {\UA} scale are chains of grains of two kinds: a
$G_2$ grain contains one carbon atom and two hydrogen atoms, while a
$G_3$ grain contains one carbon atom and three hydrogen atoms bonded
to it.

As in OPLS at the AA scale, the intra-molecular components in UA are
bonds, valence angles, and torsion angles.

As was done for the AA scale, we define UA fragments consisting solely
of $G_2$ grains. In the following, for simplicity only these UA
fragments will be considered. Thus, in addition to grains $G_2$, we
will only have to consider $G_2G_2$ bonds, $G_2G_2G_2$ valence angles,
$G_2G_2G_2G_2$ torsion angles, and inter-molecular forces between two
$G_2$ grains.

In the {\MD} system, the $G_2$ grains are represented by cyan coloured
balls, as in Fig.\ref{figure:G8} where the fragment considered is made
up of 8 grains $G_2$, 7 bonds $G_2G_2$, 6 valence angles $G_2G_2G_2$
and 5 dihedrals $G_2G_2G_2G_2$.
\begin{myfigure}
 \includegraphics[height=6cm] {\image 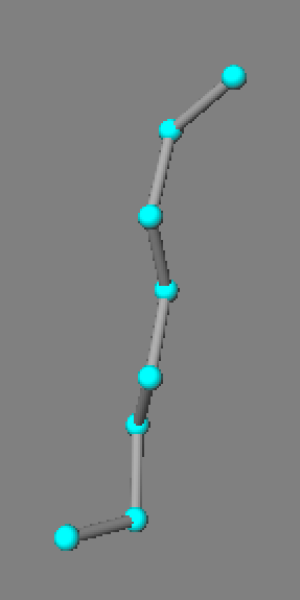}
\caption {\small UA Fragment composed of 8 grains $G_2$.}
\label {figure:G8}
\source {ua/ImageUAApp.java make image suspend}
\end{myfigure}

In accordance with the interpretation of UA grains, the mass of a UA
grain is the sum of the masses of the AA atoms it contains, so $0.014$
for $G_2$.

There is a strong geometric link between {\UA} molecules and
{\AllAtom} molecules: the UA molecule ``equivalent‘’ to an AA molecule
is constructed by ``erasing'' all the hydrogens, replacing the carbons
by UA grains and the AA components by the corresponding UA components.

Conversely, the AA molecule equivalent to a UA molecule is constructed
by replacing the UA grains with carbons, by adding hydrogens and CH
bonds to these carbons, by replacing the UA components with AA
components of the same type, and by introducing appropriate CH bonds,
CCH and HCH valence angles, and HCCH torsion angles.

Fig.\ref {figure:C8-G8} shows a molecule $\alkane 8 {16}$ at
equilibrium on the left, and the equivalent UA molecule on the right.

\begin{myfigure}
 \includegraphics[height=6cm] {\image 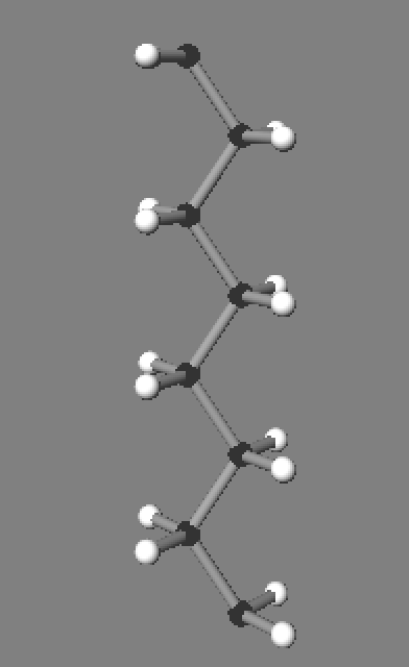}
 \includegraphics[height=6cm] {\image 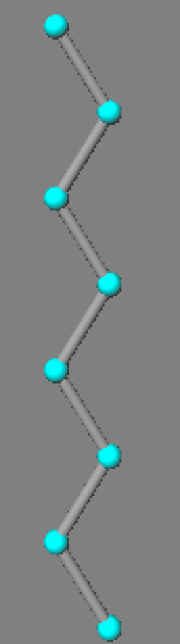}  
 \caption {\small Left: AA molecule (fragment) $\alkane 8 {16}$.
   Right: equivalent UA molecule, containing 8 grains $G_2$.}
 \label {figure:C8-G8}
 \source {right: ua/ImageUAApp.java tstep=0 make image}
\end{myfigure}

A question arises: how to define the UA potential from the AA
potential ? The answer will be given in
Chap.\ref{chapter:simulationsUA} by taking advantage of the
geometrical link existing between the two scales.

\section {{\CG} Scale}\label{echelles:cg}
On the CG scale, {\CG} grains are formed by two consecutive {\UA}
grains.  We thus have three types of CG grains which differ according
to the number of hydrogens they contain.  CG grains are denoted by
$CG_n$ where $n$ is the number of hydrogen atoms~; $n$ is therefore 4,
5 or 6.  The $CG_4$ grain is shown in Fig.\ref{figure:grain-CG4}.

\begin{myfigure}
\includegraphics[width=6cm] {\image 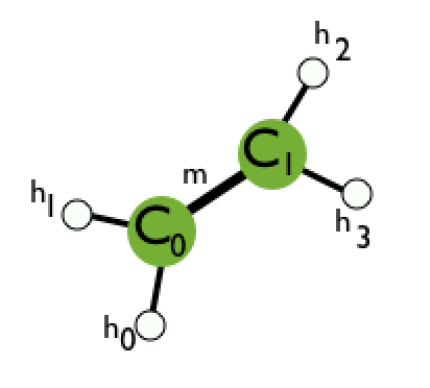}
\caption {\small $CG_4$ grain composed of two carbons and four
  hydrogens. By definition, the centre of the CG grain is the middle $m$
  of the AA bond that joins the two carbons.}
\label {figure:grain-CG4}
\end{myfigure}

The mass of a CG grain is the sum of the masses of the two carbons, plus
those of the hydrogens it contains. Thus, the mass of the $CG_4$ grain
is $0.028$.

In what follows, just as for the UA scale we only consider $G_2$
grains for the sake of simplicity, in the CG scale we will only consider
$CG_4$ grains (i.e. made up of two carbon and four hydrogen atoms).

In the {\MD} system, $CG_4$ grains are represented by yellow
balls. Fig.\ref{figure:molecule4-CG4} shows a CG molecule made up of 4
$CG_4$ grains.
\begin{myfigure}
\includegraphics[height=6cm] {\image 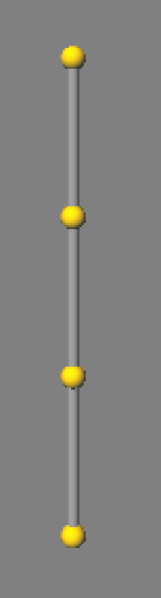}
\caption {\small CG molecule made up of four $CG_4$ grains.}
\label {figure:molecule4-CG4}
\source {cg/ImageCGApp.java make image}
\end{myfigure}

We establish a geometric link with the scale AA by considering that a
{\CG} grain is centred on the middle of the {\AllAtom} bond of the
equivalent $\alkane 2 n$ molecule.  Thus, using the notations of
Fig.\ref{figure:grain-CG4}, the AA molecule equivalent to a grain
$CG_4$ contains 5 bonds ($h_0C_0$, $h_1C_0$, $C_0C_1$, $h_2C_1$,
$h_3C_1$), 6 valence angles ($h_0C_0h_1$, $h_0C_0C_1$, $h_1C_0C_1$,
$h_2C_1C_0$, $h_3C_1C_0$, $h_2C_1h_3$) and 4 torsion angles 
($h_0C_0C_1h_2$, $h_0C_0C_1h_3$, $h_1C_0C_1h_2$, $h_1C_0C_1h_3$).

A UA molecule at equilibrium (i.e. with zero intra-molecular energy)
has for CG equivalent a molecule whose $CG_4$ grains are all aligned.
The CG molecule in Fig.\ref{figure:molecule4-CG4} is thus the CG
equivalent of the AA and UA molecules in Fig.\ref{figure:C8-G8}.

At equilibrium, in CG, the valence angles are flat (their value is $\pi$).

Let us now consider the case of torsion angles at the scale CG.  A CG
torsion angle that is not at equilibrium must contain at least one
non-flat CG valence angle (indeed otherwise, if the two CG valence
angles were flat, the CG torsion angle would be at equilibrium).

Therefore, in CG, it is possible to consider that the energy of a
torsion angle is actually {\it distributed between the two CG valence
  angles} it contains.  This choice simplifies the determination of CG
potentials: the CG torsion angles have no longer to be considered.

We have therefore chosen to define only CG bonds and CG valence angles
as components of CG molecules, without considering torsion angles on
this scale.  Thus, the CG molecule of Fig.\ref {figure:molecule4-CG4}
has only three bonds and two valence angles as components.

It should be noted that the ``absorption'' of torsion angles by the
valence angles associated with them is possible on the CG scale, but
not on the AA scale nor on the UA scale. In fact, in AA and UA a
torsion angle can have energy while the two associated valence angles
are at equilibrium, which is impossible on the CG scale.

As with UA, the basic question is how to define the CG potentials from
the AA ones, taking advantage of the geometric link between the two
scales. This question will be considered in
Chap.\ref{chapter:simulationsCG}.

\section {Complexity}
In the transition from an AA molecule to an equivalent UA or CG
molecule, the number of atoms and intra-molecular components (bonds,
valence angles, torsion angles) decreases significantly. As a result,
UA or CG simulations have a lower complexity, making it possible to
simulate longer molecules, with higher simulated time / real-time
ratios.

The difference in complexity is even more obvious in the case of
inter-molecular forces which grow with the square of the number of atoms or
grains.

Fig.\ref {figure:complexite} compares the complexities at the three scales.
\begin{myfigure}
  \includegraphics[width=12cm] {\image 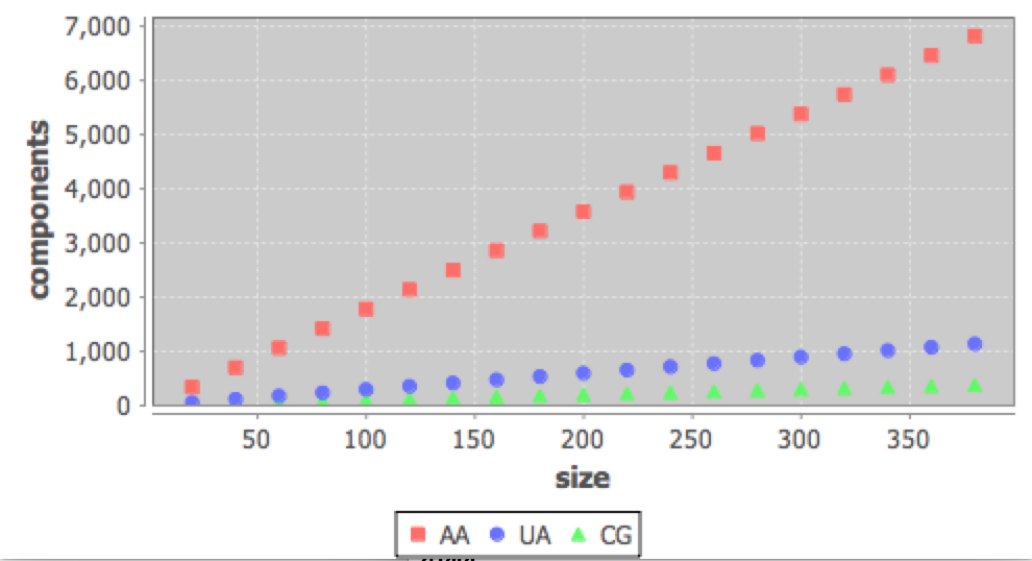}
\\
  \includegraphics[width=12cm] {\image 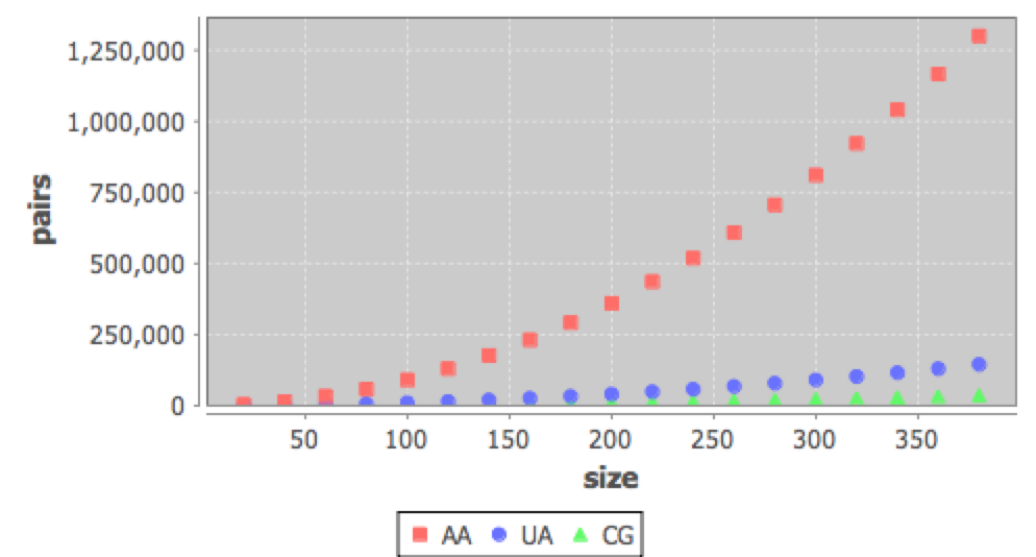}  
  \caption {\small Comparison of complexity of simulations at the three scales.
Top: intra-molecular forces; Bottom: inter-molecular forces. }
\label {figure:complexite}
\source {up: demos/ContentApp.java intra=true make content~~~down: demos/ContentApp.java intra=false make content}
\end{myfigure}

The x-axis shows the size $n$ of the molecules ($n$ carbon atoms for
AA, $n$ grains $G_2$ for UA, and $n/2$ grains $CG_4$ for CG). In the
top image of Fig.\ref {figure:complexite}, we have on the ordinate the number of intra-molecular
components (bonds, valence angles, and torsion angles) to
be simulated at each instant.

In the bottom image of Fig.\ref {figure:complexite}, we have the number of
pairs of atoms or grains that need to be analysed to determine the
inter-molecular forces between two molecules of size $n$.

To illustrate the real-time gain obtained with the UA and CG scales,
we consider a situation where two molecules are placed face-to-face
and launched at opposite speeds, one against the other.  Each molecule
will bounce against the other and then move apart.

The initial configurations at the three scales are shown in Fig.\ref
{figure:face2face}.

\begin{myfigure}
\includegraphics[width=14cm] {\image 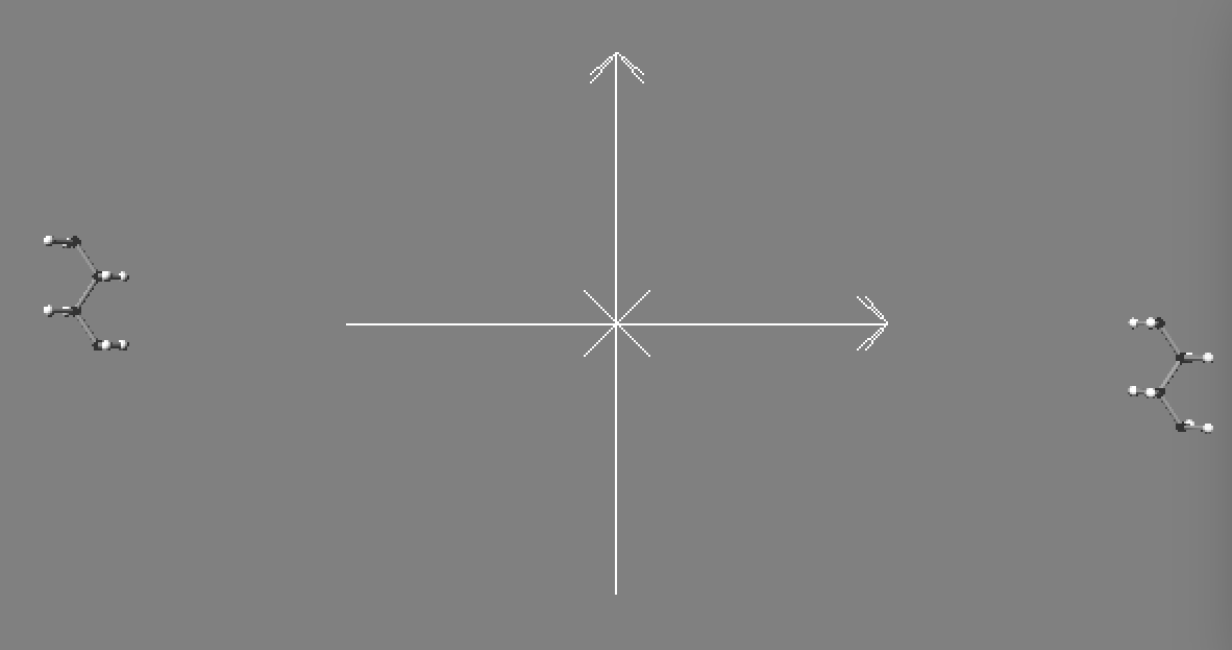}\\
\includegraphics[width=14cm] {\image 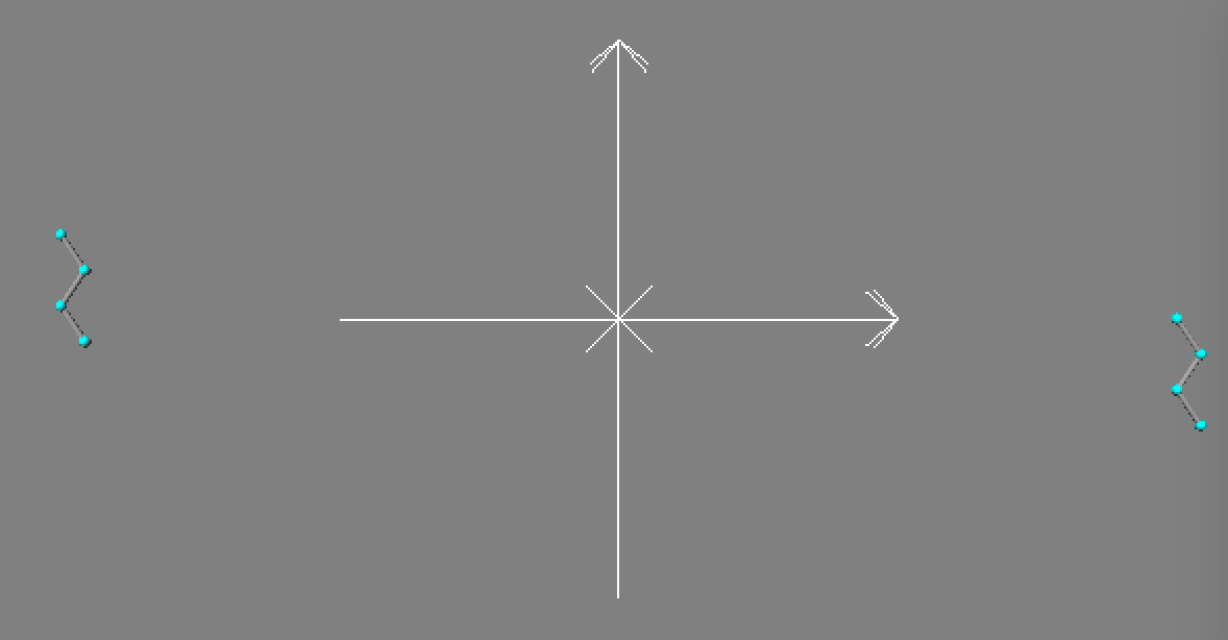}\\
\includegraphics[width=14cm] {\image 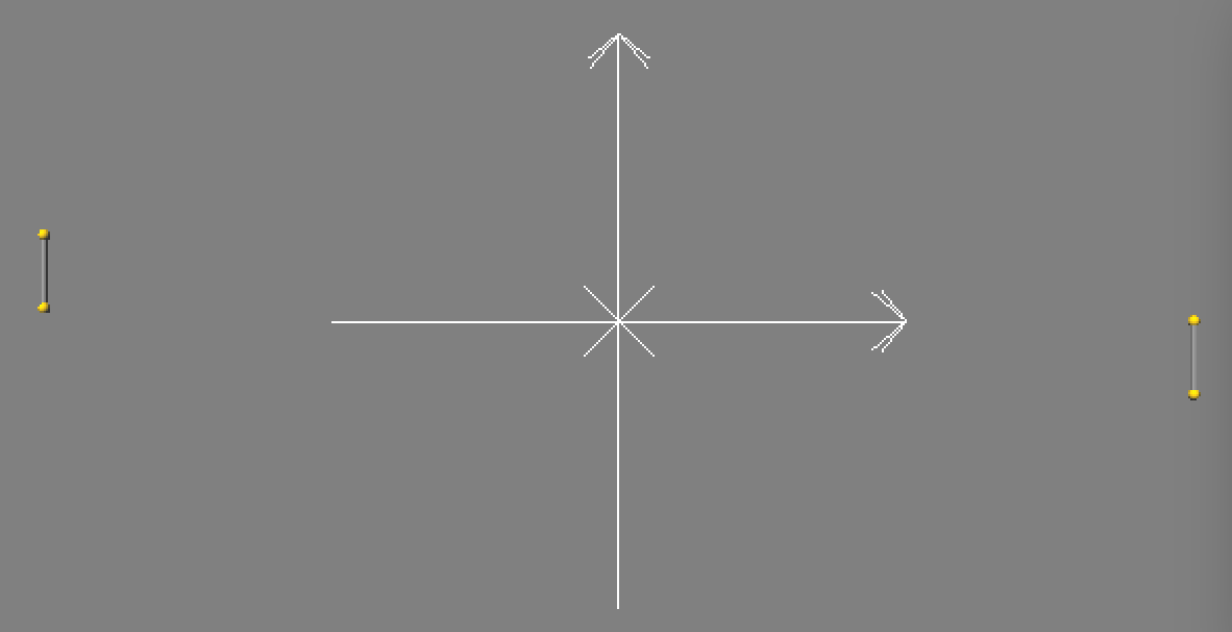}
  \caption {\small Initial configurations of simulations at the three scales.}
\label {figure:face2face}
\source {target/CollisionApp.java dev=0.3 tstep=1E-4 make collision}
\end{myfigure}

All three coordinate axes are present in the three images. At each
scale, the molecule on the left is slightly shifted upwards in $y$ (by
$0.3$ $nm$).  The time step is $10^{-4}$ $ps$ for all three
simulations. The initial distance between each molecule and the centre
of coordinates is $\Delta = 2$ $nm$ and the absolute value of the
initial velocity is $1$ $nm/ps$.

In each simulation, we measure the real-time (in milli-seconds) until
the distance, after bouncing, between the molecules and the center is
greater than $\Delta + 0.5$ $nm$.

Results are as follows: for AA, $13.108$ $ms$; for UA, $2.004$ $ms$;
for CG, $1.144$ $ms$. One observes thus a factor 11 between AA and CG.


\chapter {Inverse-Boltzmann Method}\label{chapter:boltzmannUA}
In this chapter we describe a method, called {\it
  inverse-Boltzmann}, to determine the {\UA} potentials from the ones
at the AA scale. The determination of the {\CG} potentials using the
same method is considered in Chap.\ref{chapter:boltzmannCG}.

The inverse-Boltzmann method is based on a statistical processing of
data obtained during simulations. In this approach, one determines the
potential energy of an oscillator from the probability density of
the oscillator presence in a given state.

To implement the method, we divide the space of variation into $N$
classes of equal size. Each class $C_i$ has an associated counter
$P_i$ which is incremented at each simulation step if the oscillator
value (for example, in the case of a bond, its length) belongs to the
class $C_i$.  More precisely, the formula defining the potential
energy $U_i$ associated with the class $C_i$ is:
\begin{equation}
U_i = -k_B \times T \times ln (P_i / P_0)
\end{equation}
where $k_B$ is the Boltzmann constant, $T$ is the temperature of
simulation, $P_0$ is the value of the counter of the class containing
the maximum number of elements (the ``most populated'' class), and
$P_i$ is the value of the counter of class $C_i$.

Let us now consider the case of an isolated harmonic oscillator. We
find that the maximum density of presence corresponds to the maximum
kinetic energy. Since the oscillator is an isolated system, the state
of maximal kinetic energy is also the state of minimal potential
energy because in an isolated system the sum of the kinetic and
potential energies is constant.

In the inverse-Boltzmann method, we generalise the previous case
of an harmonic oscillator to the various components of molecules.

One uses the same simulation for all molecule components. Three sets
of data will therefore be produced, one for bond lengths, one for
valence angles, and one for torsion angles.

Two points should be immediately stressed:
\begin{enumerate}
  
\item
  The obtained energies depend on the temperature $T$, which thus
  becomes a crucial simulation parameter.
  
\item 
  The inverse-Boltzmann approach only delivers relative energies:
  one needs to state the value of one of the classes (for example, the
  class with the maximal number of elements) in order to be able to
  determine the energies of the other classes.

\end{enumerate}

This approach raises two fundamental questions. First, the components
(for example, the bonds) are not isolated: they continuously exchange
energy with the others oscillators present in the molecule.  To what
extent do these exchanges disturb the method~?

Second, is the relationship between potential energy and probability
density of presence valid for all cases of oscillators~?
 
In the remainder of this chapter, we describe the application of the
inverse-Boltzmann method to the {\UA} scale.

\section{UA Intra-molecular Forces}
The treatment of bonds, valence angles and torsion angles is carried
out using the same molecular dynamics simulation of the molecule
$\alkane{6}{12}$ at the temperature of $\temp {218.152}$.

The simulation begins with $2\times10^{7}$ stabilisation steps without
filling in the classes. Then, classes are filled at the end of
each of the following $10^8$ steps.  The probabilities of presence in
each class are finally calculated at the end of the simulation.

\subsection*{UA Bonds}
The {\UA} bonds are processed by segmenting the domain
of values into 200 evenly distributed classes.
Fig.\ref{UABondBoltzmann} shows the obtained curve.
\begin{myfigure}
  \includegraphics [width=13cm] {\image 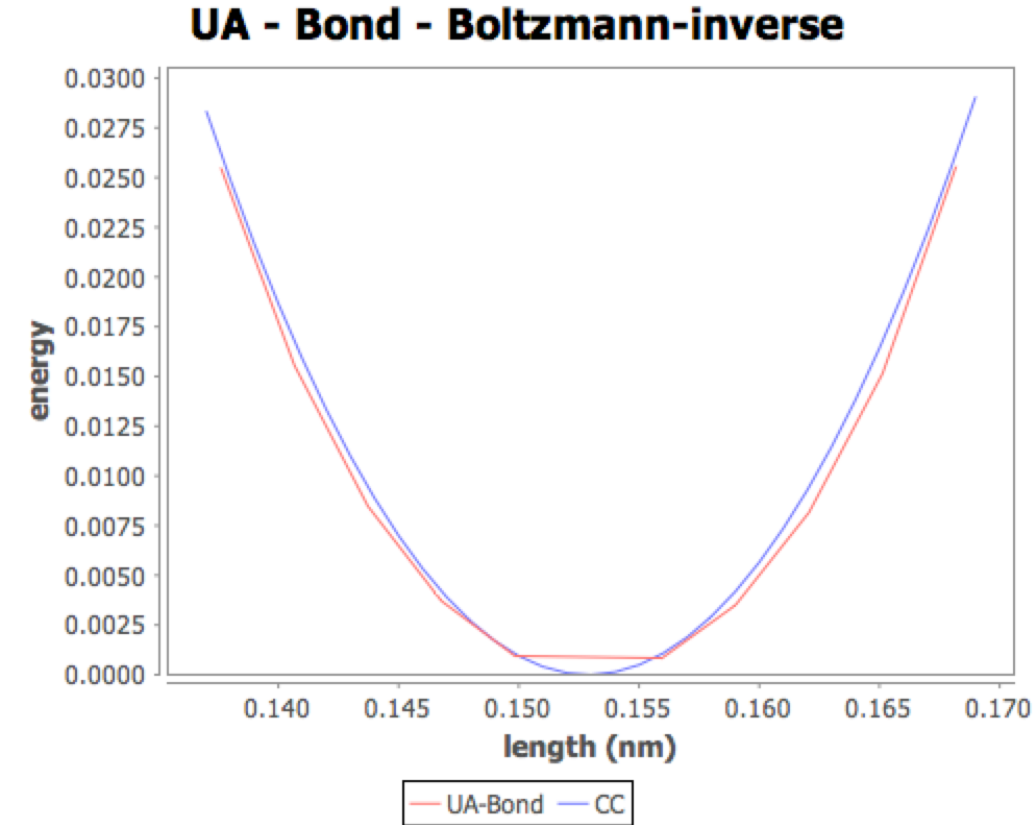}
  \caption{\small
    UA bond potential obtained by inverse-Boltzmann method from the
    molecule $\alkane{6}{12}$ at the temperature of $\temp {218.152}$
    compared with the AA potential of the bond CC.
}
\label{UABondBoltzmann}
\source{boltzmann/BoltzmannCnApp}
\source {curves/BoltzmannUABondApp.java ~~~ make boltzmannuabond}
\end{myfigure}

In the case of UA bonds, the inverse-Boltzmann method gives a good result
which can be made more accurate by increasing the number of
classes or the simulation time.

We can therefore consider that the energy exchanges of UA bonds with
the other components (bonds, valence or torsion angles) do not disturb
the approach which associates a harmonic potential to the UA bond,
very close to that of the CC bond potential at the AA scale.

\subsection*{UA Valence Angles}
For GGG valence angles, the range of variation is segmented into
180 classes. Fig.\ref{UAValenceBoltzmann} shows the obtained curve.
\begin{myfigure}
    \includegraphics [width=13cm] {\image 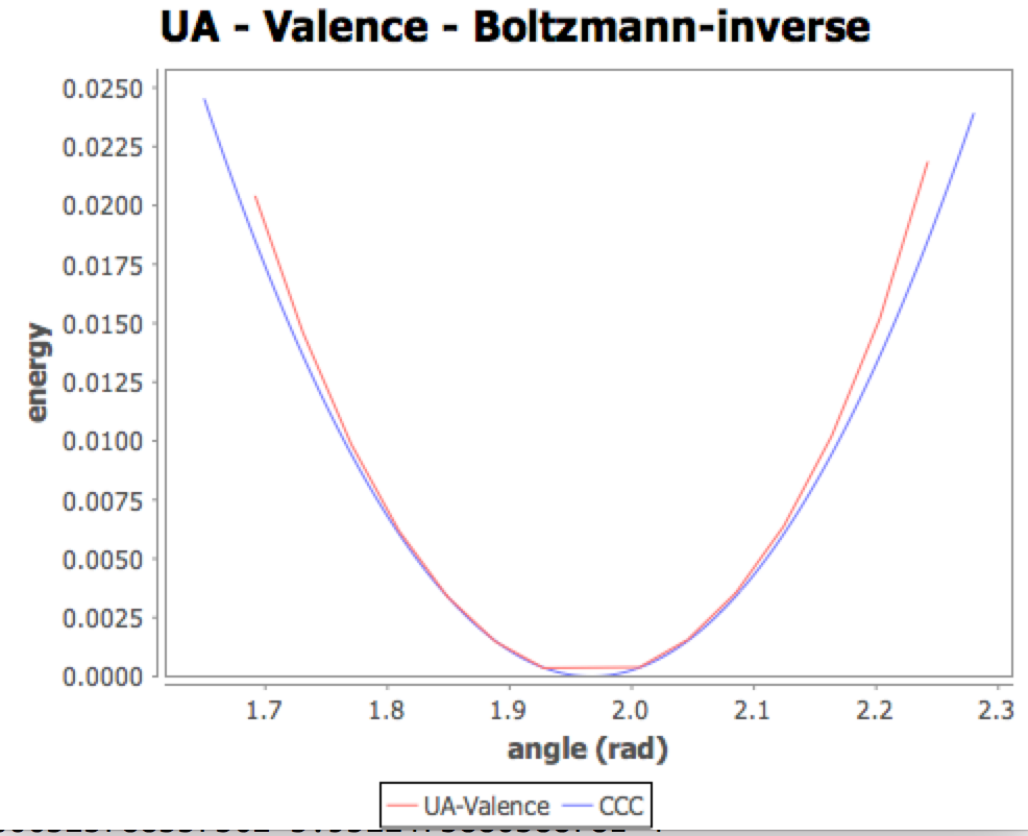} 
    \caption{\small UA valence potential obtained by inverse-Boltzmann
      from the molecule $\alkane{6}{12}$ at the temperature of
      $\temp {218.152}$, compared with the AA potential of the CCC
      valence angle.  }
\label{UAValenceBoltzmann}
\source {curves/BoltzmannUAValenceApp.java ~~~ make boltzmannuavalence}
\end{myfigure}

The result differs slightly from the CCC potential. This discrepancy
will be explained later by the effect of the CCH valence angles, which
are strongly correlated to the valence angle considered.

The exchange of energy between the UA valence angle and the other
molecule components have little effect: the method gives UA valence
angles a harmonic potential very close to that of the AA CCC valence
angle potential.

\subsection*{UA Torsion Angles}
For {\UA} torsion angles, we segment the range of variation into 180
classes.  The {\UA} torsion angle potential is in good correspondence
with the sum of the {\AllAtom} potentials of all the torsion angles
sharing the same central bond.

Fig.\ref{torsion} shows a bond CC (in red).
\begin{myfigure}
    \includegraphics [width=8cm] {\image 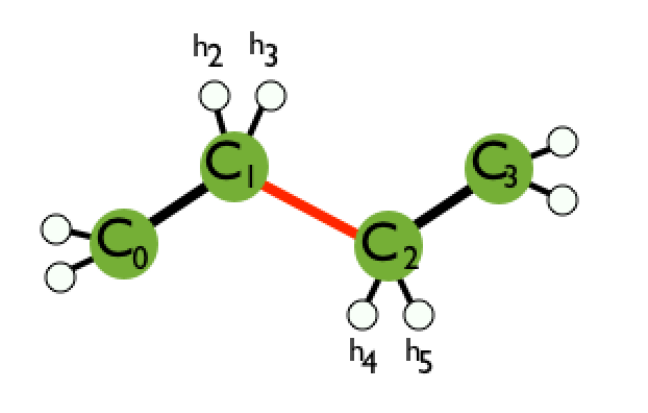} 
    \caption{\small Torsion angles sharing the same central bond (in red).}
\label{torsion}
\end{myfigure}

On Fig.\ref{torsion}, one sees that there are nine torsion angles sharing
the same central bond $C_1C_2$:
\begin{itemize}
\item[$\bullet$] the angle CCCC $C_0C_1C_2C_3$;
\item[$\bullet$] the four angles HCCH:
$h_2C_1C_2h_4$,
$h_3C_1C_2h_4$,
$h_2C_1C_2h_5$,
$h_3C_1C_2h_5$~;
\item[$\bullet$] the four angles HCCC:
$h_2C_1C_2C_3$,
$h_3C_1C_2C_3$,
$C_0C_1C_2h_4$,
$C_0C_1C_2h_5$.
\end{itemize}

Fig.\ref{UATorsionBoltzmann} shows the ptential of the UA torsion angle
obtained by inverse-Boltzmann and compares it with the sum of the
potentials of the nine torsion angles sharing the same central
bond. One observes that the two curves are in good correspondance.

\begin{myfigure}
    \includegraphics [width=13cm] {\image 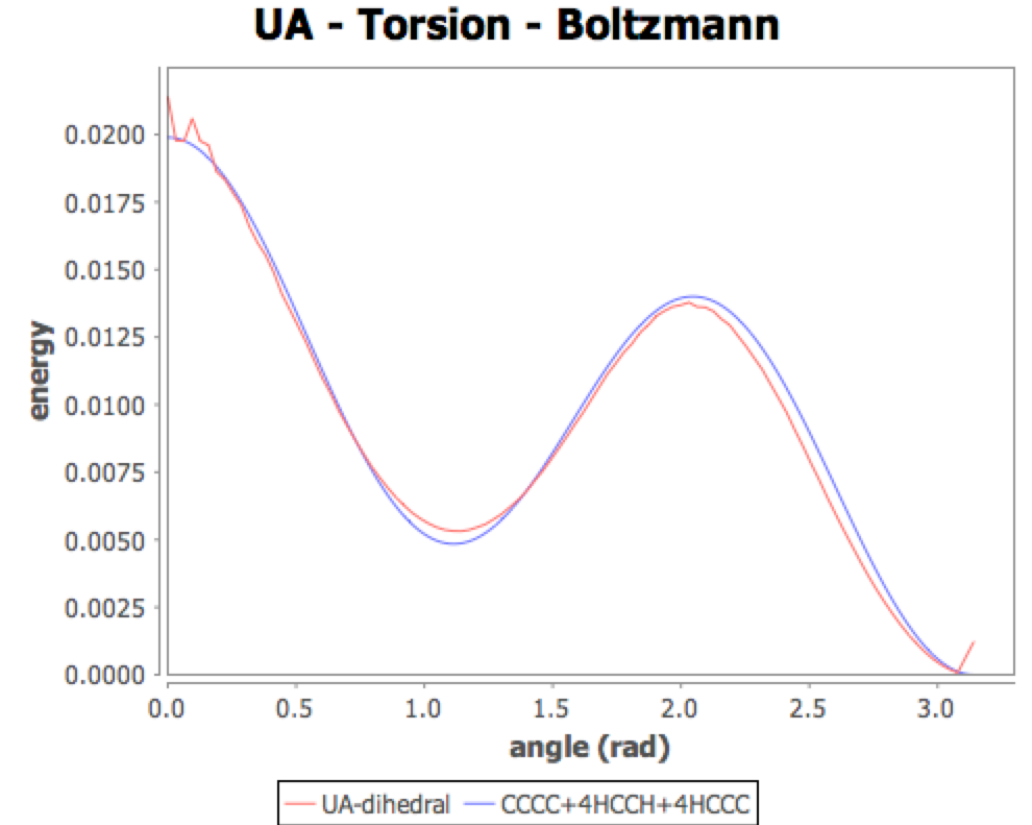} 
    \caption{\small UA torsion angle potential obtained by
      inverse-Boltzmann from the molecule $\alkane{6}{12}$ at a temperature
      of $\temp {218.152}$.  }
\label{UATorsionBoltzmann}
\source {curves/BoltzmannUADihedralApp.java ~~~ make boltzmannuadihedral}
\end{myfigure}

Thus, the energy exchanges of the torsion angle UA with the other
components of the molecule have little effect. The torsion angle UA
potential is in accordance with the sum of the AA potentials of the
torsion angles associated.

\section {UA Inter-Molecular Forces}
The data obtained with the inverse-Boltzmann method from the
simulation of two {\UA} grains is given in Fig.\ref{UALjBoltzmann}.
\begin{myfigure}
    \includegraphics [width=13cm] {\image 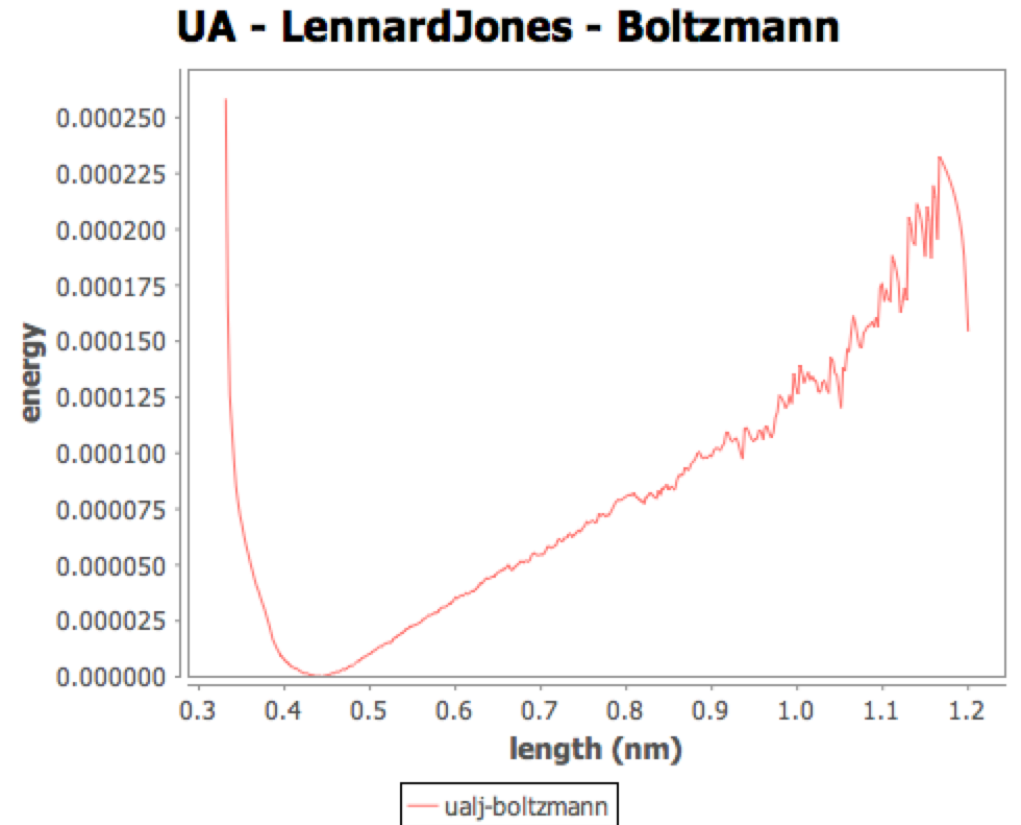} 
    \caption{\small UA inter-molecular potential obtained by the
      inverse-Boltzmann method from two molecules $\alkane{1}{2}$.  }
\label{UALjBoltzmann}
\source {curves/VisualiseUALjBoltzmannApp make ualjboltzmann}
\end{myfigure}

It can be seen that the right-hand side of the inverse-Boltzmann curve
is not asymptotic to a parallel to the $y=0$ axis but, on the
contrary, tends to increase with positive values of $y$.  This is not
consistent with the dynamics: one indeed expects that the potential
becomes weaker and weaker as the grains move further apart.

The curve obtained with the inverse-Boltzmann method is thus clearly
not that of an inter-molecular potential.

Let us try to explain what happens by considering rare
events. Actually, we have two disjoint sets of rare events: events
corresponding to the strongly repulsive domain, when the grains are
very close, on the one hand; on the other hand, events corresponding
to the weakly attractive domain, when the grains are very far apart.

Thus, in the first case the potential must be very high, while in
the second case it must be very low.  However, the inverse-Boltzmann method
only takes into account the relative rarities of events and therefore
tends to identify these events giving them the same energy, which
clearly contradicts the dynamics.

The inverse-Boltzmann method for determining the inter-molecular
potential {\UA} therefore stumbles on a major obstacle.  Looking
ahead, we see that the preceding reasoning also applies to the
determination of the {\CG} inter-molecular potential.  Actually, the
inverse-Boltzmann method seems unsuitable for the determination of
inter-molecular potentials in general.


\chapter {Inverse-Boltzmann for CG}\label{chapter:boltzmannCG}
We apply now the inverse-Boltzmann method to the {\CG} scale.

The inverse-Boltzmann method being unsuitable for the treatment of
inter-molecular forces (cf. Chap.\ref{chapter:boltzmannUA}), we will
only consider the intra-molecular {\CG} components.  The treatments of
{\CG} bonds and valence angles are based on the same molecular
dynamics simulation of the $\alkane{6}{12}$ molecule, as in
Chap.\ref{chapter:boltzmannUA}.

The simulation starts with $2\times10^{7}$ stabilisation steps,
without filling-in the classes. Then, classes are filled-in at the end
of each of $10^8$ steps. Finally, the probabilities of presence in
each class are calculated at the end of the simulation.

\section{CG Bonds}
CG bonds are processed by segmenting the range of variation of the
values into 200 evenly distributed classes.
Fig.\ref{C6H12BondBoltzmann} shows the obtained curve for a
temperature of $\temp {218.152}$.
\begin{myfigure}
  \includegraphics [width=13cm] {\image 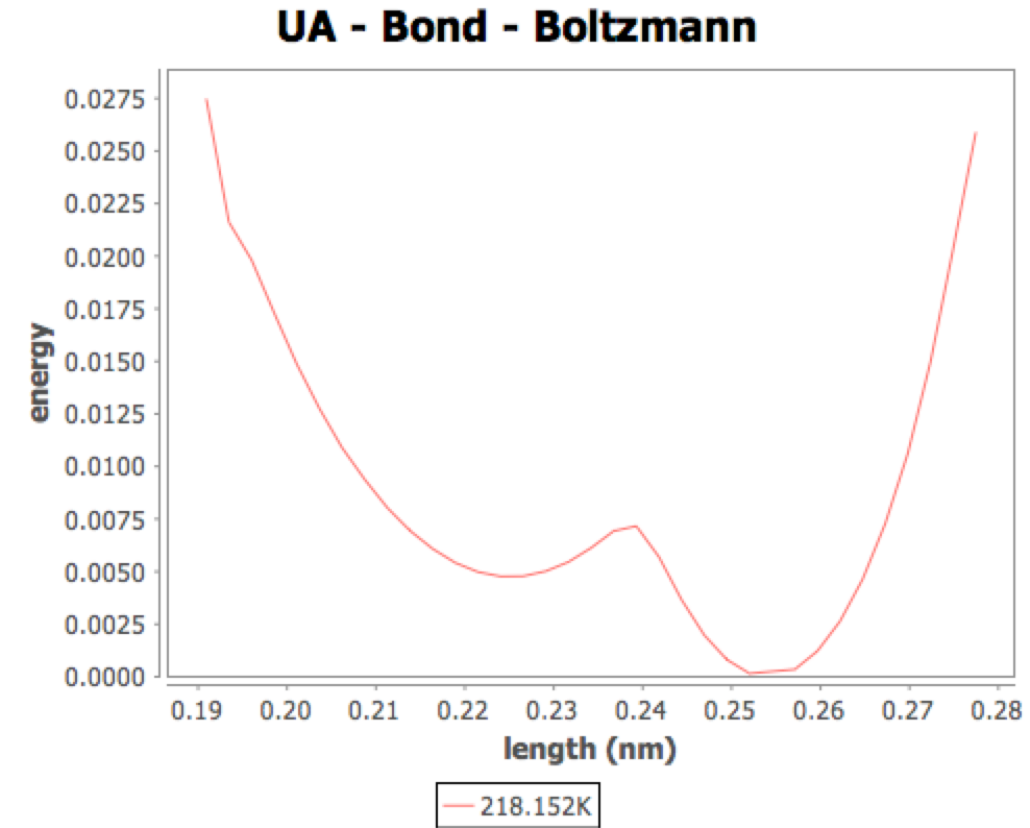}
  \caption{\small CG bond potential obtained by the inverse-Boltzmann method
    from the molecule $\alkane{6}{12}$ at the temperature of
    $\temp {218.152}$.  }
\label{C6H12BondBoltzmann}
\source {curves/VisualiseCgBondApp.java boltzmann=true make cgbond}
\end{myfigure}
The curve obtained depends on the simulation temperature, as shown in
Fig.\ref{bondSenseTemp} where three different temperatures are
considered.
\begin{myfigure}
    \includegraphics [width=13cm] {\image 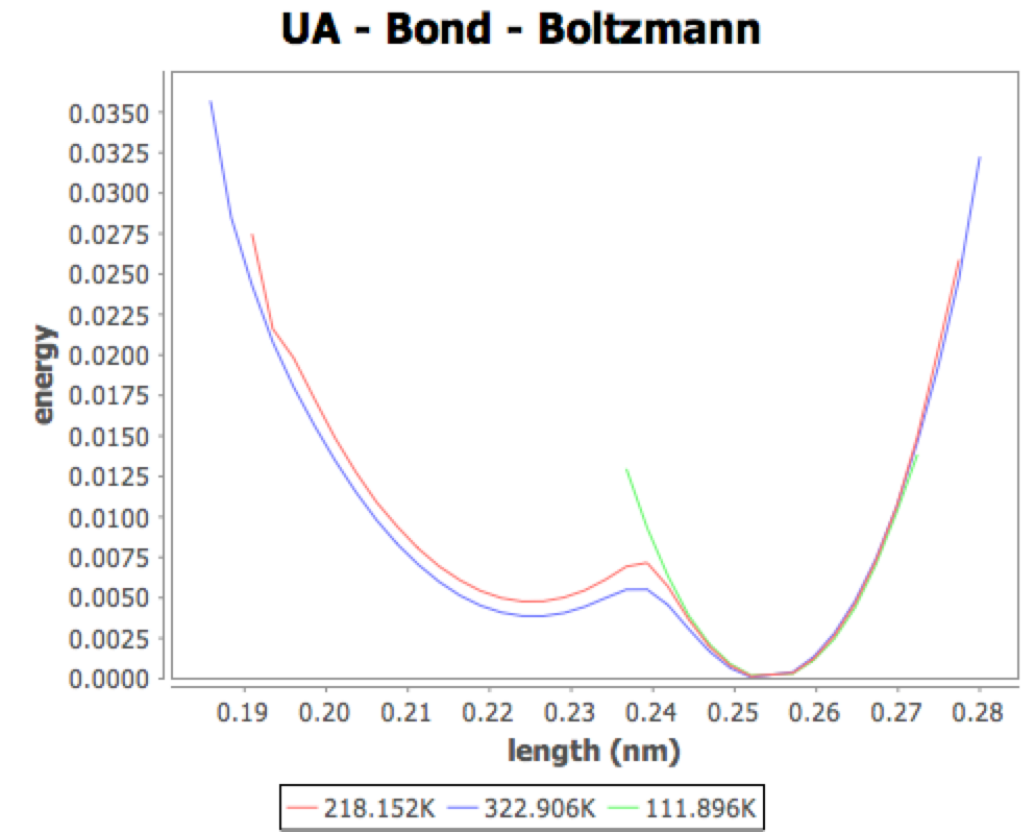} 
    \caption{\small CG bond potential determined at three different temperatures.}
\label{bondSenseTemp}
\source {curves/VisualiseCgBondApp.java boltzmann=true make cgbond}
\end{myfigure}
At the temperature of $\temp{111}$, the domain covered does not
contain the inflection point: the central torsion angle is never
triggered. At the temperature of $\temp{218}$, one observes the
torsion of the central angle.

At a temperature of $\temp{322}$ we can see that that the potential is
lower at the inflection point. At higher temperatures the inflection
becomes less and less marked.

Here, we are faced with a difficulty: what temperature should we
choose to determine the {\CG} bond potential~? The inverse-Boltzmann
method does not provide any answer to this question.

\section{CG Valence Angles}
For CG valence angles, the range of variation is segmented into 180
classes. Fig.\ref{C6H12ValenceBoltzmann} shows the curve obtained.
\begin{myfigure}
    \includegraphics [width=13cm] {\image 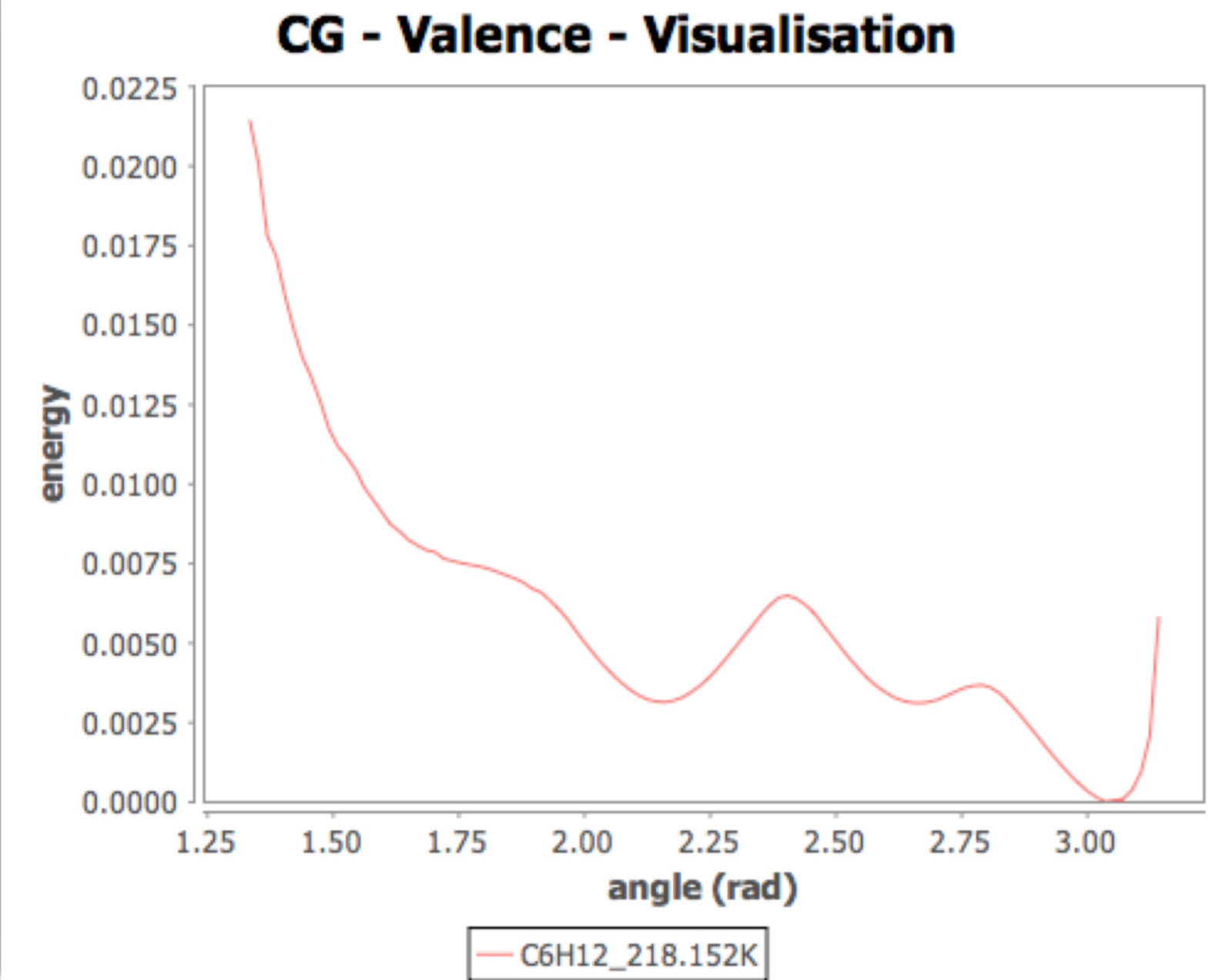} 
    \caption{\small CG valence angle potential obtained by
      inverse-Boltzmann from the molecule $\alkane{6}{12}$ at the
      temperature of $\temp {218.152}$.  }
\label{C6H12ValenceBoltzmann}
\source {curves/VisualiseCgValenceApp.java boltzmann=true make cgvalence}
\end{myfigure}
The right-hand side of the curve (angle greater than $3$ radians) is
clearly wrong. In fact, the potential energy is minimal when the CG
grains are aligned, which corresponds to a CG valence angle equal to
$\pi$ and this is not what the curve shows. This anomaly occurs
independently of the temperature.

In addition, the curve of the {\CG} valence angle varies according to
the temperature: in the same way as for {\CG} bonds, it is not clear
which temperature to choose for determining the {\CG} valence
potential.

\section* {Conclusion}
Several conclusions can be drawn concerning the inverse-Boltzmann
method:
\begin{itemize}
\item[$\bullet$] Correct handling of rare events requires large
  numbers of events, and thus simulations that take very long execution
  times.

\item[$\bullet$] When the number of events is not sufficient to
  process correctly a range of values, the potential energy may be
  underestimated.

\item[$\bullet$] The curves obtained may depend on the temperature,
  which raises the question of the choice of temperature to consider.

\item[$\bullet$] In addition, the simulation temperature limits the
  range of values analysed and therefore the range of
  definition of the potential function.

\item[$\bullet$] Finally, the method gives wrong answers in the case
  of inter-molecular potentials.
\end{itemize}
  
We will now consider another means of determining potentials: the
minimisation method.


\chapter {Minimisation Method}\label{chapter:minimisationUA}
In this chapter, we propose a method for determining the potentials of
molecular components based on a {\it constrained minimisation}
technique. This method is fundamentally based on the existence of
geometric links with the basic scale {\AllAtom}.

To determine the potential of a molecular component (for example, the
{\UA} valence angle), we choose a molecule in which this component
appears and set the value $v$ of the component.  We then calculate the
global potential $p$ of the underlying molecule (geometrical aspect)
at the scale {\AllAtom}, after performing a minimisation process which
preserves the value $v$ of the component (constrained aspect).  The
potential of the component is by definition $p$ for the component
value $v$.  By varying $v$, we determine the potential we are looking
for.

For example, for the {\UA} valence angle, we fix the angle between
three UA grains, i.e. between three carbon atoms, and we minimise all
the AA components of the molecule except this angle.

Thus, the carbon atoms of the three grains of the UA angle remain
immobile, while the other atoms, either carbons or hydrogens, move to
minimise the energy of the overall molecule.

The bonds and angles involving the hydrogen atoms are thus placed in
equilibrium positions where their energy is minimal. The same applies
to carbon atoms, with the exception of those forming the UA valence
angle. The final energy of the AA molecule after complete minimisation
is that associated with the UA valence angle.

In the remainder of this chapter, we apply the minimisation method to
the {\UA} scale. Recall that a {\UA} grain is formed by a carbon atom
and the hydrogen atoms bonded to it.  The centre of a UA grain
coincides with the carbon atom.  The minimisation applied to CG is
described in Chap.\ref{chapter:minimisationCG}.

\section{UA Bonds}
To deal with the {\UA} potential of the $G_2G_2$ bond, we consider the
equivalent molecule (in fact, a fragment) $\alkane{2}{4}$ and the
minimisations are carried out by keeping constant the distance between
the two carbons of the molecule.

The match with the OPLS potential {\AllAtom} is perfect and is
illustrated in Fig.\ref {figure:bond-validation}.
\begin{myfigure}
\includegraphics[width=13cm] {\image 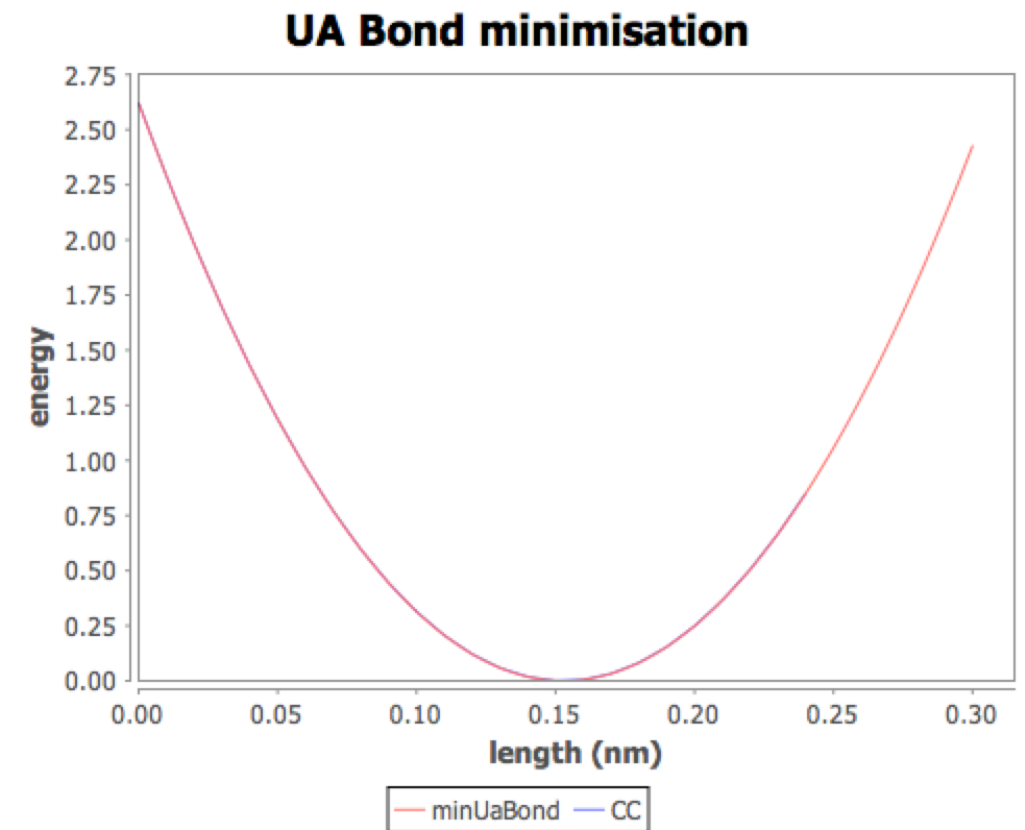}
\caption {\small Potential of the bond $G_2G_2$ obtained by 
  minimisation, compared with the OPLS potential CC.}
\label {figure:bond-validation}
\source {curves/VisualiseUABondApp.java make uabond}
\end{myfigure}

This result reflects the independence that exists between the CC bond
and the valence angles HCH and HCC in the molecule $\alkane{2}{4}$:
the HCH and HCC angles can be at their equilibrium values
independently of the bond value.

\section{UA Valence Angles}\label{section:valenceUA}
For the UA valence potential $G_2G_2G_2$, we consider the fragment
$\alkane{3}{6}$ and the minimisations are carried out keeping the
valence angle CCC constant (remark: the results are exactly the same
with the full molecule $\alkane{3}{8}$).

The minimisation result is shown in
Fig.\ref{figure:valence-validation}.  It can be seen that the curve
produced by the minimisation methord does not exactly coincide with
the CCC potential, in particular in the part corresponding to the
extension of the valence angle.

\begin{myfigure}
\includegraphics[width=13cm] {\image 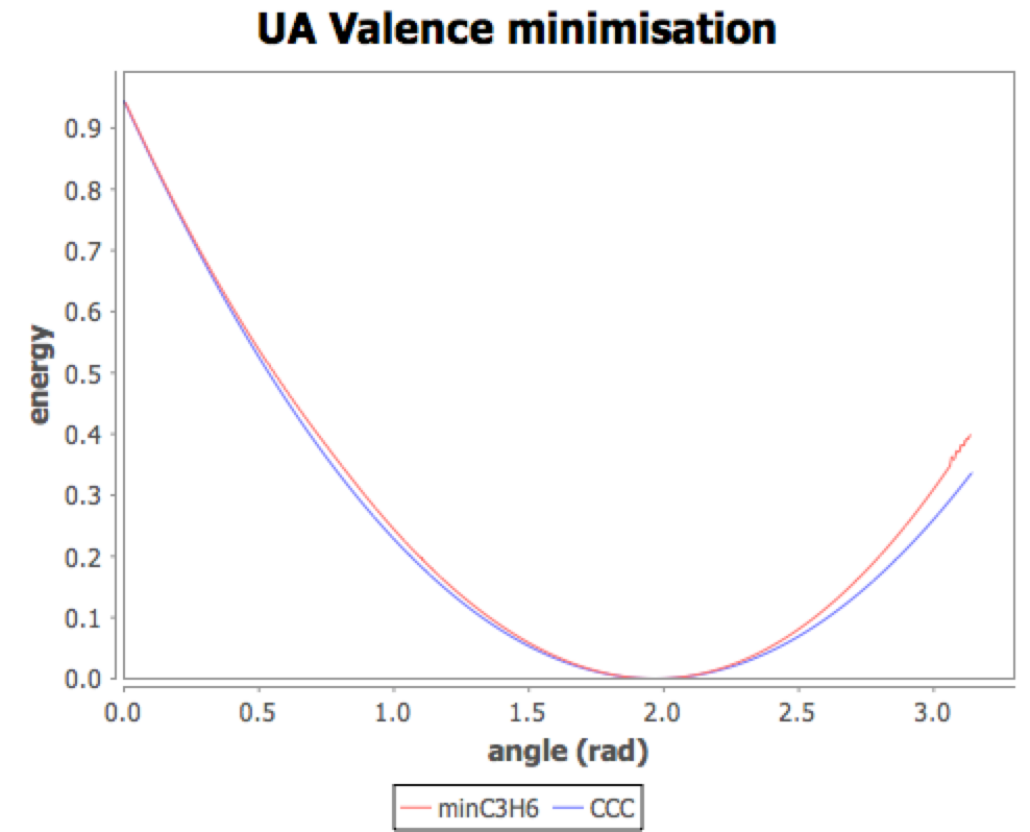}
\caption {\small Potential of the valence angle $G_2G_2G_2$ obtained by
  minimisation, compared with the OPLS potential CCC.}
\label {figure:valence-validation}
\source {curves/VisualiseUAValenceApp.java corrected=false
  make aavalence}
\end{myfigure}

The potential {\UA} for the valence angle $G_2G_2G_2$ therefore
differs from the {\AllAtom} potential of the valence angle CCC.

By removing the potential energy of the eight valence angles CCH one
exactly recovers the CCC potential, as shown in Fig.\ref
{figure:valence-validation-cch}.

\begin{myfigure}
\includegraphics[width=13cm] {\image 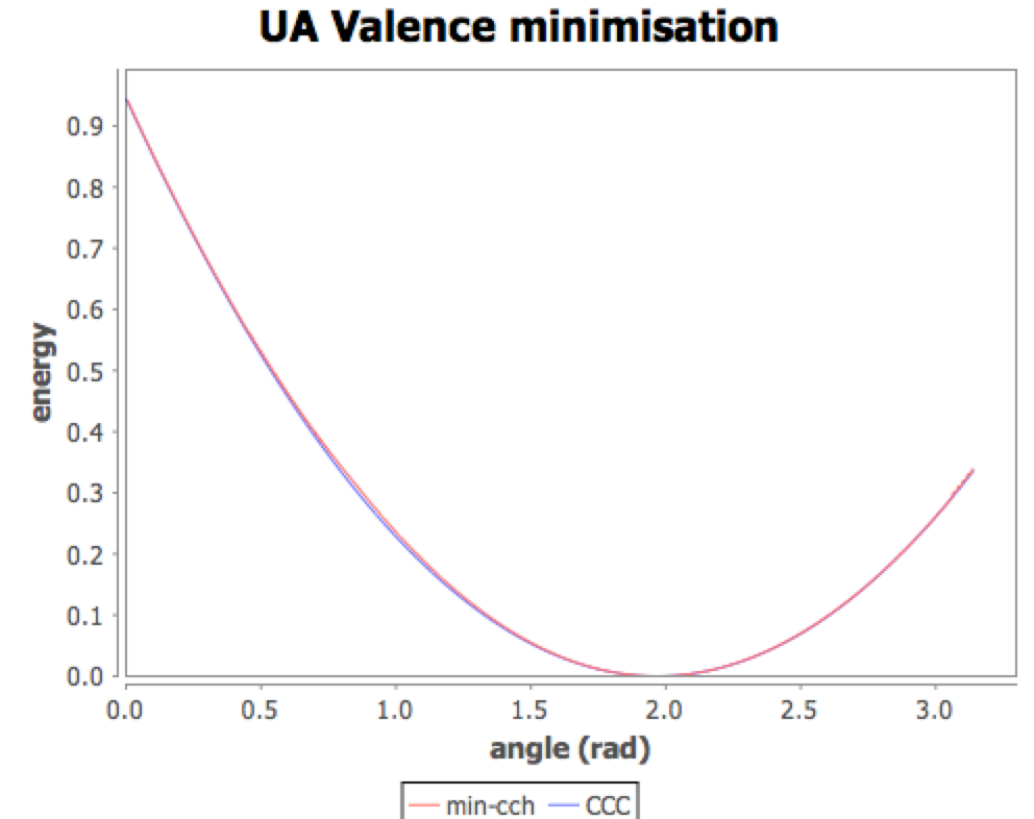}
\caption {\small Potential of the valence angle $G_2G_2G_2$ obtained
  by minimisation, removing the components of the valence angles CCH,
  compared eith the OPLS potential CCC.}
\label {figure:valence-validation-cch}
\source {curves/VisualiseUAValenceApp.java corrected=true
  make aavalence}
\end{myfigure}

The difference with the valence potential CCC lies in the partial
correlation (actually, only in extension of the valence angle) between
the angle CCC and the angles CCH.

The potential $G_2G_2G_2$ is in fact the sum of two functions, the
first $f_{inf}$ defined for angles smaller than the angle of
equilibrium, the second $f_{sup}$ defined for bigger angles. The
$f_{inf}$ function coincides on its definition space with the harmonic
function $CCC$, while $f_{sup}$ multiplies $CCC$ by a factor of
$1.142$.

The two functions $f_{sup}$ and $f_{inf}$ can be both approximated by
a single harmonic function which is the harmonic function $CCC$
multiplied by a factor of $1.1$.  We will use this approximation in
the following; the UA valence potential $G_2G_2G_2$ will be considered
as being a harmonic potential equal to the valence potential $AAA$
multiplied by the factor $1.1$.

\section{UA Torsion Angles}\label{section:miniUA:torsion}
To deal with the torsion potential $G_2G_2G_2G_2$, we start from the
fragment $\alkane{4}{8}$ and carry out the minimisations keeping the
torsion angle CCCC constant (results are exactly the same with the
molecule $\alkane{4}{10}$).

The CCCC torsion angle is in fact totally correlated with the HCCH and
HCCC torsion angles centred on the middle of the central CC bond.
Four HCCH torsion angles and four HCCC torsion angles are thus
involved.  Fig.\ref {figure:torsion-validation} shows the potential
obtained by minimisation compared to the sum CCCC+4HCCH+4HCCC.  It can
be seen that the two curves exactly correspond to each other.

\begin{myfigure}
\includegraphics[width=13cm] {\image 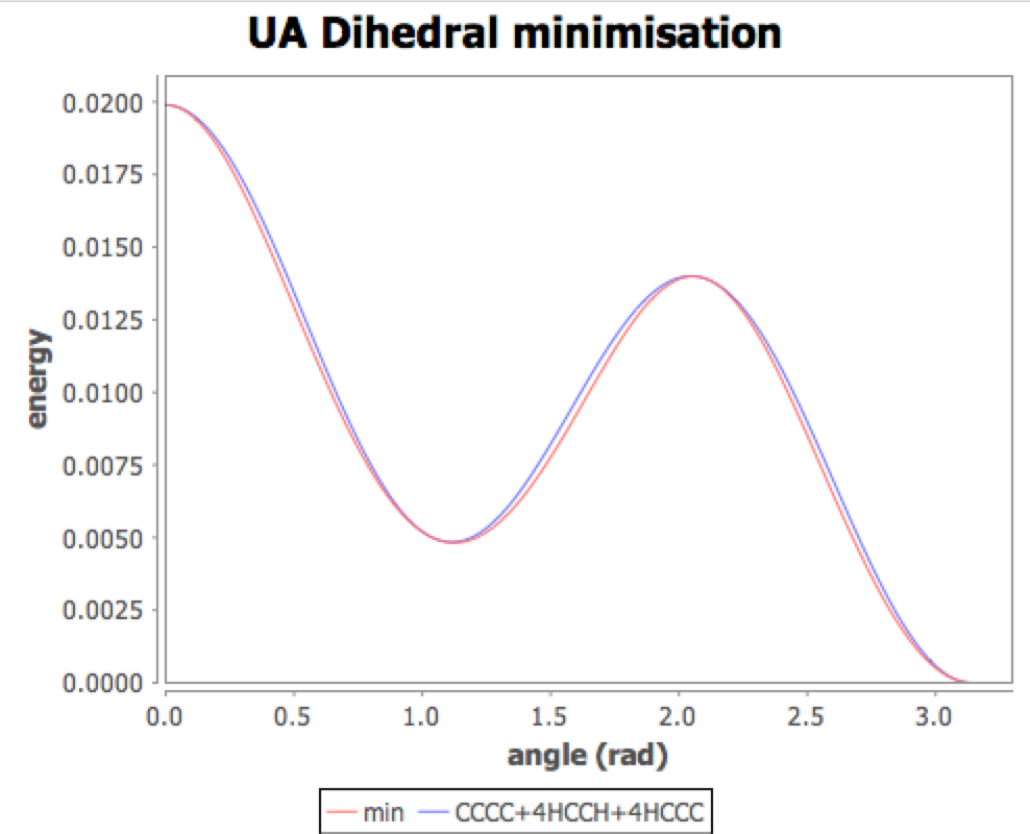}
\caption {\small Potential of the torsion angle $G_2G_2G_2G_2$
  obtained by minimisation, compared with the sum of OPLS potentials
  CCCC+4HCCH+4HCCC.}
 \label {figure:torsion-validation}
\source {curves/VisualiseUADihedralApp.java make uadihedral}
\end{myfigure}

\section{UA Inter-molecular Forces}\label{section:miniUA:intermol}
To determine the Lennard-Jones potential between two {\UA} grains, we
use two $\alkane 1 2$ molecules placed face-to-face. The curve
obtained by varying the distance between the two molecules is shown on
Fig.\ref{figure:lj-minimisation}.

\begin{myfigure}
\includegraphics[width=13cm] {\image 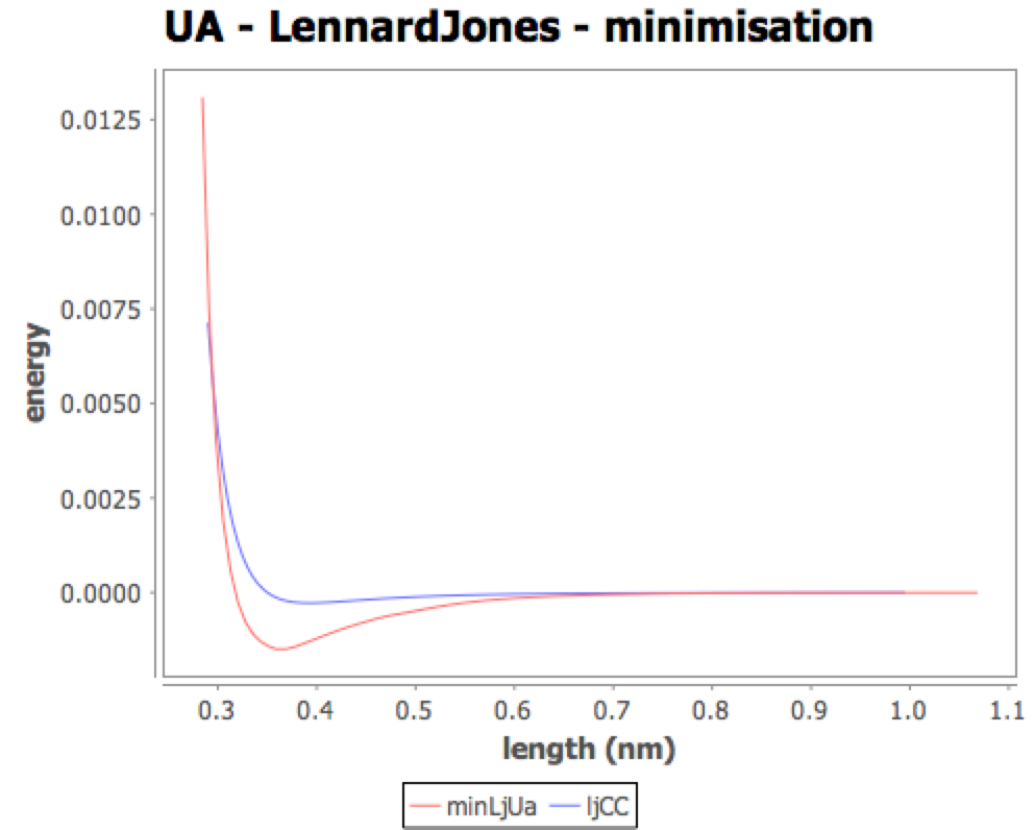}
\caption {\small Inter-molecular potential between two UA grains
  obtained by minimisation, compared with the OPLS Lennard-Jones
  potential between two carbon atoms.}
\label {figure:lj-minimisation}
\source{curves/VisualiseUAljApp.java aa=true make ualj}
\end{myfigure}

We obtain a Lennard-Jones potential which takes into account the
presence of hydrogens, as can be seen on
Fig.\ref{figure:lj-ua-approx6-12}.

\begin{myfigure}
\includegraphics[width=13cm] {\image 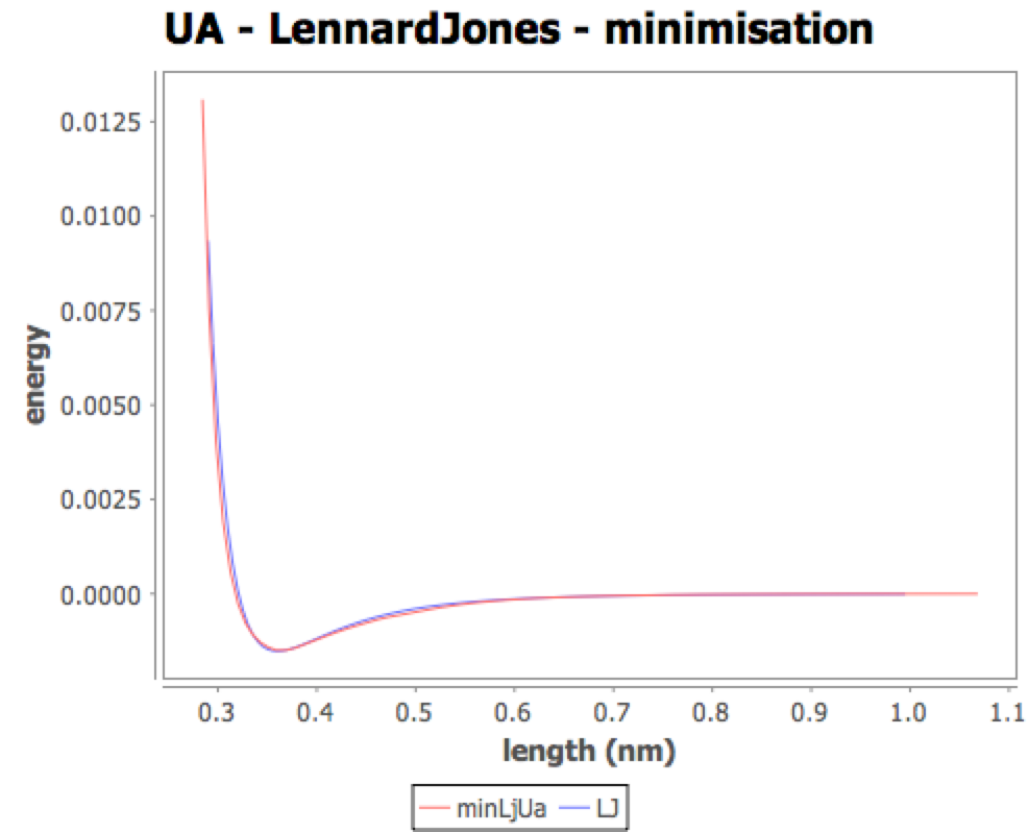}
\caption {\small Inter-molecular potential between two UA grains
  obtained by minimisation, compared with a Lennard-Jones curve.}
\label {figure:lj-ua-approx6-12}
\source{curves/VisualiseUAljApp.java approx=true make ualj}
\end{myfigure}

In Fig.\ref {figure:lj-ua-approx6-12}, the parameters of the
Lennard-Jones potential are $\epsilon = 5.5 \times \epsilon_{CC}$ and
$\sigma = 0.917 \times \sigma_{CC}$.

\section* {Conclusion}
The fragment $\alkane 2 4$ is used to calculate the {\UA} bond and we
obtain exactly the same potential as that of the {\AllAtom} CC bond
(Fig.\ref{figure:bond-validation}). In other words, the bond
potentials {\AllAtom} and {\UA} coincide and are both harmonic.

The fragment $\alkane 3 6$ is used to determine the {\UA} valence
angle potential.  (Fig.\ref{figure:valence-validation}). The {\UA}
valence potential is different (and non-harmonic) but nevertheless
very close to the {\AllAtom} valence potential.

From the fragment $\alkane 4 8$ we obtain the curve in
Fig.\ref{figure:torsion-validation} as the potential of the {\UA}
torsion angle.  The {\UA} potential of a torsion angle $\theta$ is
therefore the sum of the CCCC potentials of the $\theta$ angle, plus
four times the HCCH potential of $\theta$, plus four times the
HCCC potential of $\theta$.

Finally, Fig.\ref{figure:lj-minimisation} shows the potential {\UA}
between two $G_2$ grains.  This potential has three main
characteristics: it has the form of a Lennard-Jones potential; it
coincides with the inter-molecular potential AA when the grains are
very close or sufficiently far apart; its central part is clearly more
``excavated'' than the inter-molecular potential AA.

Preservation of the Lennard-Jones shape indicates that the UA
inter-molecular potential behaves globally like the inter-molecular
potential AA, i.e. the presence of hydrogen atoms does not change the
shape of the interactions between UA grains.

The ``excavated'' aspect is linked to the hydrogen atoms
which, at a short distance, contribute significantly to the attractive
forces between grains.


\chapter {Minimisation for CG}\label{chapter:minimisationCG}
As we have seen in Sec.\ref{echelles:cg}, the determination of the CG
potential by minimisation should be broken down into three aspects
only: binding potential, valence potential and inter-molecular
potential; no torsion potential should be considered at the {\CG}
level.

\section {CG Bonds}
To determine the potential of  {\CG} bonds, one uses the fragment
$\alkane{4}{8}$ of  Fig.\ref{CG4}.

\begin{myfigure}
  \includegraphics [width=10cm] {\image 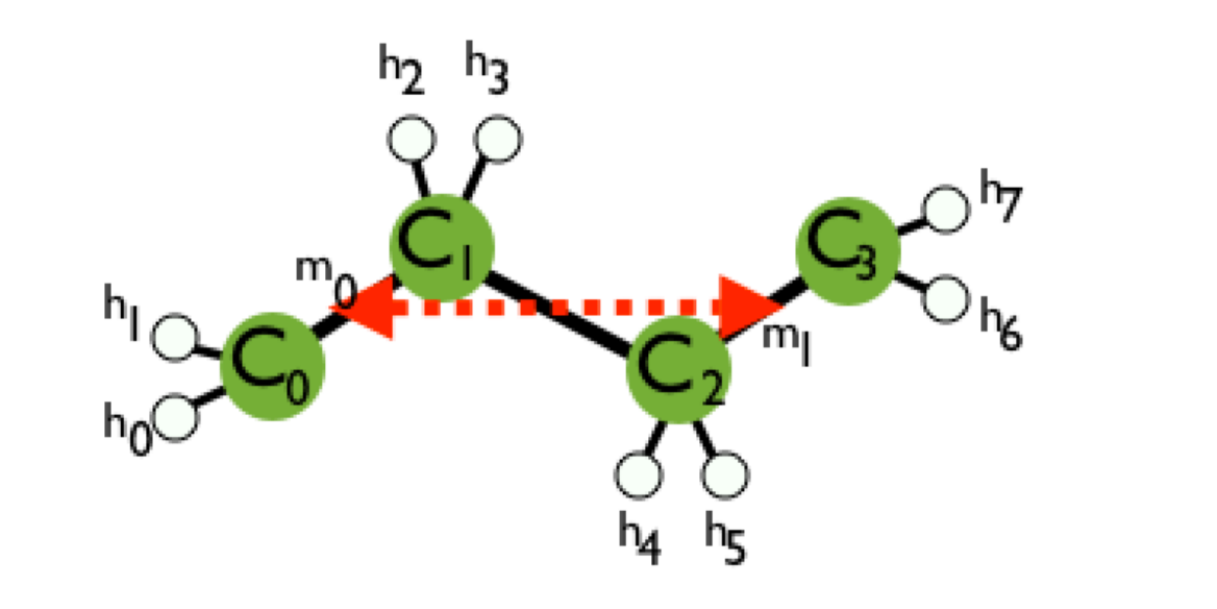}
    \caption{\small Molecule $\alkane{4}{8}$.The CG bond is in red.}
    \label{CG4}
    \source{}
\end{myfigure}

This molecule is considered to be made up of two grains: the first
comprises the two carbons $C_0$ and $C_1$ and the hydrogens attached
to them, and the second comprises the two carbons $C_2$ and $C_3$ and
their attached hydrogens.  The first grain is centred on the middle
$m_0$ of the AA bond between $C_0$ and $C_1$. The second grain is
centred on the middle $m_1$ of the AA bond between $C_2$ and $C_3$.
The CG bond links the two middles $m_0$ and $m_1$.

The CG bond potential is calculated by minimising the energy of the
molecule while keeping constant the length $l$ between $m_0$ and $m_1$
(constrained minimisation).

We begin by defining the energy $eg_0$ of the $G_0$ grain centred on
$m_0$ as the sum of the energies of the internal components of the
grain:

\begin{itemize}
\item[$\bullet$]  bonds (5):
  $h_0C_0$,
  $h_1C_0$,
  $C_0C_1$, 
  $C_1h_2$,
  $C_1h_3$;
\item[$\bullet$]  valence angles (6):
  $h_0C_0C_1$, $h_0C_0h_1$,
  $h_1C_0C_1$, 
  $h_2C_1C_0$, $h_2C_1h_3$,
  $h_3C_1C_0$;
\item[$\bullet$]  torsion angles (4):
  $h_0C_0C_1h_2$, $h_0C_0C_1h_3$,
  $h_1C_0C_1h_2$, $h_1C_0C_1h_3$.
\end{itemize}

We define in the same way the energy $eg_1$ of the grain $G_1$
centered on $m_1$.

The binding energy $el$ between the two grains is the sum of the
energies of the components connecting the two grains:

\begin{itemize}
\item [$\bullet$] bond (1):
     $C_1C_2$;
  \item[$\bullet$]  valence angles (6):
    $C_0C_1C_2$,
    $h_2C_1C_2$,
    $h_3C_1C_2$,
    $C_1C_2C_3$, $C_1C_2,h_4$, $C_1C_2h_5$;
    
  \item[$\bullet$]  torsion angles (13):
    $h_0C_0C_1C_2$,
    $h_1C_0C_1C_2$,
    $C_0C_1C_2C_3$, $C_0C_1C_2h_4$, $C_0C_1\-C_2h_5$,
    $h_2C_1C_2h_4$, $h_2C_1C_2h_5$, $h_2C_1C_2C_3$,
    $h_3C_1C_2h_4$, $h_3C_1C_2h_5$, $h_3C_1C_2C_3$,
    $C_1\-C_2C_3h_6$,
    $C_1C_2C_3h_7$.
  \end{itemize}

  The (potential) energy $ep$ of the molecule $\alkane{4}{8}$ is the
  sum of the energies of its 50 AA components (bonds, valence angles
  and torsion angles). This energy must be equal to the sum of $eg_0$,
  $eg_1$ and $el$:
  \begin{equation}
  ep = eg_0 + eg_1 + el
  \end{equation}

  \section {CG Valence Angles}
  To determine the CG valence potential, we use the molecule
  (fragment) $\alkane{6}{12}$ made up of three grains (Fig.\ref{CG6}).
  The first grain comprises the carbons $C_0$ and $C_1$, the second
  grain the carbons $C_2$ and $C_3$, and the third grain the carbons
  $C_4$ and $C_5$.
  
  The first grain is centered on the middle $m_0$ of the AA bond
  $C_0C_1$; the second grain on the middle $m_1$ of the AA bond
  $C_2C_3$; and the third grain in the middle $m_2$ of the AA bond
  $C_4C_5$.
  
\begin{myfigure}
  \includegraphics [width=11cm] {\image 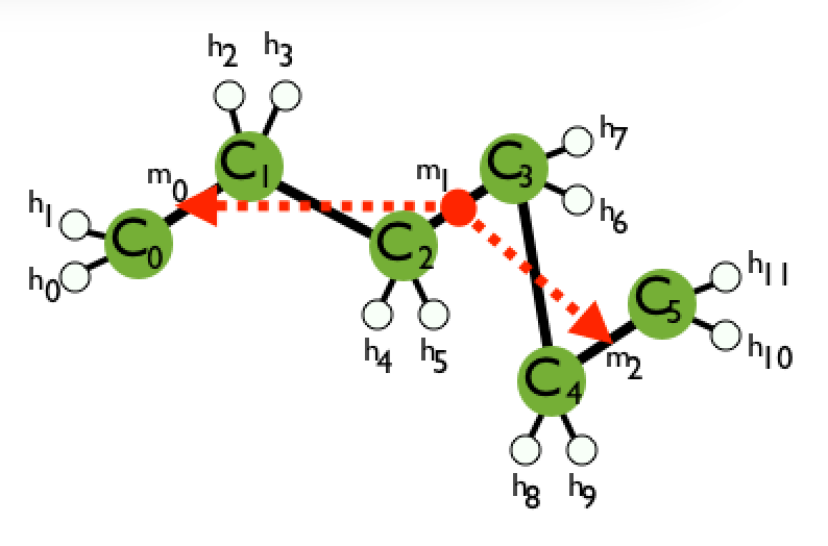}
  \caption{\small Molecule $\alkane{6}{12}$. The CG valence angle is
    the angle formed by the two red arrows.}
  \label{CG6}
\end{myfigure}

The CG valence potential is calculated by minimising the energy of the
molecule while keeping constant the angle formed by the three centers
$m_0$, $m_1$, and $m_2$ (so this is also a case of constrained
minimisation).

We define the angle energy $ea$ of the three grains as that of the
central torsion angle $C_1C_2C_3C_4$.

The potential energy $ep$ of the molecule $\alkane{6}{12}$ is the sum
of the energies of its 86 AA components (bonds, valence angles, and
torsion angles).

The potential energy must be equal to the sum of the energies of the
grains $eg_0$, $eg_1$, $eg_2$, plus the two binding energies $el_1$
and $el_2$, plus the angle energy $ea$:
  \begin{equation}
  ep = eg_0 + eg_1 + eg_2 + el_1 + el_2 + ea
  \end{equation}

\section {Ponderation}
The potential energy of a CG molecule is the sum of the energies of
the grains, plus the energies of the CG bonds and those of the CG
valence angles making up the molecule. This sum must be equal to the
potential energy of the AA molecule.  To obtain this result, it is
necessary to {\it weight} the energies of the grains, in order to
avoid to count these energies several times.  The weighting factor $P$
chosen is $1/5$~:
\begin{equation}
  P = 1/5
\end{equation}

To justify the value of $P$, let us consider the CG molecule in
Fig.\ref{grains}, more precisely the grain $G_2$ in it. This grain
participates in the two bonds $G_1G_2$ and $G_2G_3$, and also in the
three valence angles $G_1G_2G_3$, $G_0G_1G_2$ and $G_2G_3G_4$.

\begin{myfigure}
     \includegraphics [width=10cm] {\image 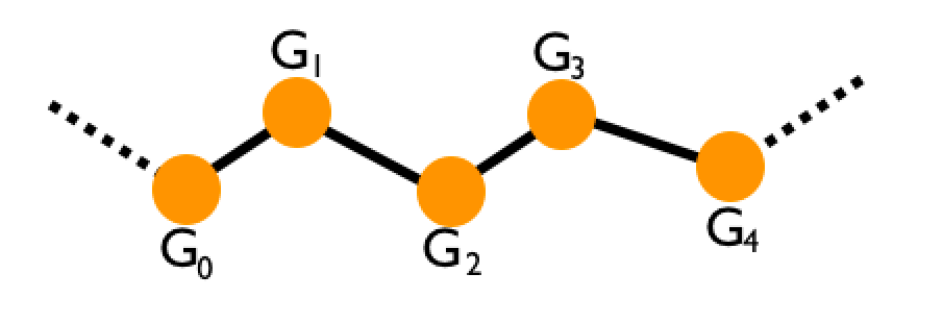} 
     \caption{\small CG fragment.
}
\label{grains}
\end{myfigure}

These are the only five CG components in which $G_2$ appears. The same
applies to each of the CG grains, with the exception of those near the
two ends of the molecule, which should be treated differently,
i.e. with different weightings.

For the sake of simplicity, in the following we will only consider
grains far away from extremities (i.e. grains similar to the grain
$G_2$ in Fig.\ref{grains}) and the only weighting to be used will be
$P=1/5$\footnote{ For a complete treatment of alkanes, it would be
  necessary to distinguish three types of grains: $E$ grains at the
  extremities; $F$ grains connected directly to extremities; and the
  others grains, of type $G$.  We would have then to define the
  ponderations of the bonds $EF$, $FG$, $GG$, and those of the valence
  potentials $EFG$, $GGF$, and $GGG$.  }.

 When calculating the CG potentials, the minimisations are performed
with perturbations steps ranging from $10^{-3}$ $nm$ to $10^{-9}$
$nm$.
  
\section{Potential of CG Bonds}
The potential of a {\CG} bond is the sum of the weighted energies of
the two grains, plus the energy of the link between them:
\begin{equation}
  eb = P \times eg_1+ P \times eg_2  + el
\end{equation}

For the potential of CG bonds we obtain the curve shown in
Fig.\ref{C4_MIN_CGBOND-curve.data}.

\begin{myfigure}
    \includegraphics [width=13cm] {\image 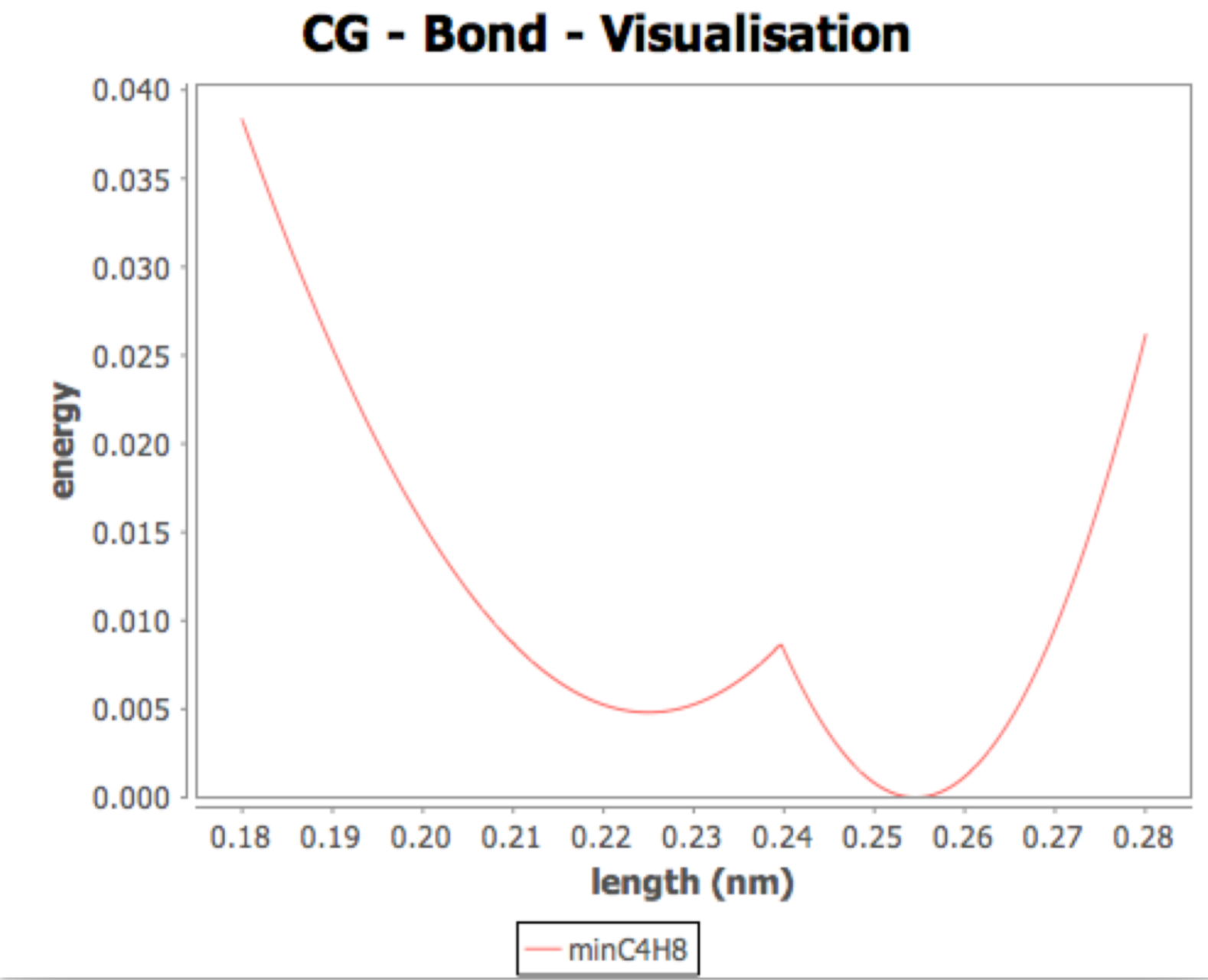} 
    \caption{\small Potential of a CG bond between two $\alkane{2}{4}$
      grains.  The potential is obtained by minimisation of the
      $\alkane{4}{8}$ fragment.  }
\label{C4_MIN_CGBOND-curve.data}
\source{curves/VisualiseCgBondApp.java minimisation=true make cgbond}
\end{myfigure}

The curve obtained does not depend on the size of the molecule.  It
shows an inflection point at around $0.24$ $nm$. This inflection comes
from the triggering of the central torsion angle $C_0C_1C_2C_3$.

To calculate the potential of the CG bond, the size of the CG bond is
decremented up to $0.18$ $nm$, then increased up to $0.28$ $nm$, in
steps of $10^{-4}$ $nm$.

\section{Potential of CG Valence Angles}\label{minimisationCG:valence}
The CG valence angle potential is the sum of the weighted energies of
the three grains making up the valence angle, plus the energy of the
angle:
  \begin{equation}
    ev = P \times eg_1 + P \times eg_2 + P \times eg_3 + ea
 \end{equation}
 The CG valence potential corresponds to the curve of
 Fig.\ref{C6_MIN_CGVALENCE-curve.data}.
 
\begin{myfigure}
     \includegraphics [width=13cm] {\image 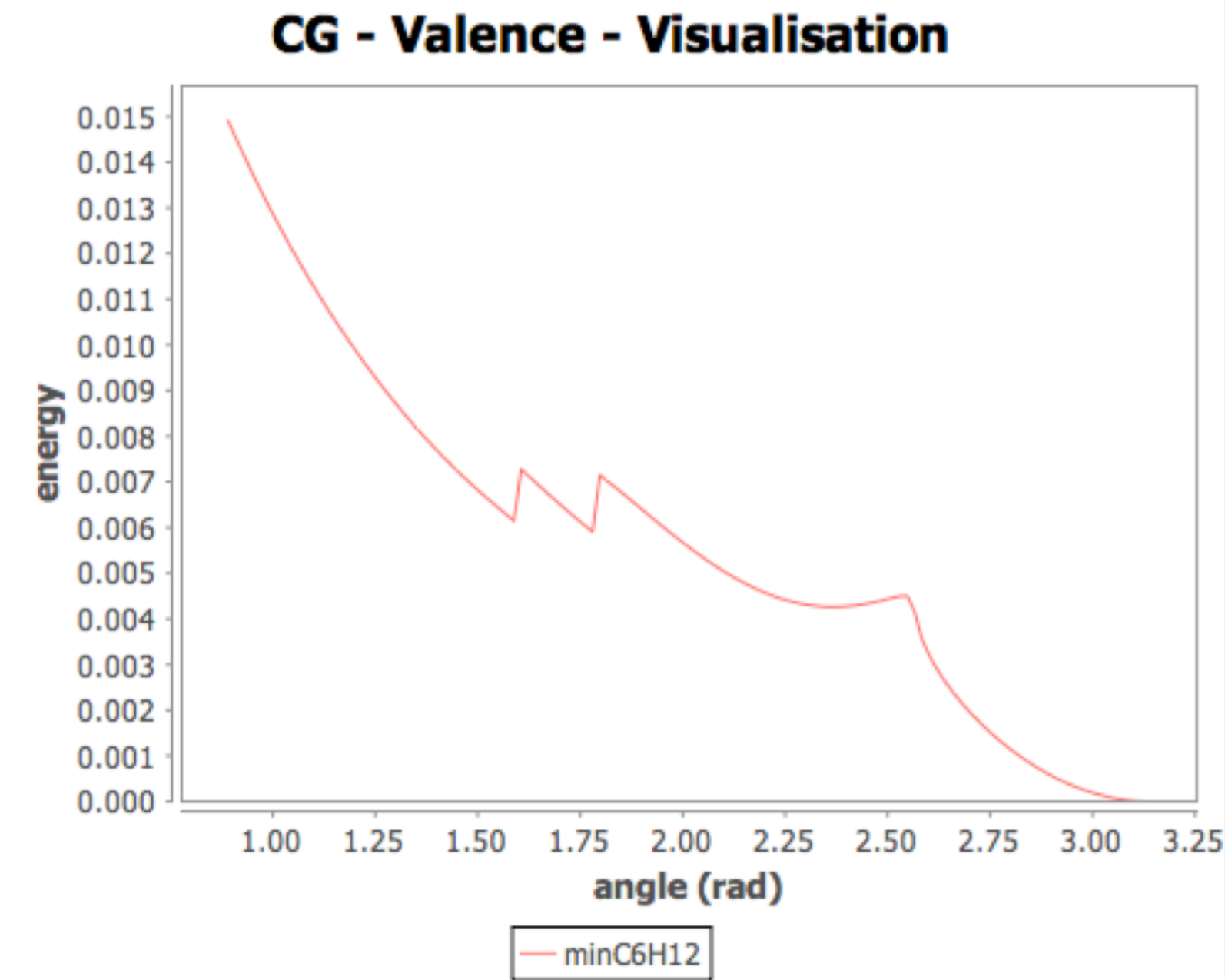} 
     \caption{\small Valence potential {\CG} between three grains
       $\alkane{2}{4}$.  The potential is obtained by minimisation of
       the molecule $\alkane{6}{12}$.  The minimum energy value is
       zero and corresponds to a valence angle of value $\pi$.  }
\label{C6_MIN_CGVALENCE-curve.data}
\source{curves/VisualiseCGValenceApp.java minimisation=true make cgvalence}
\end{myfigure}

As with the binding potential, the curve obtained does not depend on
the size of the molecule.  The curve obtained shows three inflection
points. The two inflection points on the left correspond to the
triggers of the two torsion angles $C_0C_1C_2C_3$ and
$C_2C_3C_4C_5$. The right-hand inflection point corresponds to the
triggering of the central torsion angle $C_1C_2C_3C_4$.

In the implementation, the CG valence angle is decremented from $\pi$
radians to $0.9$ radians, then increased up to $\pi$, in steps of
$0.017$ radians (262 measurements).

\section {CG Inter-molecular Forces}
To determine the force exerted between two CG grains, we consider two
$\alkane{2}{4}$ molecules at a fixed distance $d$.  The potential is
calculated by minimising the energy of the two grains, while keeping
constant the distance $d$ (conditioned minimisation).

The $ep$ energy is the value of the potential between the two grains
obtained after minimisation maintaining the distance $d$. By varying
the parameter $d$, one obtains the CG van Der Waals potential curve
shown in Fig.\ref{figure:vdwCG}.

\begin{myfigure}
\includegraphics[width=13cm] {\image 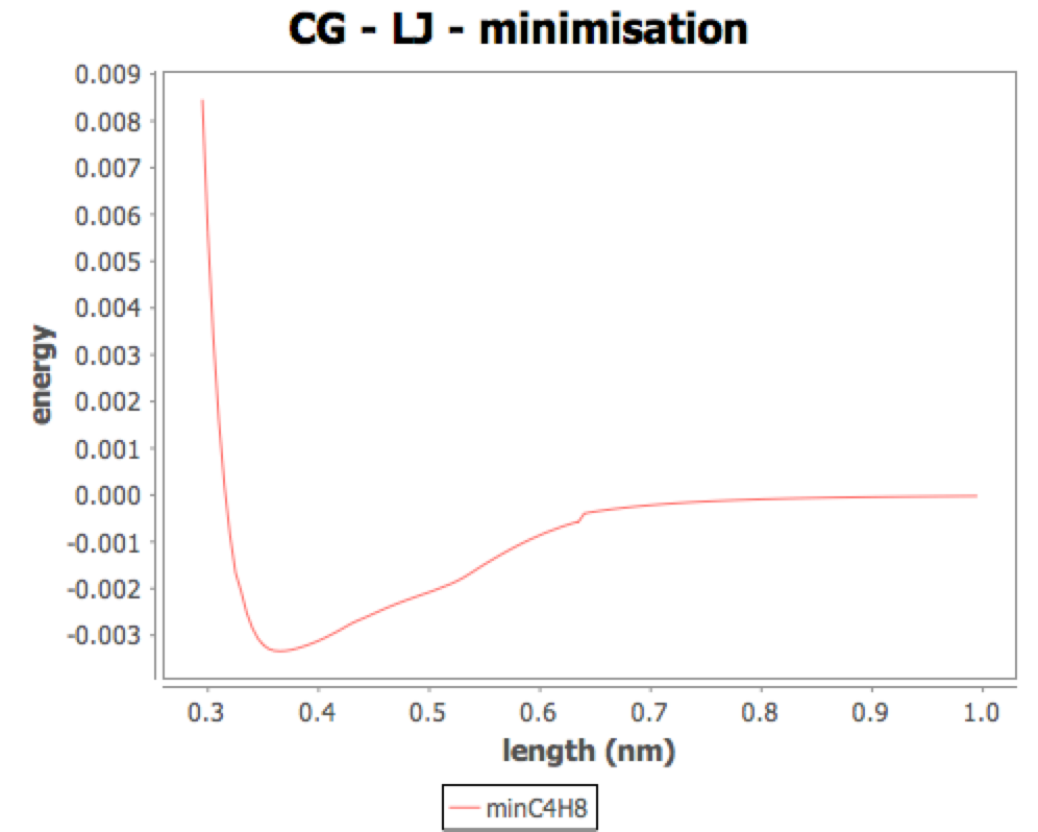}
\caption {\small Inter-molecular potential between two CG grains
  obtained with the minimisation method.}
\label {figure:vdwCG}
\source{curves/VisualiseCgBondApp.java minimisation=true make cglj}
\end{myfigure}

The $ep$ energy is the sum of the thirty-six energies of the
Lennard-Jones potentials associated with the pairs of atoms between
the two grains (the AA Lennard-Jones potentials are represented in
Fig.\ref{figure:opls-lj}).

The potential obtained by minimisation in Fig.\ref{figure:vdwCG} does
not have the form of a Lennard-Jones potential, defined by the
equation Eq.\ref{equation:lj-1}. In
Fig.\ref{figure:vdw6-12-cg-compare} the obtained potential is compared
with a Lennard-Jones potential
($\epsilon = 12.5 \times \epsilon_{CC}$,
$\sigma = 0.9 \times \sigma_{CC}$): it can be seen that the potential
obtained differs over a significant range (between $0.4$ $nm$ and
$0.7$ $nm$) with the Lennard-Jones potential.

\begin{myfigure}
\includegraphics[width=13cm] {\image 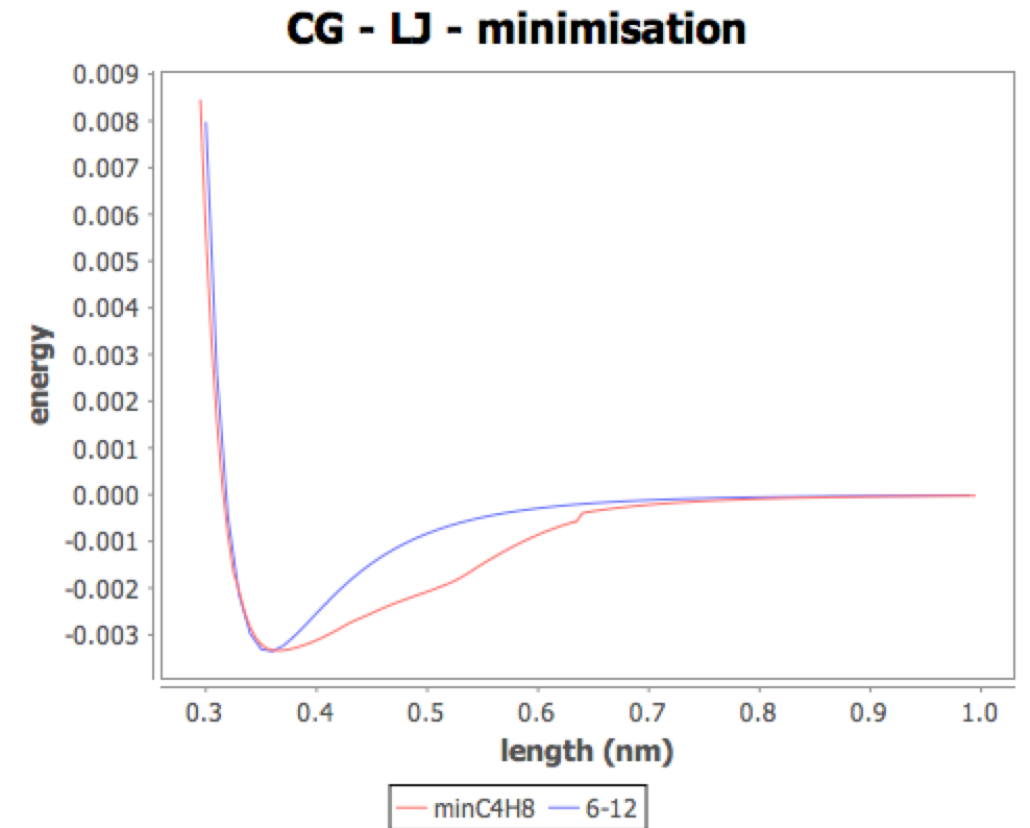}
\caption { {\small Potential of van Der Waals force between two CG
  grains obtained with the minimisation method, compared with a
  Lennard-Jones curve.}}
\label {figure:vdw6-12-cg-compare}
\source{curves/VisualiseCgBondApp.java minimisation=true lj6-12=true make cglj}
\end{myfigure}

The inter-molecular potential CG is linear in the domain where it does
not coincide with a Lennard-Jones potential. In this domain,
minimisation has the effect of reorganising the AA components in the
two CG grains in such a way as to minimise their energies. Within each
grain, the energy exchanges between these components induce a linear
evolution of the overall grain energy.

Outside the linear zone, the CG inter-molecular potential behaves like
a standard Lennard-Jones potential, which means the absence of
reorganisation of the CG grains in the zones of close proximity of the
two grains or of long distance between them.

\section {Comparison with the Inverse-Boltzmann Approach}
To compare the inverse-Boltzmann and the minimisation approaches in
the calculation of the {\CG} bond potential, we superimpose in
Fig.\ref{compareBond} the curves in Fig.\ref{C4_MIN_CGBOND-curve.data}
and Fig.\ref{C6H12BondBoltzmann}.

\begin{myfigure}
    \includegraphics [width=13cm] {\image 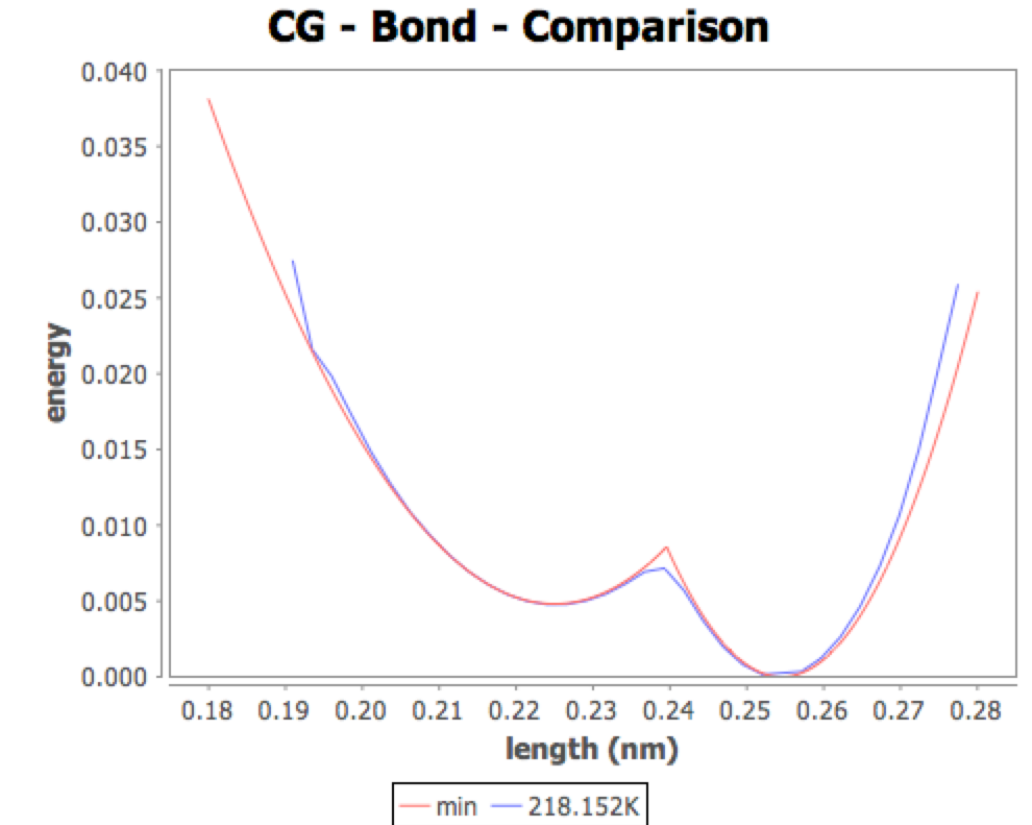} 
    \caption{\small CG bond potential obtained with the
      inverse-Boltzmann method from molecule $\alkane{6}{12}$ at
      temperature of $\temp {218.152}$ compared with the potential
      obtained with the minimisation method.}
\label{compareBond}
\source{curves/VisualiseCgBondApp.java compare=true make cgbond}
\end{myfigure}

There is good agreement between the curves, except for the inflection
point, at around $0.24$ $nm$. This point corresponds to the triggering
of the central dihedral, a rare event to which, therefore, is
associated a potential energy lower than it should be (this can be
seen as an intrinsic weakness of the inverse-Boltzmann approach).

To compare the inverse-Boltzmann and minimisation approaches in case
of the CG valence potential, we superimpose in
Fig.\ref{compareValence} the curves of
Fig.\ref{C6_MIN_CGVALENCE-curve.data} and
Fig.\ref{C6H12ValenceBoltzmann}.

\begin{myfigure}
    \includegraphics [width=13cm] {\image 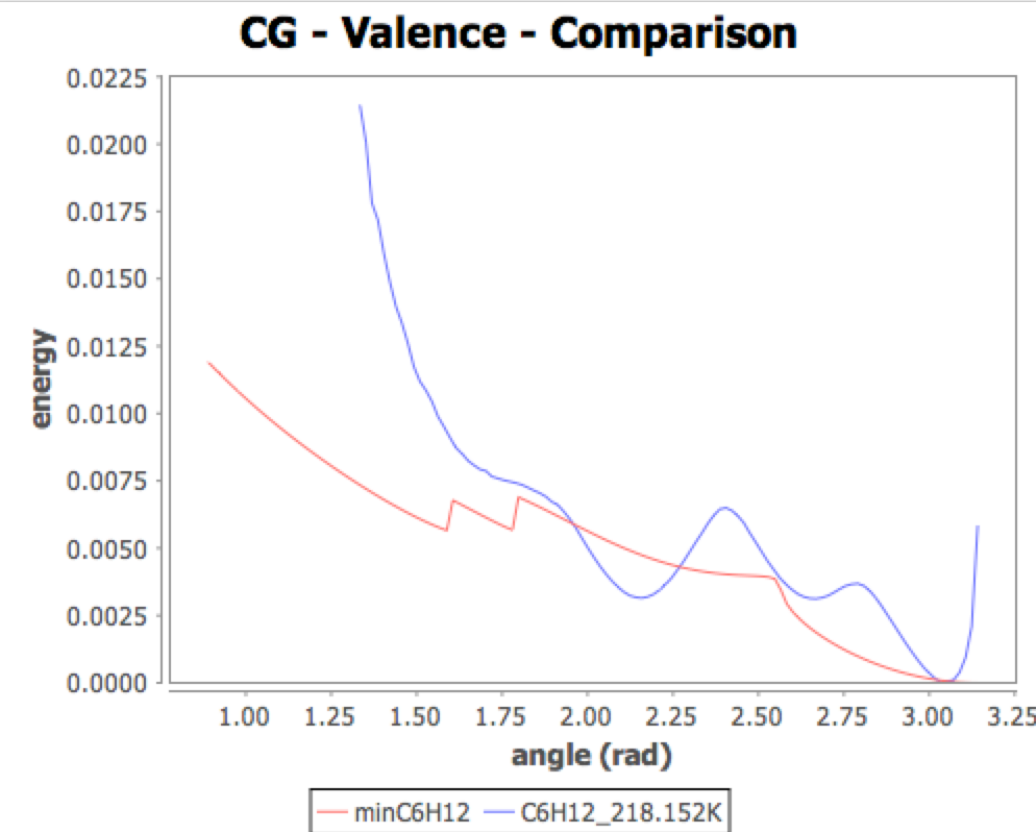} 
    \caption{\small CG valence potential obtained with
      inverse-Boltzmann from molecule $\alkane{6}{12}$ at temperature
      of $\temp {218.152}$ (in blue) compared with the potential
      obtained with minimisation (in red). }
\label{compareValence}
\source{curves/VisualiseCgValenceApp.java comparison=true make cgvalence}
\end{myfigure}

The inverse-Boltzmann curve is systematically higher than the
minimisation curve for angles less than $1.9$ radians. This is due to
the temperature being too low for the angular exploration below $1.9$
to be really significant.

We can consider that the part of the inverse-Boltzmann curve situated
below the minimisation curve results from a lack of precision due to a
too small set of events.

To increase the number of events taken into account, longer molecules can
be used, for example the $\alkane{30}{60}$ molecule at the temperature
of $\temp{206}$. The result is shown in Fig.\ref{compareSize} where it
can be seen that the part of the inverse-Boltzmann curve situated
under the minimisation curve is greatly reduced.

\begin{myfigure}
    \includegraphics [width=13cm] {\image 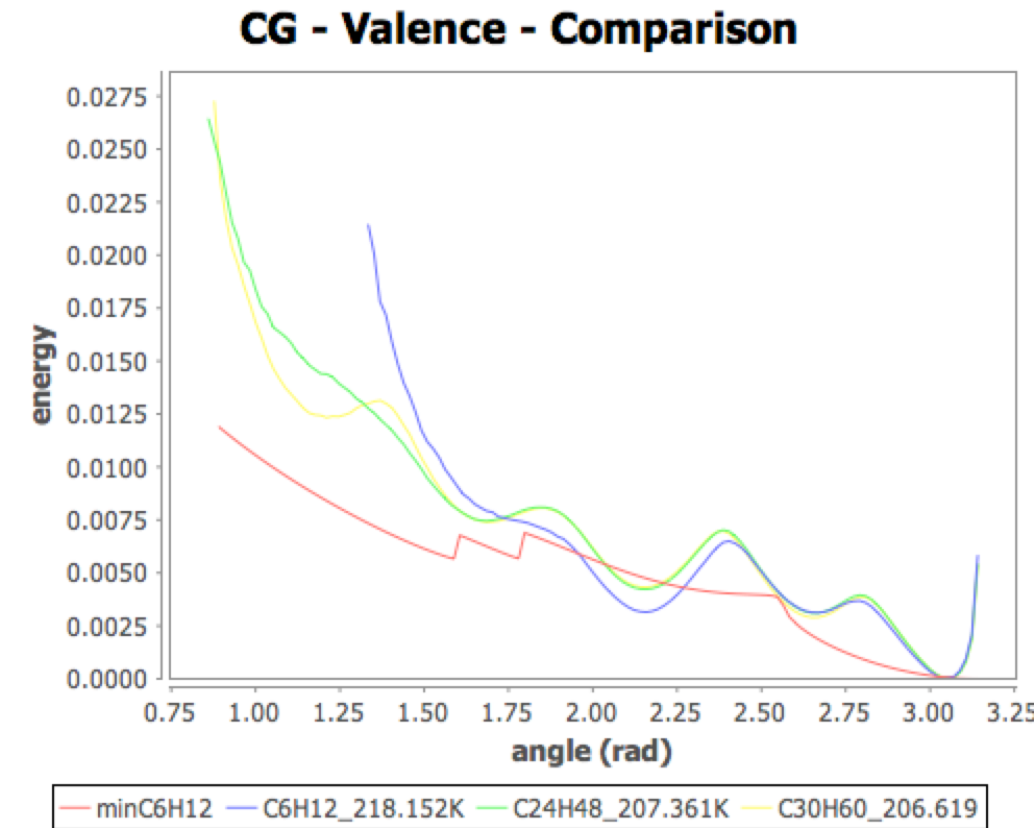} 
    \caption{\small CG valence potential obtained with
      inverse-Boltzmann from the molecule $\alkane{6}{12}$ at
      temperature of $\temp {218.152}$ (in blue) compared with the one
      obtained with the molecul $\alkane{30}{60}$ at the temperature
      of $\temp{206.619}$ (in green).  }
\label{compareSize}
\source{curves/VisualiseCgValenceApp.java comparison=true make cgvalence}
\end{myfigure}

With a sufficiently large number of events, we conjecture that the
inverse-Boltzmann curve will always be above the curve obtained by
minimisation. This could be seen as a (indeed partial) validation of
the minimality of the curve obtained by minimisation, which is lower to all
those obtained from actual simulations.


\chapter {Simulations in UA}\label{chapter:simulationsUA}
Using the minimisation method, we have determined the potentials of
the components of UA molecules formed from $G_2$ grains, which makes
it possible to simulate molecules at this scale.  The UA potentials
obtained from the AA potentials are summarised in
Fig.\ref{table:parameters-ua}.

\begin{myfigure}
  \begin{eqnarray*}
    &k_{G_2G_2} = k_{CC}~~~~~r_{G_2G_2} = r_{CC}
\\
    &k_{G_2G_2G_2 }= 1.1 \times k_{CCC} ~~~~~~~
    \theta_{G_2G_2G_2 }= \theta_{CCC}
\\
      &A1_{G_2G_2G_2G_2} = A1_{CCCC} ~~~~~
    A2_{G_2G_2G_2G_2} = A2_{CCCC} ~~~~~
\\
    &A3_{G_2G_2G_2G_2} = A3_{CCCC} + 4 \times A3_{CCCH}  + 4 \times A3_{HCCH}\\
    &\epsilon_{G_2G_2} = 5.5 \times \epsilon_{CC}~~~~~~\sigma_{G_2G_2} = 0.917 \times \sigma_{CC}
  \end{eqnarray*}
  \caption{\small UA parameter values associated with grains $G_2$,
    expressed in internal units (AA OPLS parameters are given in
    Fig.\ref{table:opls}).}
\label{table:parameters-ua}
\end{myfigure}

The potential of the valence angle $G_2G_2G_2$ is approximated by a
harmonic potential whose strength is that of $CCC$ multiplied by
$1.1$, as we saw in \ref{section:valenceUA}.  The torsion angle
potential was determined in \ref{section:miniUA:torsion} and the
inter-molecular potential in \ref{section:miniUA:intermol}.

In the {\MD} system, the $G_2$ grains (i.e. made up of one carbon atom
and two hydrogen atoms) appear as cyan-coloured balls.  Fig.\ref
{figure:ua-equilibrium} shows a UA molecule composed of 8 $G_2$ grains
at equilibrium (its energy is zero).

\begin{myfigure}
  \includegraphics[height=7cm] {\image uaG8-equilibrium.pdf}
  \caption {\small UA molecule at equilibrium made up of 8 grains
    $G_2$.}
\label {figure:ua-equilibrium}
\source {ua/ImageApp.java moleculeNumber=1 initialShift=0 tstep=0 make image}
\end{myfigure}

Simulations of UA molecules are very stable, as can be seen in
Fig.\ref {figure:ua}.

\begin{myfigure}
  \includegraphics[height=7cm] {\image uaG8.pdf}
  \includegraphics[width=10cm] {\image 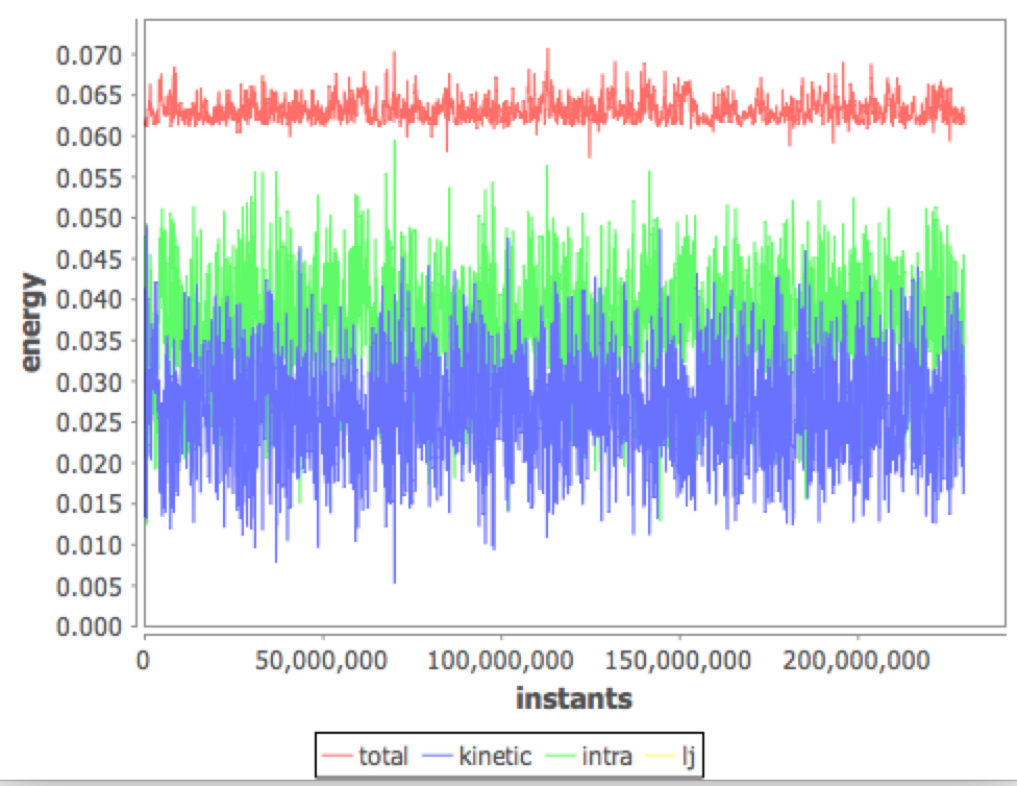}  
\caption {\small Simulation of one UA molecule made up of 8 grains $G_2$.}
\label {figure:ua}
\source{ua/SimpleUAApp.java moleculeNumber=1 initialShift=0.02 tstep=1E-4 make simple}
\end{myfigure}

On the left, a molecule with eight grains $G_2$ is shown during
simulation. Initially the molecule is given an initial energy obtained
by shifting the grains towards the positive $y$ (vertical axis).

On the right, one can see the evolution of energies over time.  The
red curve represents the total energy of the molecule, which remains
stable, as we can see.

Fig.\ref {figure:uax2} describes the simulation of two UA molecules
with 8 grains, placed face-to-face at a distance of $1.8$
$nm$. Initially the two molecules are at equilibrium. The energy of
inter-molecular forces is shown in yellow, the kinetic energy in blue,
the intra-molecular energy in green, and the sum of the energies
in red.

\begin{myfigure}
  \includegraphics[height=6cm] {\image 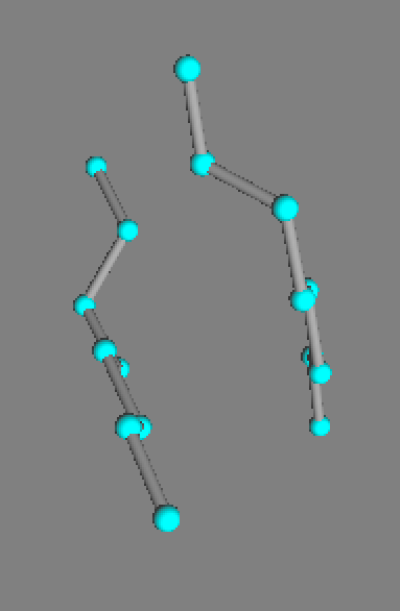}\\
  \includegraphics[width=11cm] {\image 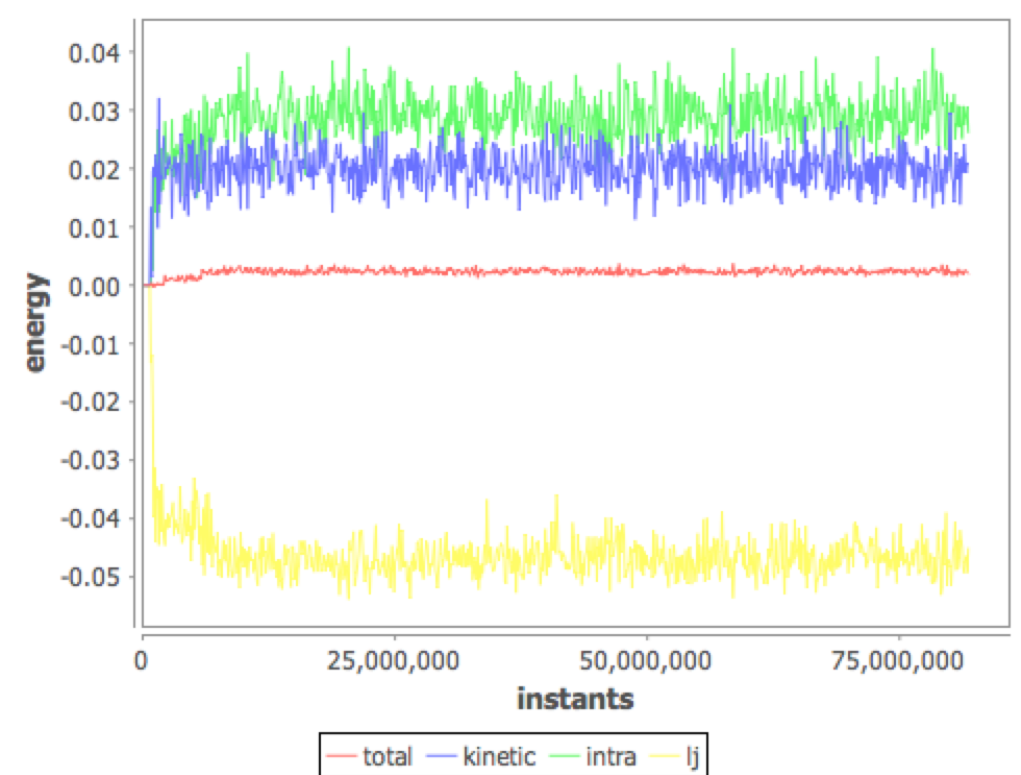}  
  \caption {\small Simulation of two UA molecules made up of 8 grains
    $G_2$.}
\label {figure:uax2}
\source{ua/SimpleUAApp.java moleculeNumber=2 initialShift=0 tstep=1E-4 make simple}
\end{myfigure}

UA scale simulations are more efficient than AA scale ones: the number
of degrees of freedom to be simulated is smaller and the time-step can
be increased. We will show this increased efficiency on an example.

We first simulate an AA $\alkane {8}{16}$ molecule during $10^7$
instants with a time step of $0.1$ femto-second.  The real-time of
simulation\footnote{ with a MacBook Pro, i7 processor at 2.6GHz, with
  16GB of memory} is $1053$ seconds. This corresponds to a simulated
time of $0.5$ nano-second. The energies are shown on the left-hand
side of Fig.\ref{figure:comparisonc8-g8-energy}.

We now simulate the corresponding UA molecule, made up of 8 grains
$G_2$ with a time-step of $0.5$ femto-second.  The simulation lasts
$175$ seconds which corresponds to a simulated time of $2.5$
nano-seconds.  Energies are shown on the right-hand side of
Fig.\ref{figure:comparisonc8-g8-energy}.

Compared with the previous AA simulation, the real-time of simulation
is therefore reduced by a factor 6, while the simulated time is
increased by a factor 5. The advantage of UA over AA is therefore
real.

 \begin{myfigure}
   \includegraphics[width=10cm] {\image 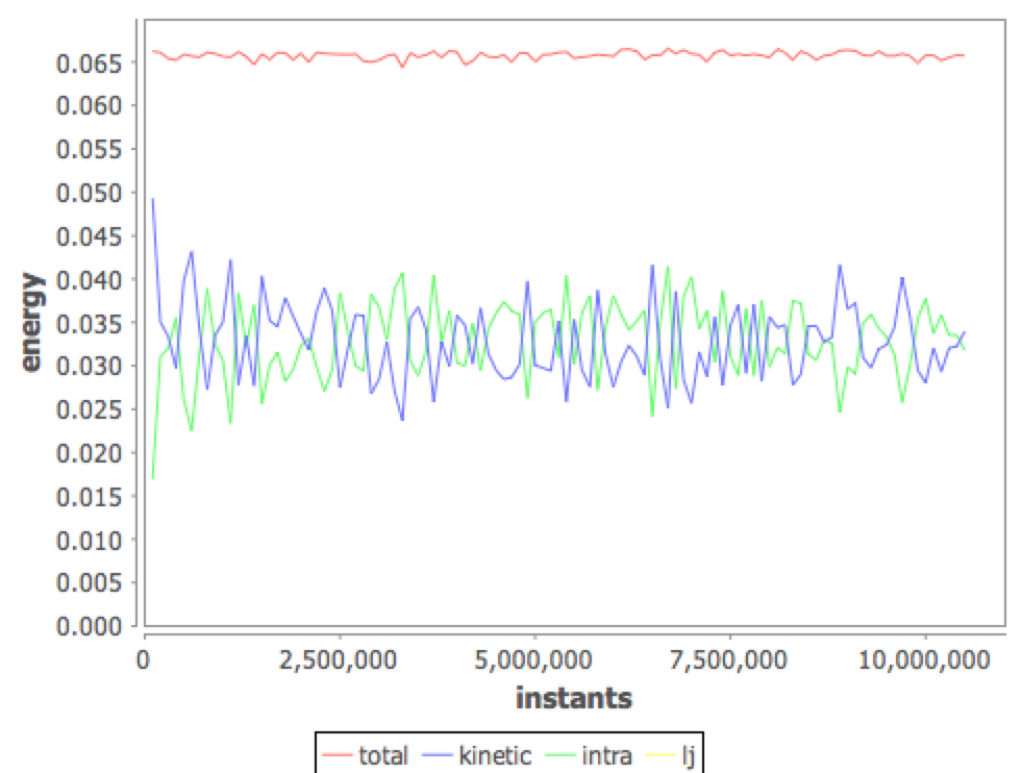}\\
     \includegraphics[width=10cm] {\image 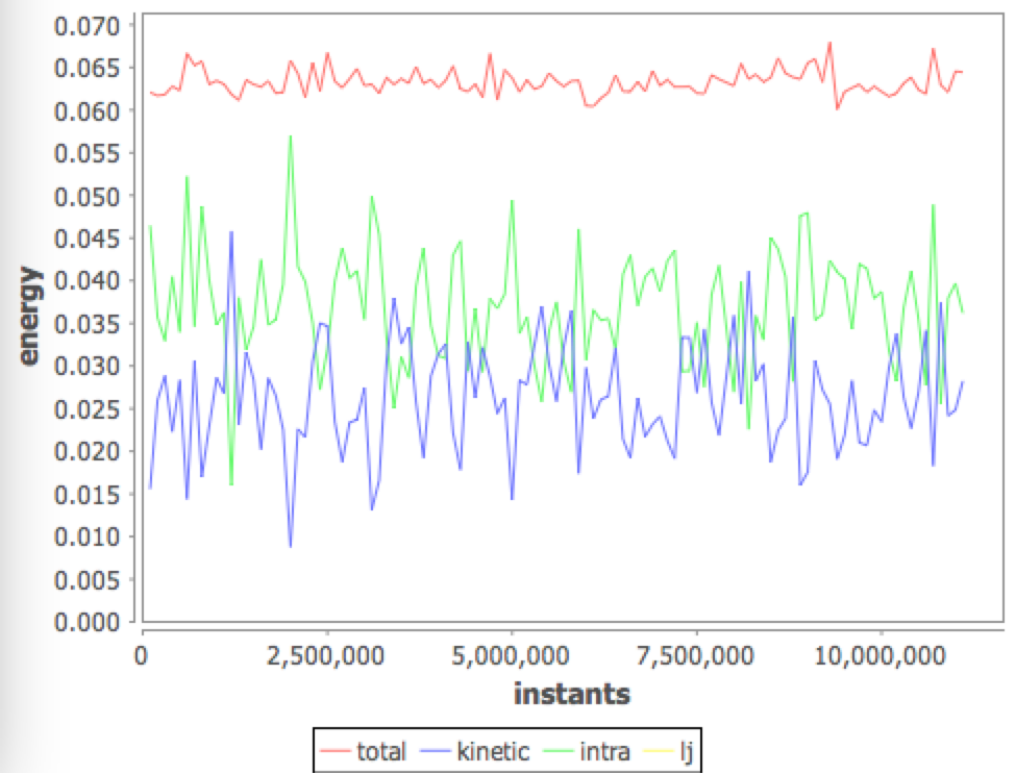}  
\caption {\small Energies during 10 millions of instants. Top,
  molecule AA $\alkane{8}{16}$. Bottom, molecule UA made up of 8 grains.}
\label {figure:comparisonc8-g8-energy}
\source{left: aa/SimpleAAApp.java one=true initialShift=0.01 make
  simple~~~right: ua/SimpleUAApp.java initialShift = 0.01 tstep = 5E-4
  make simple}
  
\end{myfigure}
 
 



\chapter {Simulations in CG}\label{chapter:simulationsCG}
Unlike the AA and UA potentials previously described, CG potentials do
not have standard forms as harmonic, ``three-cosine'', or
Lennard-Jones functions.  We must therefore first precisely define
what are these non-standard forms of CG potentials.

\section {Determination of the CG potential}

\subsection*{Bonds}
The CG bond potential appears to be made up of two harmonic functions,
applied on either side of a split value ($split = 0.241$ $nm$). One
first defines a functional that produces harmonic functions as:
\[
   harm: k,r,d \rightarrow k \times (r - d)^2
\]
where $k$ is the strength, $r$ is the equilibrium value and $d$ is a
value parameter (a length in $nm$ in the case of bonds, and an angle
in radians in the case of valence angles).

The CG bond potential can now be defined using the functional $harm$
as the following function $cgBond$:
\begin{eqnarray*}
  &split = 0.241\\
  &h_1: d \rightarrow harm (16.5,0.225,d)\\
  &h_2: d \rightarrow harm (38.5,0.255,d)\\  
  &cgBond: d \rightarrow
    {\bf if}~~d < split~~
    {\bf then}~~ h_1  (d) + 0.0049 ~~
    {\bf else}~~h_2 (d)
\end{eqnarray*}

On left of $0.241$ (that is, for values less than $0.241$ $nm$), the
function $h_1$ defined by $k=16.5$ and $r=0.225$ applies.  Moreover, a
constant shift of $0.0049$ in $y$ is added to the result.

For values greater than $0.241$ $nm$, the harmonic function $h_2$ with
parameters $k=38.5$ and $r=0.255$ applies.

Fig.\ref{figure:determinecgbond} compares the CG potential obtained by
minimisation of the $\alkane 4 8$ molecule (in red) and the previous
$cgBond$ function, constructed ``by cases'' using the two harmonic
functions $h_1$ and $h_2$ (in blue).

\begin{myfigure}
  \includegraphics[width=13cm] {\image 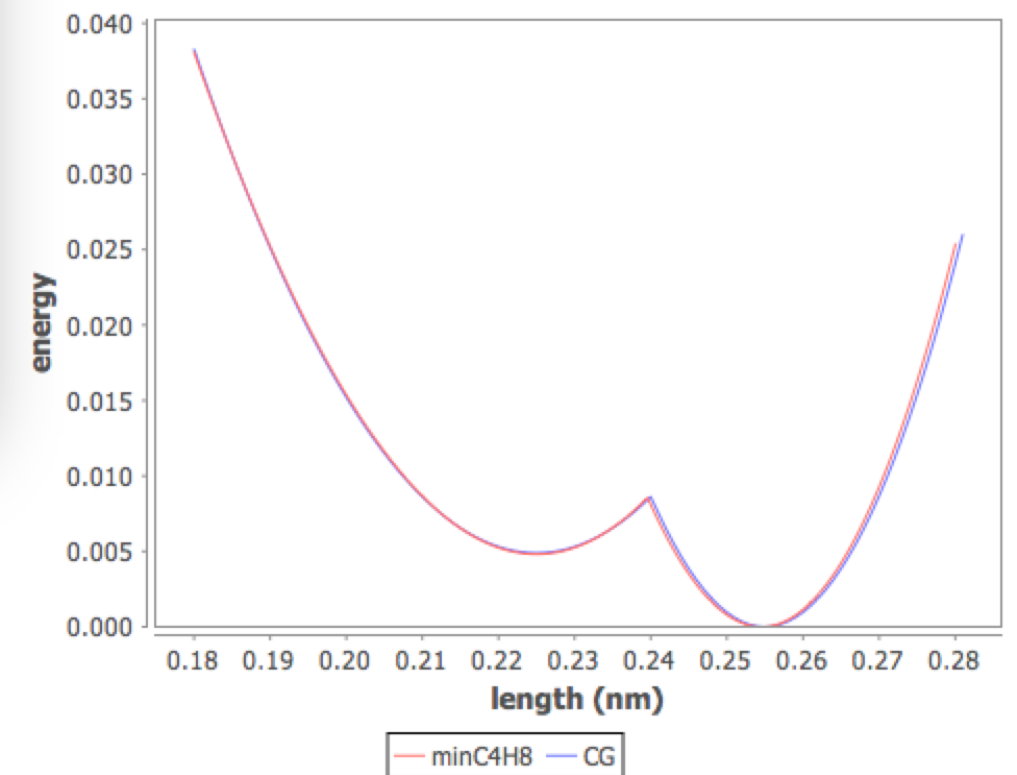}
  \caption {\small The CG potential (in blue) is built ``by case''
    with the aid of two harmonic functions.}
\label {figure:determinecgbond}
\source{cg/DetermineCgBondApp.java make detbond}
\end{myfigure}

One observes the exact match of the two curves, the one obtained by
minimisation, and the other built ``by cases'' with the aid of two
auxiliary harmonic functions.

\subsection*{Valence Angles}
Building up the CG valence potential needs four functions
$h_0, h_1, h_2, h_3$ defined using the functional $harm$ as follows:
\begin{eqnarray*}
  &split_0 = 2.56  &h_0: a \rightarrow harm (0.0095, \pi, a)\\
  &split_1 = 1.79  &h_1: a \rightarrow harm (0.0055, split_0, a) + 0.004\\
  &split_2 = 1.6  &h_2: a \rightarrow harm (0.041, split_1, a) + 0.0056\\
  &&h_3: a \rightarrow harm (0.0059, 2.0, a) + 0.0047\\
\end{eqnarray*}

The $cgValence$ function is then defined ``by cases'' as:
\begin{eqnarray*}
  cgValence: a \rightarrow
    &&{\bf if}~~a > split_0~~ {\bf then}~~ h_0  (a) ~~\\
    &&{\bf else}~{\bf if} ~~ a > split_1  ~~{\bf then} ~~ h_1 (a) ~~\\
    &&{\bf else}~{\bf if} ~~ a > split_2  ~~{\bf then} ~~ h_2 (a) ~~\\
    &&{\bf else} ~~ h_3 (a)
\end{eqnarray*}

The $cgValence$ function is represented by the blue curve in Fig.\ref
{figure:determinecgvalence}, the red curve being the result obtained
by minimisation.

\begin{myfigure}
  \includegraphics[width=13cm] {\image 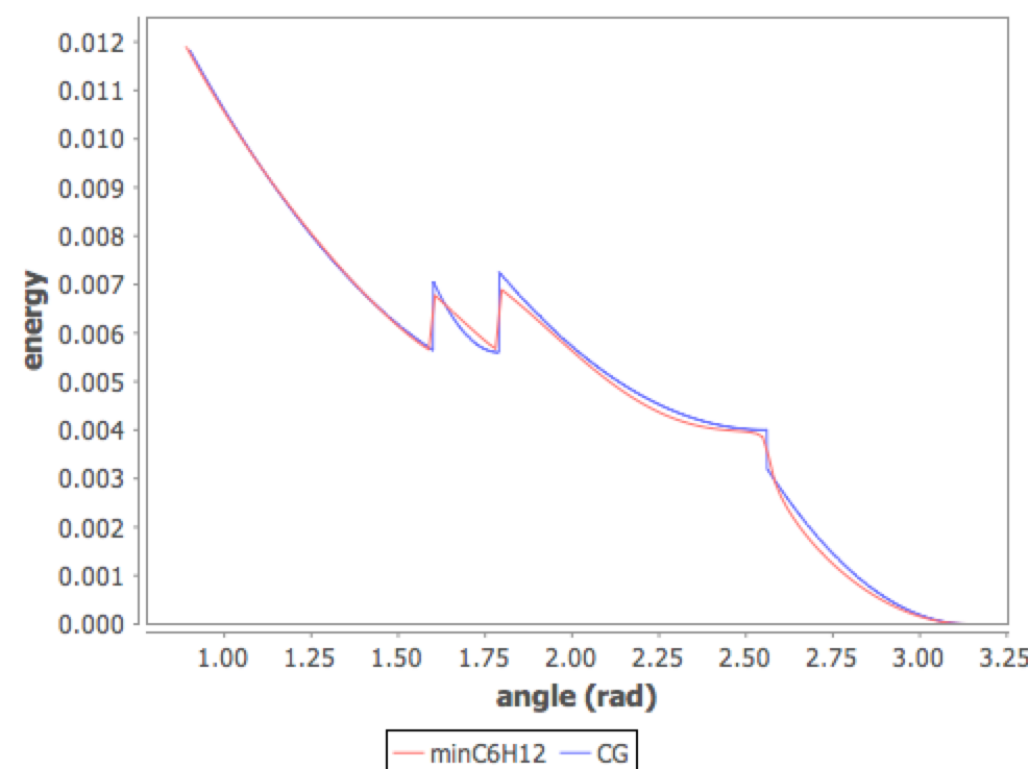}
  \caption {\small The CG valence potential (in blue) is built ``by
    cases'' with the aid of four harmonic functions.}
\label {figure:determinecgvalence}
\source{cg/DetermineCgValenceApp.java make detvalence}
\end{myfigure}

As for CG bonds, one gets an exact match of the two curves, the one
obtained by minimisation, and the other built ``by cases'' with the
aid of four auxiliary harmonic functions.

\subsection*{Inter-molecular Forces}
The inter-molecular (van Der Waals) forces are often represented by
Lennard-Jones functions that can be produced using the functional
$lj6\_12$ defined by:
\[
  lj6\_12: \epsilon, \sigma, r \rightarrow 4 \epsilon \times ((\sigma/r)^{12} - (\sigma/r)^6)
\]

To build the CG inter-molecular potential, we start by defining a
particular Lennard-Jones function $lj$ as follows:
\begin{eqnarray*}
  &\epsilon = 0.00336\\
  &\sigma = 0.318\\
  &lj: d \rightarrow lj6\_12 (\epsilon , \sigma , d)
\end{eqnarray*}

The function $lj$ is represented by the green curve in
Fig.\ref{figure:determinecglj}.

To completely determine the CG Lennard-Jones potential, one needs to
define two points: the point $p_1 = (0.365,-\epsilon)$ whose
coordinate in $y$ is minimal and the point $p_2 = (0.68,-000195)$.

Between these two points, the potential is globally a straight line,
corresponding to the affine function joining the two points.

The definition of the CG inter-molecular potential $cgLj$ is then:
\begin{eqnarray*}
&cgLj: d \rightarrow\\
    &{\bf if}~~d < 0.365~~{\bf or}~~d > 0.68~~
    {\bf then}~~ lj  (d) ~~
    {\bf else}~~line (d)
\end{eqnarray*}
where the function $line$ is the affine function linking the two
points $p_1$ et $p_2$.

The CG potential is represented by the blue curve in
Fig.\ref{figure:determinecglj}. The red curve is the CG potential
obtained with the minimisation method.

\begin{myfigure}
  \includegraphics[width=13cm] {\image 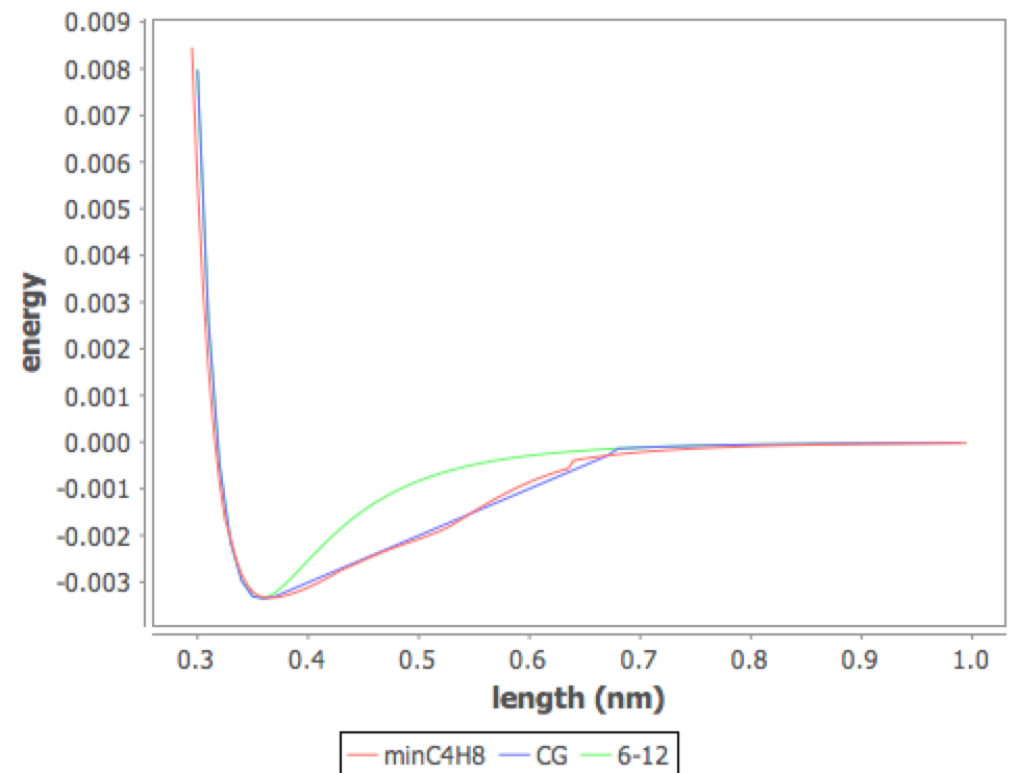}
  \caption {\small The inter-molecular potential CG (in blue) is
    defined with the aid of a Lennard-Jones function (in green) and an
    affine function. In red, the potential obtained by minimisation.
  }
\label {figure:determinecglj}
\source{cg/DetermineCgLjApp.java make detlj}
\end{myfigure}

Here again, we observe the good matching between the curve obtained
with the minimisation method and the previous curve built ``by
cases''.

\section {CG Execution}
To illustrate the CG execution, we first consider an isolated CG
molecule made up of six $CG_4$ grains.  The $x$ coordinates are
represented horizontally, the $y$ coordinates vertically, and those in
$z$ are represented in height.

Some energy is initially supplied to the molecule at equilibrium by
shifting its first grain in the direction of increasing $y$.  Thus,
the first CG bond is stretched and it is the energy due to this
stretching that is initially supplied to the molecule.

The right-hand side of Fig.\ref{figure:CG6bond} shows the evolution of
the various energy components over time.  Measurements are taken every
$10^5$ instants. The kinetic energy is in blue, the intra-molecular
energy is in green and the total energy in red (the inter-molecular
energy is always zero).

The molecule is shown during execution on the left of the figure and
we can see that it remains linear. This is consistent with the fact
that the forces here are solely due to the initial bond stretching.

\begin{myfigure}
  \includegraphics[height=8cm] {\image 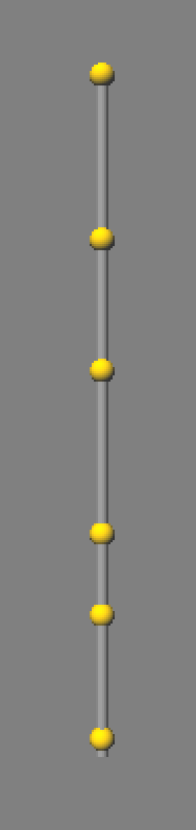}
  \includegraphics[width=10cm] {\image 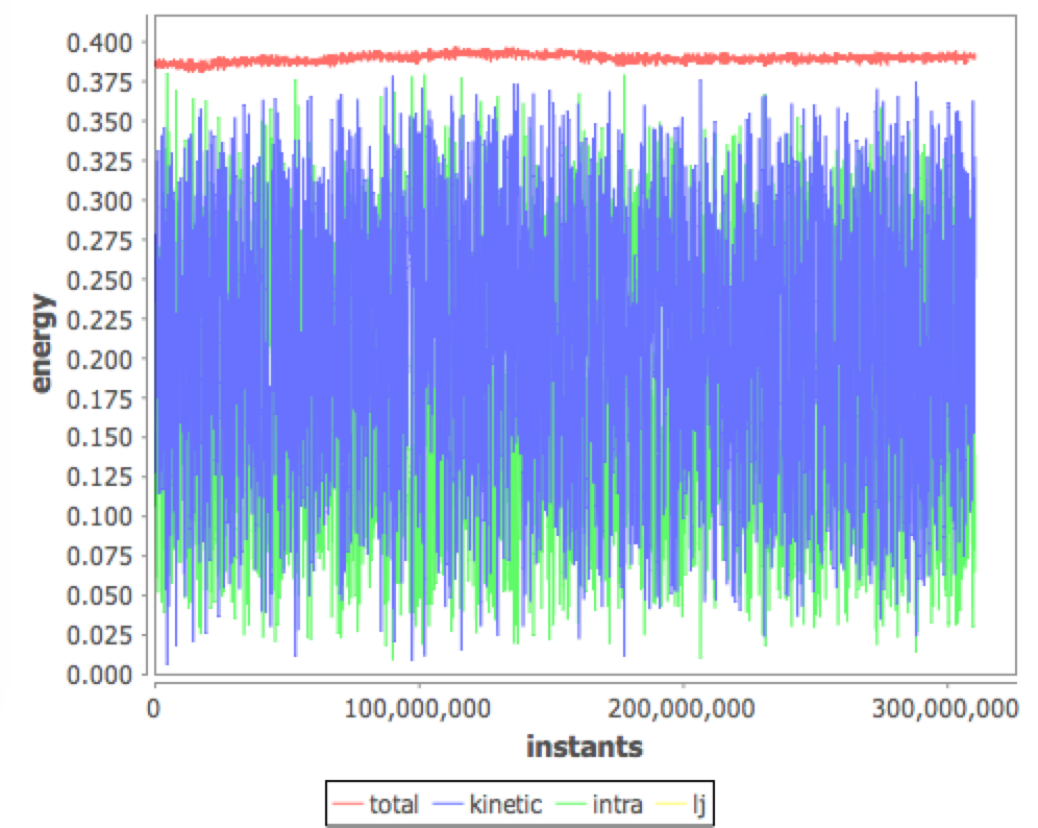}  
  \caption {\small Simulation of a CG molecule made up of 6 grains
    $CG_4$. Initially, only the first bond has some energy. On the
    left, the molecule after $3\times10^{8}$ instants.}
  \label {figure:CG6bond}
  \source{cg/SimpleCGApp.java bond=true make simple}
\end{myfigure}

We now change the initial conditions by rotating the first grain
around the second, while remaining in the same plane.  The initial
energy is then only that of the first valence angle, formed by the
first three grains.

The execution results are shown in Fig.\ref{figure:CG6valence} where
we can see that the molecule is no longer linear but always remains in
the same plane: in fact, all the forces exerted during execution are
co-planar forces.

\begin{myfigure}
  \includegraphics[height=8cm] {\image 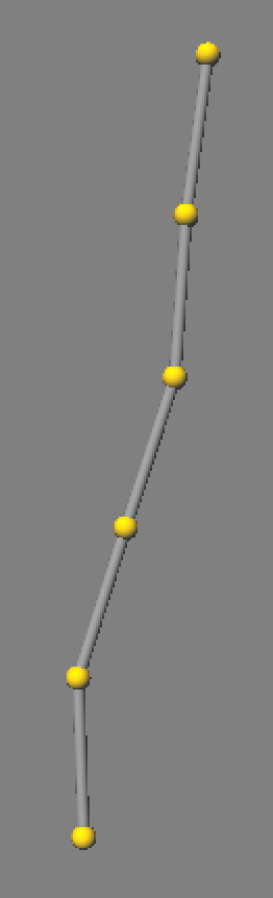}
  \includegraphics[width=10cm] {\image 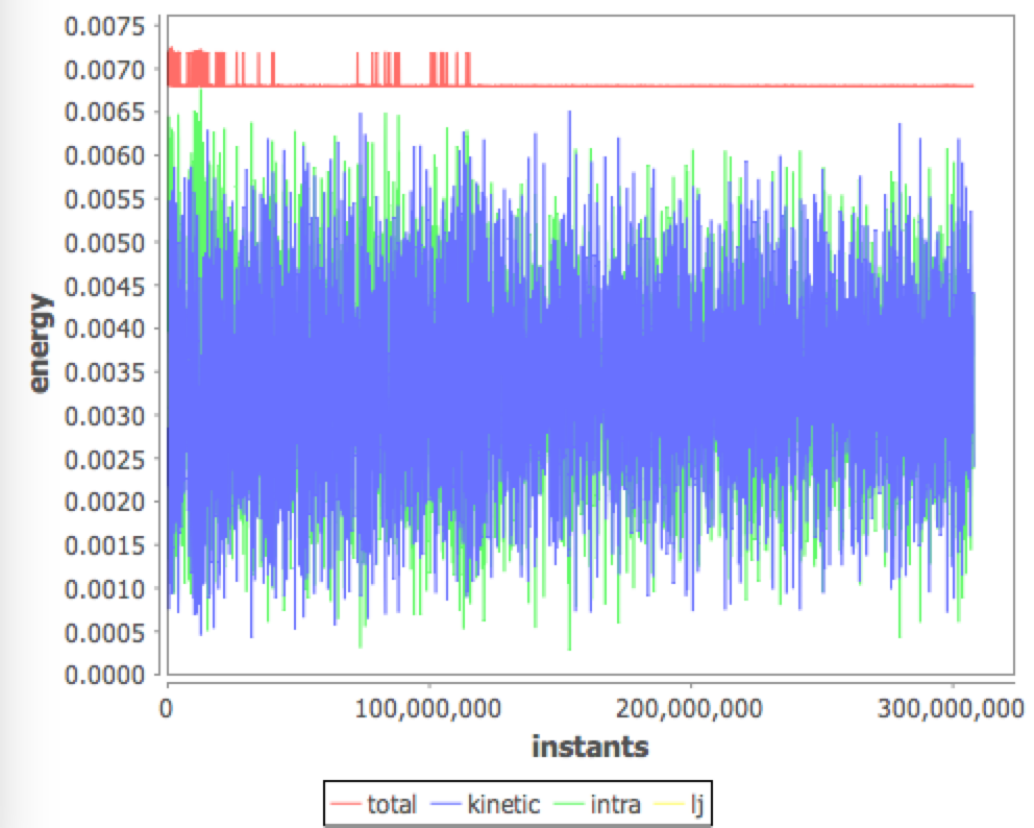}  
  \caption {\small Simulation of a CG molecule made up of 6 grains
    $CG_4$.  Initially, only the first valence angle contains some
    energy. On top image, the molecule after $3\times10^{8}$ instants.}
  \label {figure:CG6valence}
  \source{cg/SimpleCGApp.java valence=true make simple}
\end{myfigure}

Finally, we introduce a twist into the molecule, shifting the first
grain in the $x$ coordinate and the last one in the $z$
coordinate. These two shifts mean that the molecule is no longer
planar.  The simulation is shown in Fig.\ref {figure:CG6dih}.

\begin{myfigure}
  \includegraphics[height=7cm] {\image 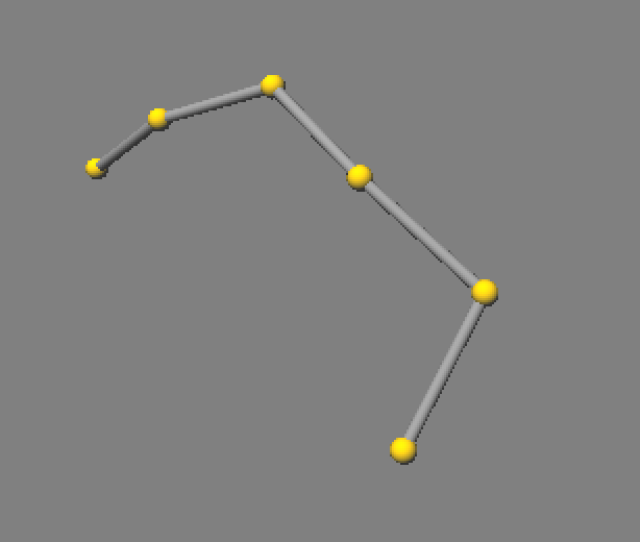}\\
  \includegraphics[width=12cm] {\image 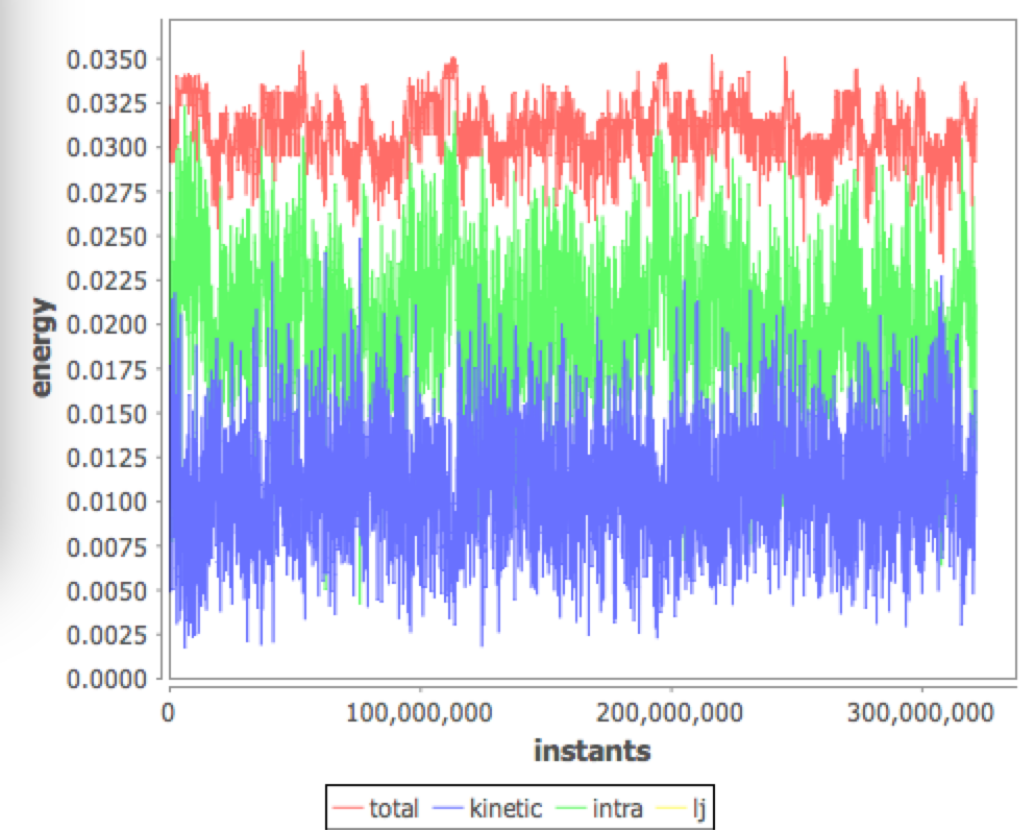}  
  \caption {\small Simulation of a CG molecule made up of 6 grains
    $CG_4$.  Initially, the first grain is shifted in $x$ and the last
    in $z$. The molecule is no longer planar.  On top image, the
    molecule after $3\times10^{8}$ instants.}
  \label {figure:CG6dih}
  \source{cg/SimpleCGApp.java dih=true make simple}
\end{myfigure}

We will now consider the case of inter-molecular forces. We consider a
simulation where two identical molecules in equilibrium are placed
face-to-face. The two molecules are linear and co-planar. Each
molecule is made up of six grains.

The right-hand side of Fig.\ref{figure:g6x2} shows the various
energies: in blue, the kinetic energy; in green, the intra-molecular
energy; in yellow, the inter-molecular energy, and in red the total
energy. The simulation lasts $10^8$ instants, corresponding to $25$
$ns$ (the time-step is $10^{-4}$ $ps$).

\begin{myfigure}
  \includegraphics[height=7cm] {\image 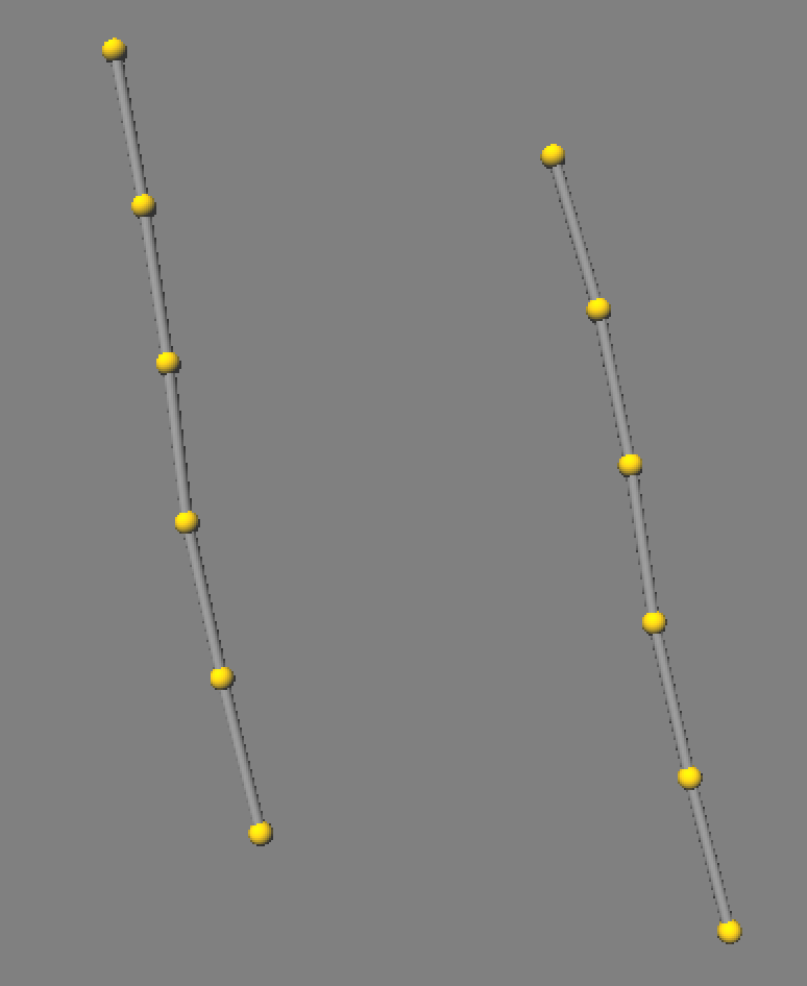}\\
  \includegraphics[width=12cm] {\image 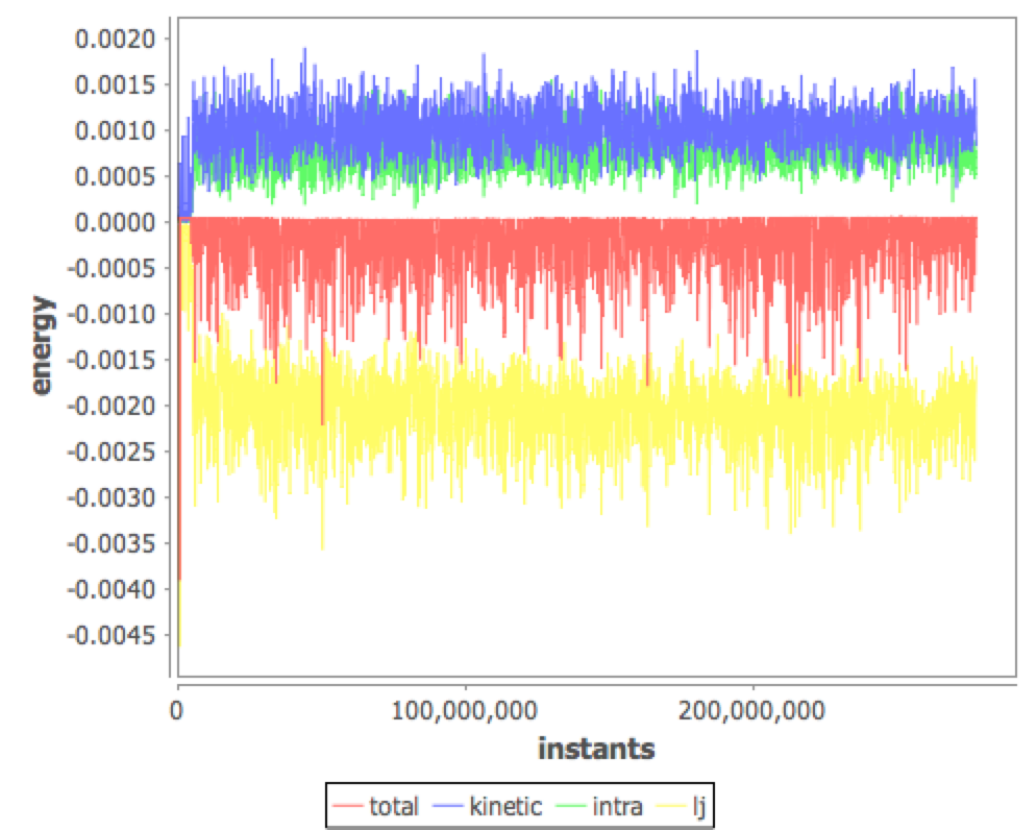}  
  \caption {\small Top: simulation of two molecules CG made up of 6
    grains $CG_4$ and initially co-planar. Bottom: variation of the
    energies.}
\label {figure:g6x2}
\source{cg/SimpleCGApp.java num=2 dx=dy=dz=0 tstep=1E-4 make simple}
\end{myfigure}

The two molecules, initially co-planar, remain so
during the execution, as shown on the left of the figure. The van Der
Waals forces remain in the same plane as the molecules.

This is not the case when the molecules are not co-planar. To
illustrate this, we repeat the previous simulation but after having
rotated one of the two molecules around the $x$ axis, by an angle of
one degree.  This means that the two molecules are no longer initially
co-planar. After a short time, the molecules begin to describe 3D trajectories.

\begin{myfigure}
  \includegraphics[height=6cm] {\image 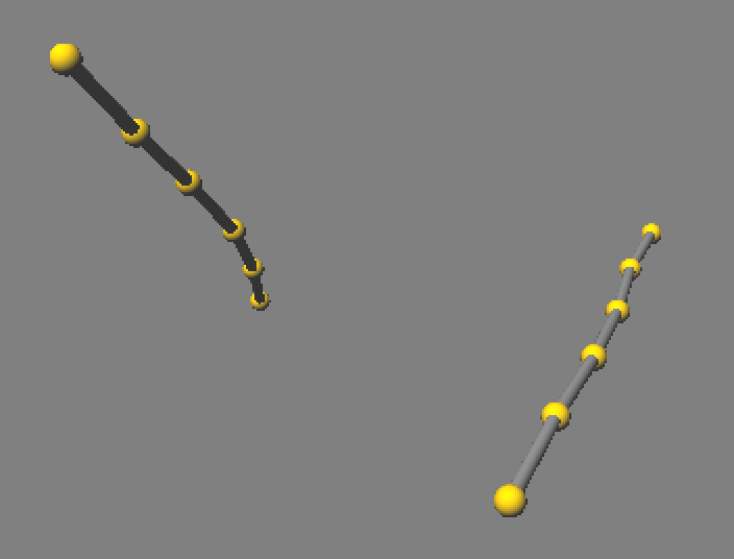}
  \caption {\small Simulation of two CG molecules made up of 6 grains
    $CG_4$, initially not co-planar, after $20^6$ instants.}
\label {figure:g6x2dev}
\source{cg/SimpleCGApp.java num=2 dx=dy=dz=0 tstep=1E-4 make simple}
\end{myfigure}

CG molecules that are longer than those shown in the previous images
may show foldings that can be very accentuated. This is the case, for
example, for the molecule made up of 40 grains in
Fig.\ref{figure:cg-rigid}.

\begin{myfigure}
  \includegraphics[width=13cm] {\image 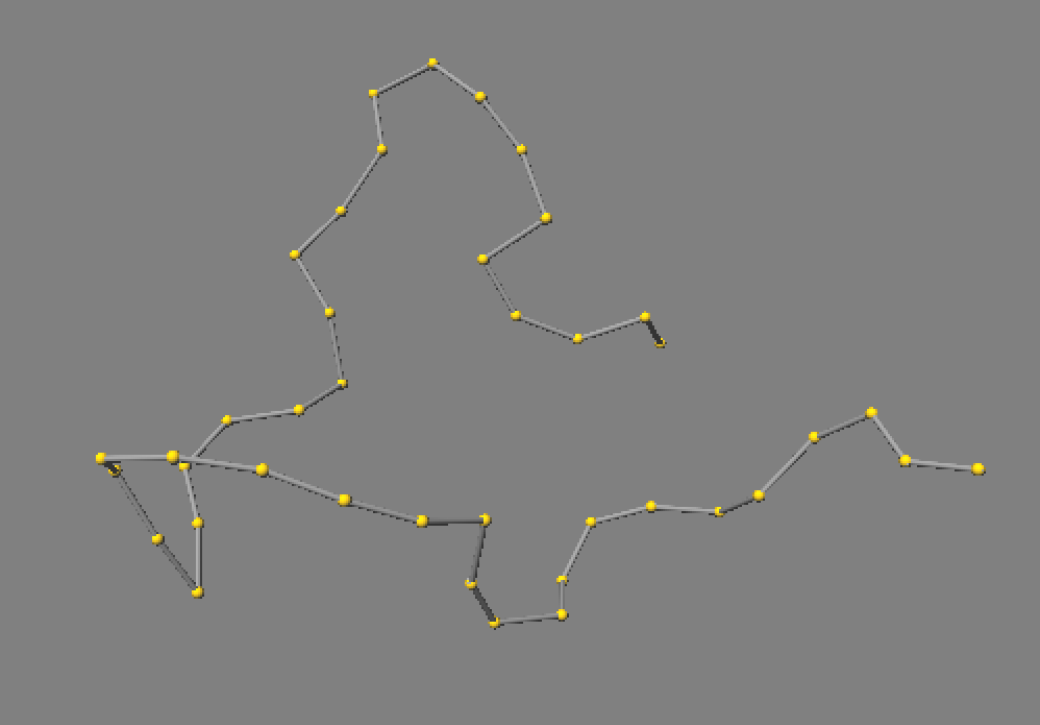}
  \caption {\small CG molecule made up of 40 grains $CG_4$, initially
    not co-planar, during simulation.}
\label {figure:cg-rigid}
\source{cg/RigidCGApp.java num=20 tstep=1E-4 make rigid}
\end{myfigure}


\chapter {Reconstructions}\label{chapter:reconstructions}
Molecule reconstructions make it possible to change {\it during the same simulation} the scale at which molecules are simulated. Typically, when a molecule is isolated, we prefer to simulate it at the UA or CG scale, much more efficient than the AA scale. However, when the molecules come close to one another, it may become necessary to go back to the AA scale, to take into account the interactions between them.

To manage reconstructions, several issues need to be addressed. The first is that, during a reconstruction, the reconstructed molecule must ``immediately'' replace the initial molecule. This means that during the simulation, the two molecules are not allowed to coexist (their interactions would be meaningless). In the reactive approach, this replacement, i.e. the destruction of the initial molecule and the introduction of the reconstructed molecule, is naturally carried out between instants.

A second question concerns the very possibility of reconstruction. It is difficult to implement reconstruction mechanisms that never cause any pathological increase in the energy of the reconstructed molecule. Reconstruction mechanisms need to be evaluated with this aspect in mind.

A third issue is that of interactions between molecules at different scales. In this respect, we can introduce automatic reconstruction mechanisms into the simulation, triggered by the proximity of the molecules, in order to bring the interacting molecules to the same scale.

\subsection*{Reconstructions between AA and UA}
The AA$\rightarrow$UA reconstruction consists of constructing a UA scale molecule from an AA scale molecule. In fact, it amounts to `` erasing'' the hydrogen atoms of the AA scale, by ``absorbing'' them into the grains of the UA scale according to the equation: \begin{equation}
  C + H + H \rightarrow G_2 \label{reconstrAA->UA}
\end{equation}

The mass is preserved, since the sum of one carbon with two hydrogens is $0.014$ in internal units (the mass of C is $0.012$ and that of H is $0.001$), which is also the mass of the grain $G_2$. The grain $G_2$ is positioned exactly on the atom $C$, which has the effect of eliminating the potential energies of the two CH bonds and that of the HCH angle, which disappear in the reconstruction.

The sum of the potentials of the AA molecule is then always greater than the sum of the UA potentials of the reconstructed molecule, resulting in a systematic loss of potential energy during the AA$\rightarrow$UA reconstruction.

The reverse reconstruction UA$\rightarrow$AA consists of constructing a molecule AA from a UA one. It involves replacing a $G_2$ grain by a carbon atom with two hydrogen atoms linked to it, according to the equation:
\begin{equation}
  G_2 \rightarrow C + H + H
\end{equation}

As in the equation (\ref{reconstrAA->UA}), masses are consistent. We place the carbons exactly on the $G_2$ grains and then try to place the hydrogen atoms. Ideally, these should be placed in such a way that the added energy (issued from CH bonds, HCH angles and HCC angles) is minimal. To achieve this, a minimisation phase of the AA molecule preserving the positions of the carbons should be launched before resuming the simulation.

Such a global minimisation phase is clearly very cumbersome, which is why we have chosen not to implement it, preferring instead to adopt a pragmatic but unsafe technique.

In this technique, each carbon is grafted with a pair of CH bonds in equilibrium with the two hydrogens forming a HCH angle also in equilibrium (see Fig.\ref {figure:reconstr-1}). Thus the added CH bonds and HCH valence angle do not introduce any additional energy.

\begin{myfigure}
  \includegraphics[height=1.3cm] {\image 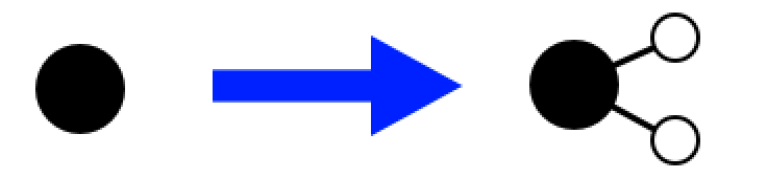}
  \caption {\small Introduction of hydrogen atoms, of bonds HC, and of the valence angle HCH.}
\label {figure:reconstr-1}
\source{reconstruction.key}
\end{myfigure}

For the valence angles HCC and the torsion angles HCCH and HCCC, we proceed in the same way as when creating a molecule at equilibrium, by alternating the pairs of hydrogens in parallel planes (see Fig.\ref {figure:reconstr-2}).

\begin{myfigure}
  \includegraphics[height=1.6cm] {\image 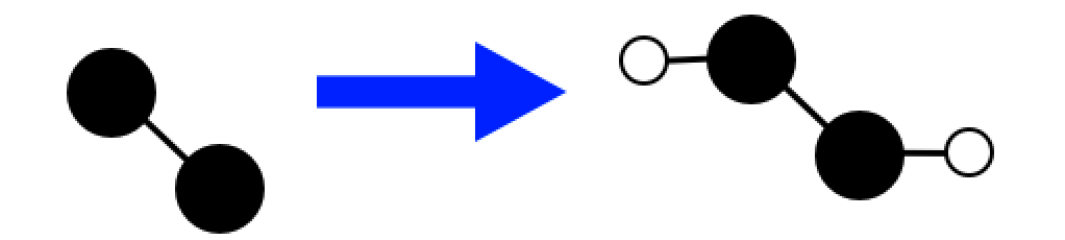}
  \caption {\small Introduction of valence angles HCC and of torsion angles HCCH and HCCC.}
\label {figure:reconstr-2}
\source{reconstruction.key}
\end{myfigure}


Thus, contrary to what happens with the AA$\rightarrow$UA reconstruction, additional energy can appear, corresponding to the energies of the HCC, HCCH and HCCC angles introduced. This extra energy can even lead to an explosion in the simulation, particularly when, due to a torsion of the molecule, hydrogen atoms are too close together.

\subsection*{Reconstructions between UA and CG}
The UA$\rightarrow$CG reconstruction consists of constructing a CG scale molecule from a UA scale one. In fact, it amounts to combining two UA grains into a single
CG grain according to the equation:
\begin{equation}
  G_2 + G_2 \rightarrow CG_4 \label{UA->CG}
\end{equation}

The mass of a grain $CG_4$ is 0.028. The $CG_4$ grain is positioned in the middle of the bond between the two $G_2$ grains and a CG bond is introduced between two consecutive $CG_4$ grains.

The inverse reconstruction CG$\rightarrow$UA consists of constructing a UA scale molecule from a CG scale one. Each CG grain corresponds to two UA grains:
\begin{equation}
  CG_4 \rightarrow  G_2 + G_2 \label{CG->UA}
\end{equation}

The two UA grains are placed on either side of the CG grain, so that the CG grain is in the middle of the two UA grains. A UA bond is introduced between the two closest UA grains. We adopt a similar technique to that used for the UA$\rightarrow$AA reconstruction, in the placement of the UA grains, with the same risk of incorrectly positioning the UA grains by introducing additional energy, which could cause the simulation to ``over-explode''.

\subsection*{Reconstructions between AA and CG}
The reconstructions between AA and CG can be considered as a sequence of two reconstructions, one between AA and UA and the other between UA and CG.  The AA$\rightarrow$CG reconstruction links two reconstructions which systematically lower the energy, and therefore does not pose a problem of energy explosion. However, this is not the case for the reverse CG$\rightarrow$AA reconstruction, where an explosion can occur both when the UA grains are introduced and when the hydrogen atoms are introduced, during the UA$\rightarrow$AA reconstruction.

\subsection*{Manual Reconstructions}
Fig.\ref {figure:reconstr-manual} shows a simulation in which the user, by clicking in the control panel, can change the scale of the molecule being simulated (a $\alkane{10}{20}$ molecule, initially supplied with some potential energy).  The change from the AA molecule to the UA molecule occurred in response to the click in the $\rightarrow$ua part of the panel. The change from the UA molecule to the CG molecule occurred in response to the click in $\rightarrow$cg. Clicks produce SugarCubes events which are instantly broadcast to all parts of the program.

\begin{myfigure}
  \includegraphics[width=6.4cm] {\image 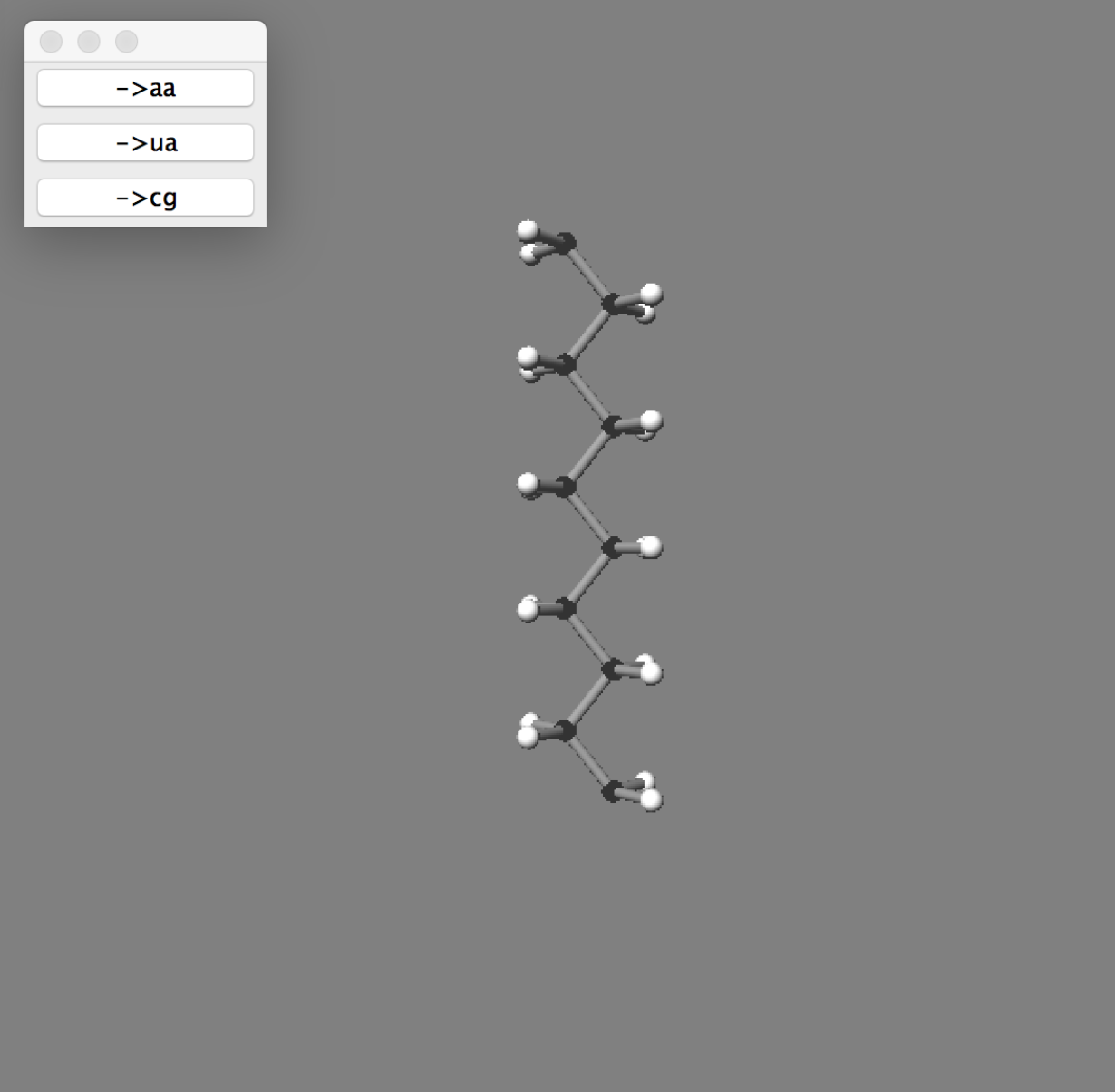}
  \includegraphics[width=6.4cm] {\image 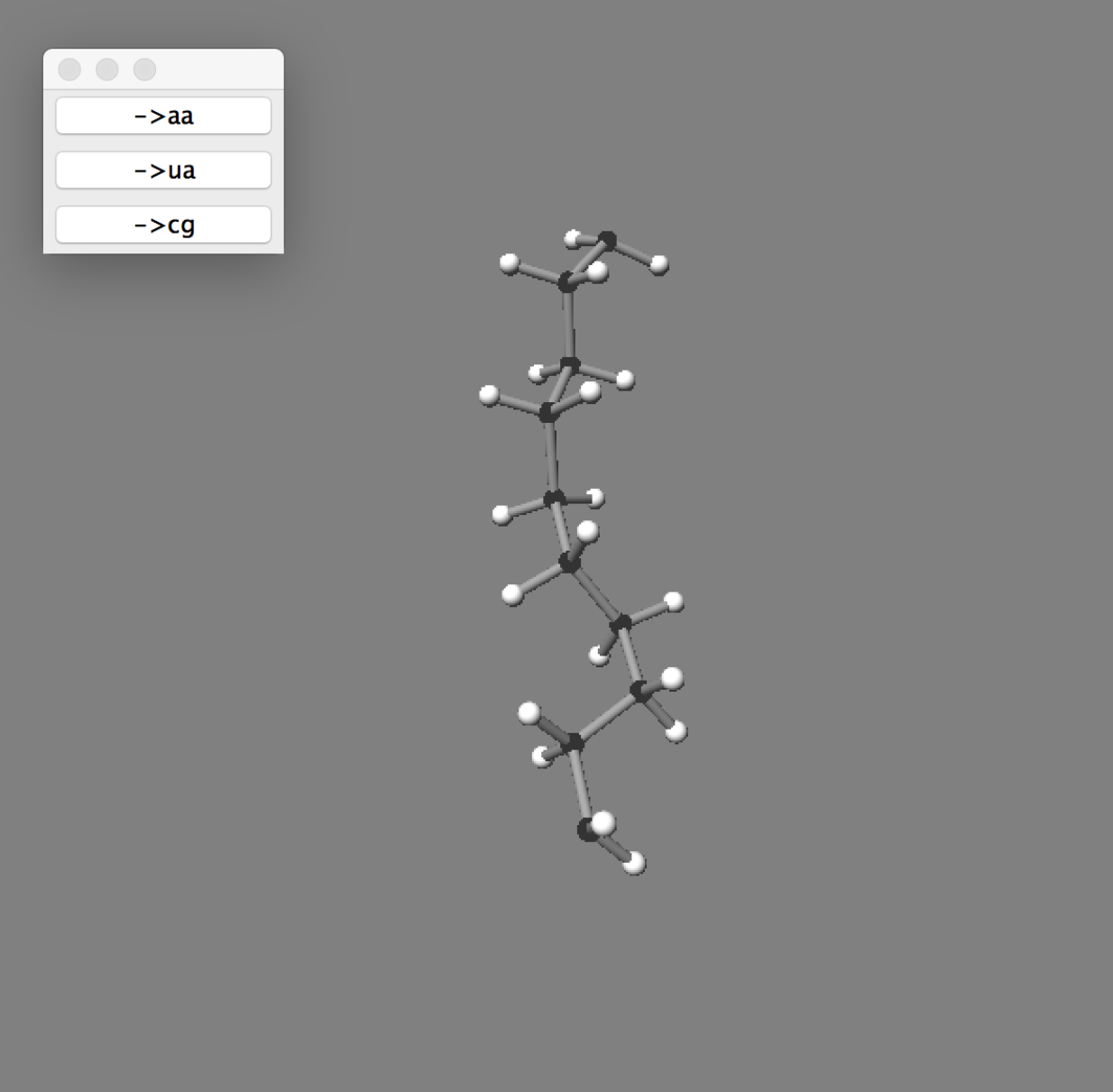}
  \\
  \includegraphics[width=6.4cm] {\image 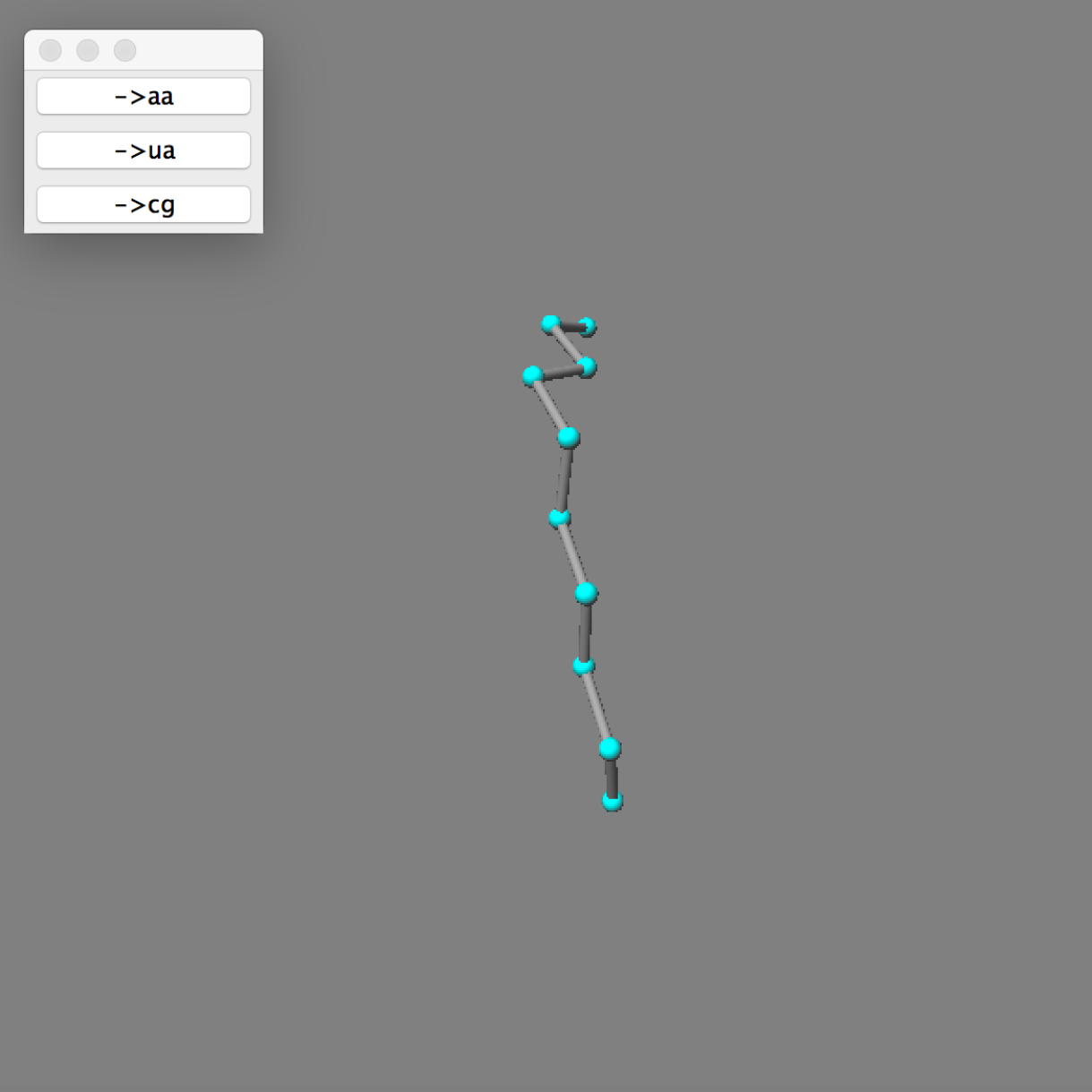}
  \includegraphics[width=6.4cm] {\image 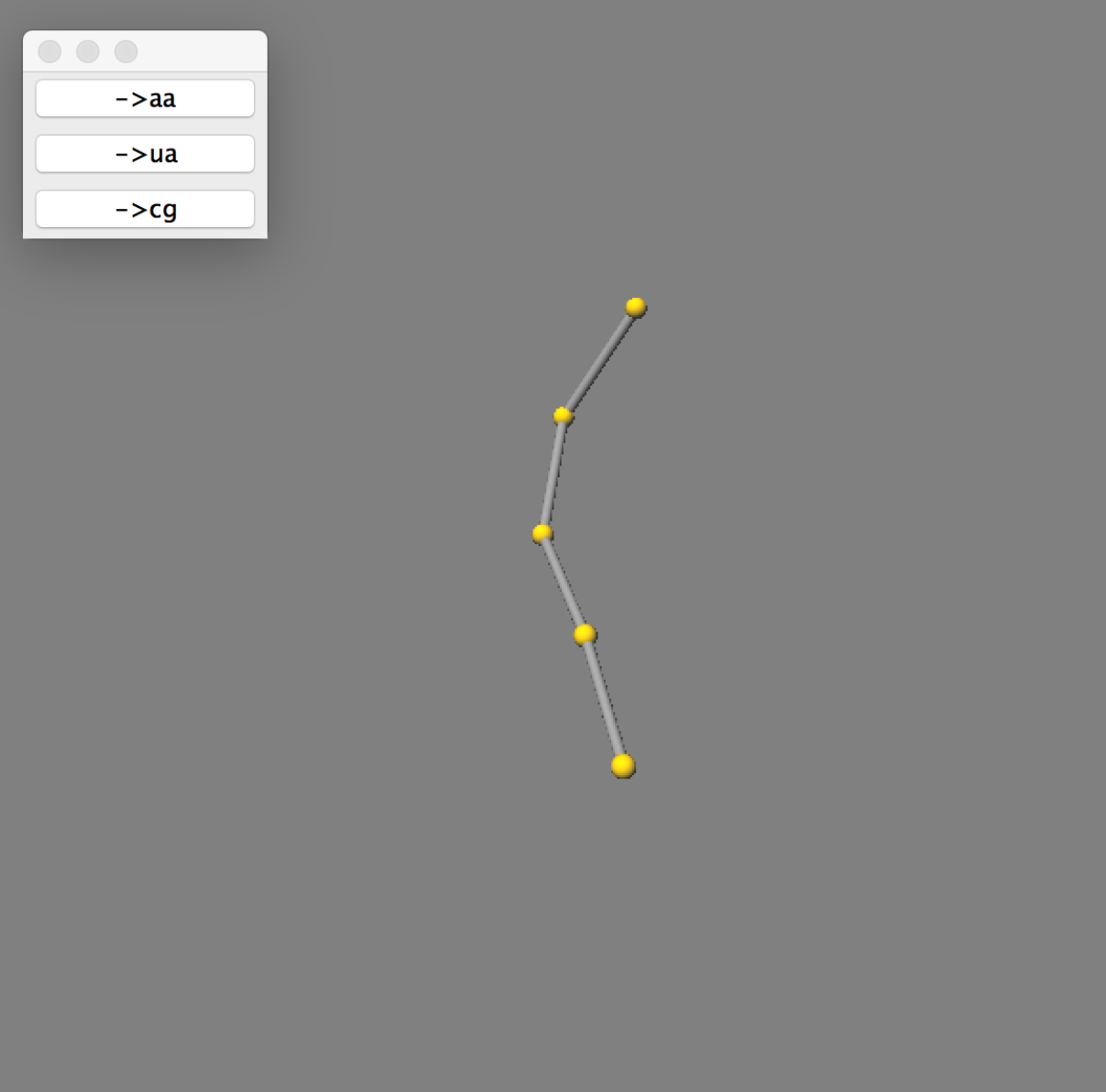}  
  \caption {\small 
    Top/left: initial AA molecule.
    Top/right: AA molecule after several instants.
    Bottom/left: after reconstruction of the AA molecule into UA.
    Bottom/right: after reconstruction of the UA molecule into CG.
  }
\label {figure:reconstr-manual}
\source{reconstruction/ManualApp.java initialShift=0.02 make manual}
\end{myfigure}




\section {Inter-molecular Forces}\label{section:reconstruction:lj}
The treatment of inter-molecular forces is tricky during reconstructions for two reasons: the first is that the introduction of hydrogens during UA$\rightarrow$AA or CG$\rightarrow$AA reconstructions can lead to explosions, as they come too close together. A similar issue occurs with UA grains in CG$\rightarrow$UA reconstructions. The second reason is that the equilibrium distances between molecules are not the same at the various scales, which means that the repulsive inter-molecular forces can increase sharply when molecules that are too close are reconstructed. We will now consider this case.

The equilibrium distance between two atoms (or grains) governed by a Lennard-Jones potential is:
\begin{equation}
  rn = 2^{1/6} \sigma
  \end{equation}

  where $\sigma$ is the distance at which the potential is zero.  In fact, the force cancels out in equation \ref {eq:lj:fa} when $2 {(\frac {\sigma} {r})}^{12} - {(\frac {\sigma} {r})}^{6} = 0$ which means, assuming $X= (\sigma/r)^6$, that $2X^2 - X = 0$. Hence $X(2X-1) = 0$, whose only valid solution is $X=1/2$, i.e. $r = 2^{1/6} \sigma$.

  At the equilibrium distance, the repulsive and attractive forces between the two atoms or grains balance out and the potential has the minimum value $-\epsilon$. Here are the values of $rn$ for the inter-atomic potentials associated with alkanes:

\begin{center}
  \begin{tabular}{c}
$rnCC = 0.39286171690828053$ \\
$rnCH = 0.33202427388991257$ \\
$rnHH = 0.28061551207734325$ \\
$rnUA = 0.36025419440489326$ \\
$rnCG = 0.35694293136238064$ \\
\end{tabular}
\end{center}

The value of $rnCG$ is obtained from that of the Lennard-Jones curve in Fig.\ref{figure:determinecglj} whose minimum coincides with that of the CG inter-molecular potential.

The distances at which the repulsive and attractive forces balance out are not the same,
which can be problematic in certain cases, for example when two AA molecules are placed face-to-face, as in Fig.\ref {figure:aa-equ} at a distance of $0.436$ $nm$, the distance at which the overall repulsive and attractive forces cancel each other out.

\begin{myfigure}
  \includegraphics[width=5cm] {\image 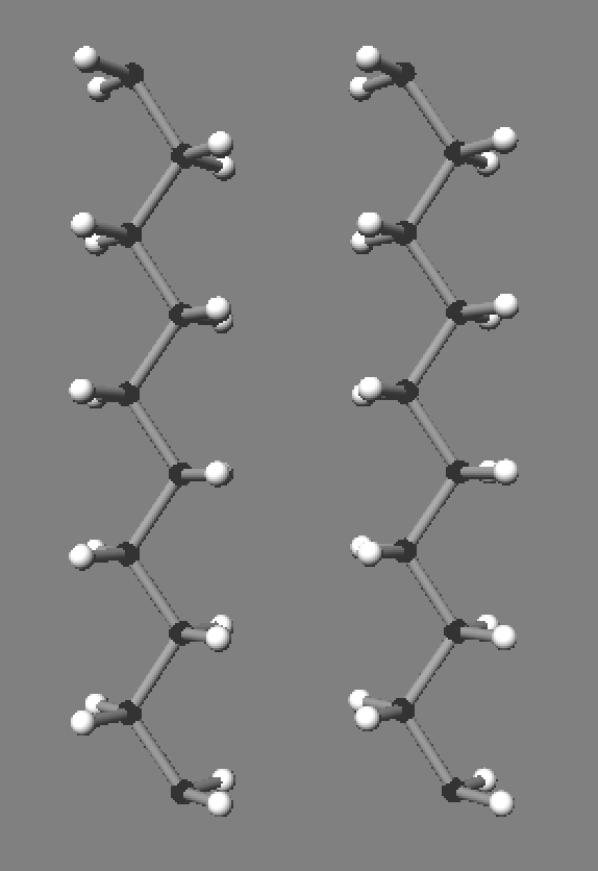}
  \caption {\small Two molecules $\alkane{10}{20}$ at a distance of $0.436$ $nm$.
    The two molecules are at equilibrium: the sum of attractive forces is equal to the sum of repulsive forces. }
\label {figure:aa-equ}
\source{reconstruction/CheckApp.java adjustAA=1.15 make check}
\end{myfigure}

The AA$\rightarrow$CG reconstruction destroys the balance of forces and results in a repulsive force that pushes the two CG molecules apart at high speed, which certainly does not correspond to reality.

 \section {Automatic Reconstructions}
 We now introduce a mechanism for automatic reconstructions triggered according to the proximity of molecules. For simplicity, we will only consider CG$\rightarrow$UA and UA$\rightarrow$AA reconstructions, as other types of reconstruction can be treated in a similar way.
 
 The triggering mechanism is implemented as a loop executed at each instant that measures distances to other molecules and generates a reconfiguration event when a molecule is detected at a distance below a specified threshold. This mechanism is added to the molecules that can be automatically reconstructed.

 The detection thresholds are chosen empirically to avoid energy explosions as far as possible:
 
\begin{center}
  \begin{tabular}{c}
    $detectUA = 0.39$\\
    $detectCG = 0.71$
\end{tabular}
\end{center}

A UA molecule which detects that the nearest carbon atom of an AA molecule is at a distance less than $detectUA$ automatically reconfigures itself into an A molecule.  A CG molecule which detects that the nearest carbon atom of an AA molecule is at a distance less than $detectCG$ reconfigures itself as an AA molecule. Similarly, when the CG molecule detects that the nearest grain of a UA molecule is at a distance less than $detectCG$, it reconfigures into a UA molecule.

It should be noted that the reconstructions can only concern a subset of the simulated molecules, which means that it is necessary to define the inter-molecular potentials between atoms and grains at different scales. To do this, we choose the highest scale (these being ordered by AA $<$ UA $<$ CG), which results in the following table:
\begin{center} \begin{tabular} {l|lcr}
        & AA         & UA     & CG      \\\hline
  AA &  AA        &  UA    &  CG      \\
  UA &  UA        &  UA    &  CG      \\
  CG &  CG        &  CG    &  CG      \\
\end{tabular}
\end{center}
  
Fig.\ref {figure:show-initial} and Fig.\ref {figure:show} show a simulation of four CG molecules, each composed of two grains.

\begin{myfigure}
  \includegraphics[width=9cm] {\image 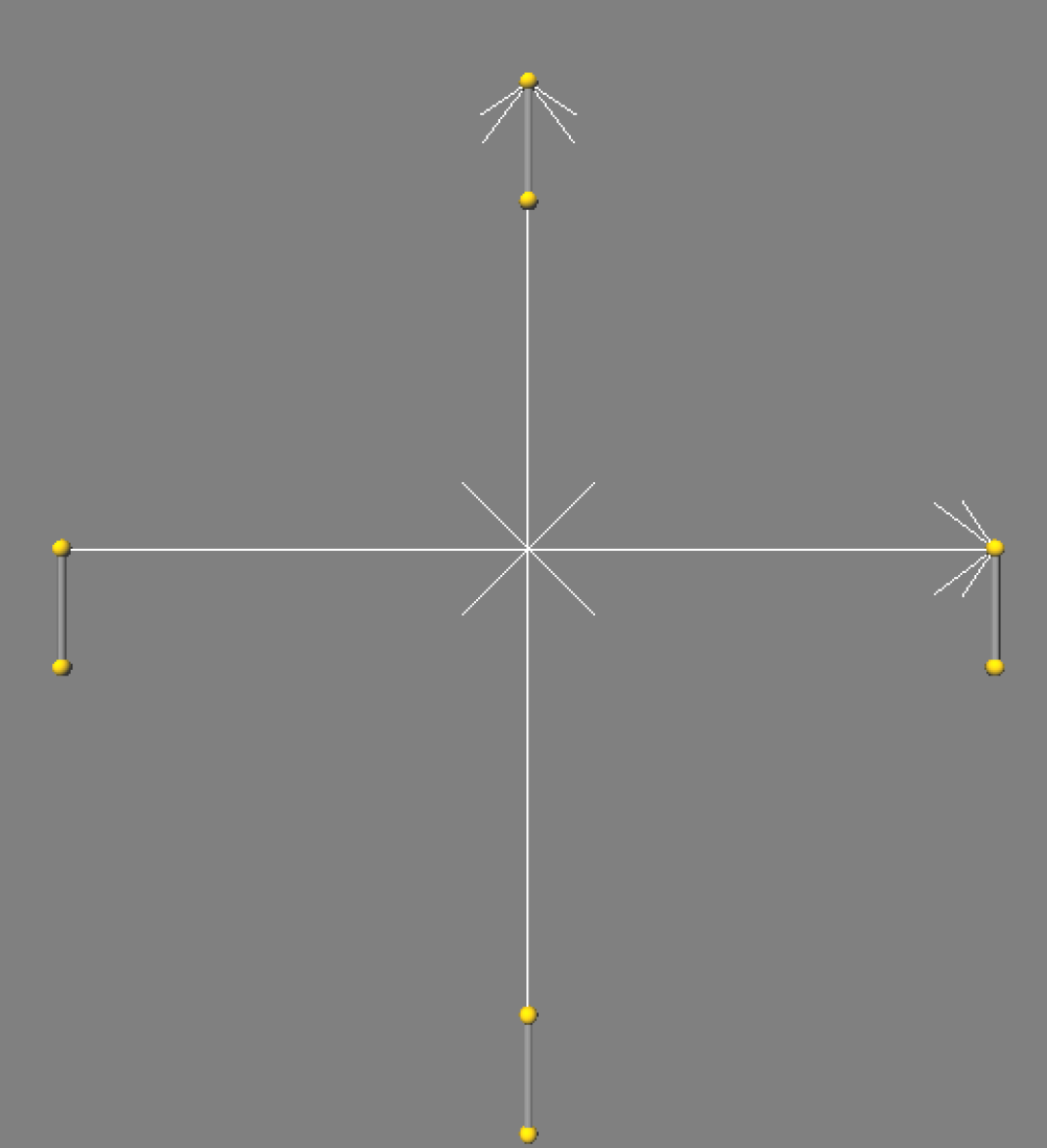}
 \caption {\small Initial configuration: four CG molecules placed at the four vertices of a square.}
\label {figure:show-initial}
\source{reconfiguration/Autom3App.java tstep = 0 make autom3}
\end{myfigure}

The initial situation is that of Fig.\ref {figure:show-initial} where four CG molecules are present. The four molecules reconfigure to UA (left/top image of Fig.\ref {figure:show}). Quite quickly, the UA molecules are detected and reconfigure themselves into AA molecules (right/top image).

The last two images (bottom) show the evolution of the four AA molecules. The last image (bottom/right) shows that the molecules are moving definitively away from the centre of the figure, grouped together in pairs.

\begin{myfigure}
  \includegraphics[width=6.5cm] {\image 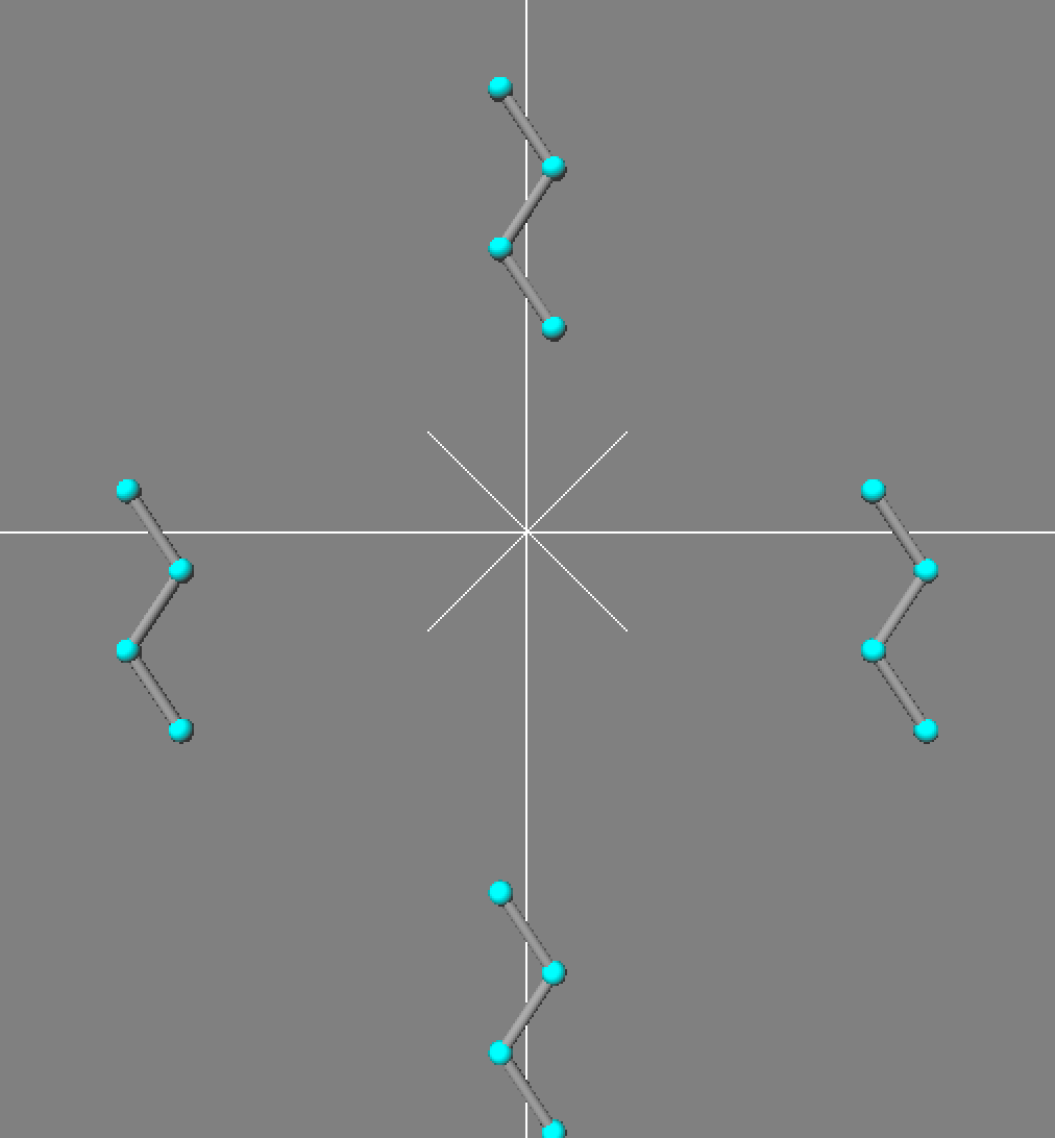}      
  \includegraphics[width=6.5cm] {\image 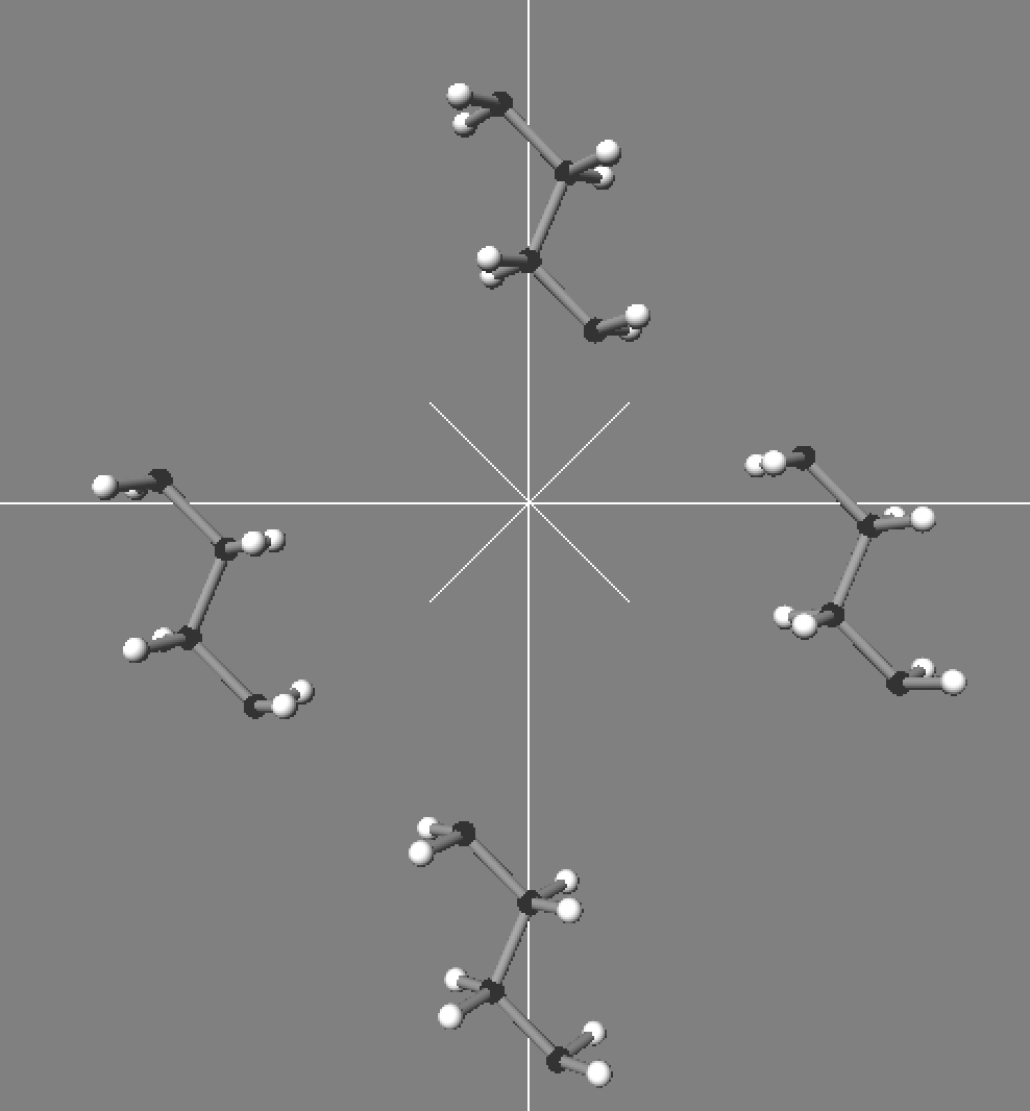}
  \\
  \includegraphics[width=6.5cm] {\image 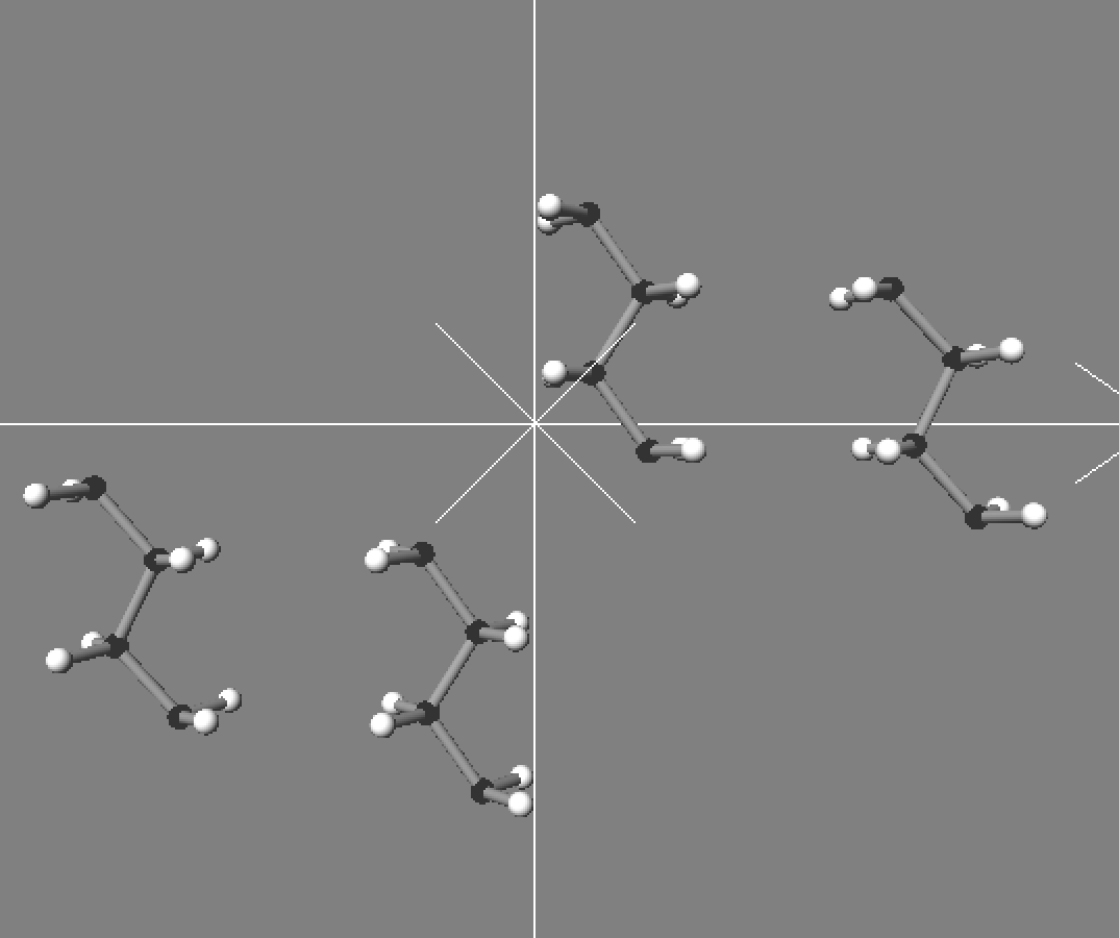}
  \includegraphics[width=6.5cm] {\image 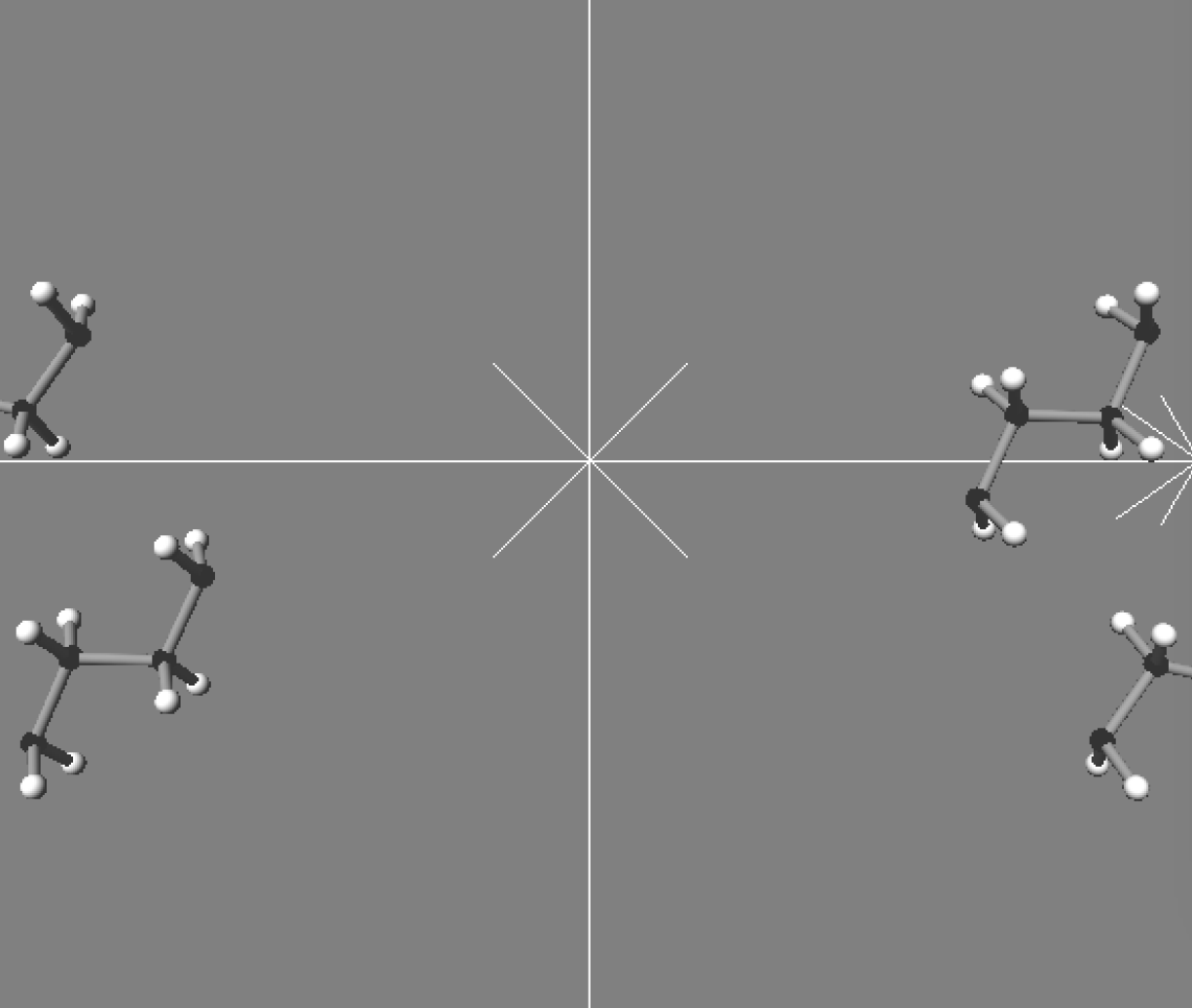}
  \caption {\small 
    Scenario of the sequence of reconfigurations of molecules initially placed as in
    Fig.\ref {figure:show-initial}.}
\label {figure:show}
\source{reconfiguration/Autom3App.java make autom3}
\end{myfigure}

 \section* {Conclusion}
 As far as reconstruction is concerned, we have adopted a pragmatic approach in which effectiveness is given priority over safety (i.e. non-exploding simulations).

 More precisely, the reconstructions presented are not reliable in the sense that they can fail by introducing too much energy into the reconstructed molecule.  This is the case when grains of the same molecule are very close, leading to grains or atoms being placed in positions that are too close.  A reconstruction can also fail when grains of different molecules are too close, which forces the grains or atoms introduced by the reconstruction to produce inter-molecular forces that are too strong.

 On the contrary, the reconstructions presented are effective in the sense that, when they are successful, the energy of the initial molecule and that of the reconstructed molecule generally differ very little. This is particularly the case for molecules with low intra-molecular energy, since in this case the process of reconstruction is
 very close to the process of molecule creation at equilibrium.

 The reconstructions described are therefore not suitable for molecules with high intra-molecular energy or, concerning inter-molecular forces, for molecules that are too close together.

 It therefore seems reasonable to reserve the reconstructions for cases where the energy of the molecules is not too large and when the molecules are sufficiently far apart.

 The alternative to the approach presented, to make reconstructions safe, would be to introduce a phase of minimisation of the reconstructed molecule, a process that can take up a lot of simulation time.
 
 Automatic reconstructions AA$\rightarrow$UA and UA$\rightarrow$CG could be envisaged, when the distances are sufficiently large, which would be additional means of speeding up simulations. This type of reconstruction has not been implemented in the current system, but it could be done without too much difficulties.


\chapter {Conclusion}\label{chapter:conclusion}
The reactive approach provides a good framework for programming MD
systems. It has a clear and precise formal semantics
(Chap.\ref{chapter:implementation}).  It makes it possible to deal
with aspects related to temporal resolution, modularity of
programming, determinism, and dynamicity (non-fixed structure of
systems, which can evolve dynamically during simulations).

The question of determining potentials on the UA and CG scales for
alkane molecules was raised.  The aim was to ``derive'' the UA and
CG potentials from the AA potentials.

The determination of coarse-grain potentials based on an
inverse-Boltz\-mann statistical physics method, using data obtained
from simulations, comes up against a number of problems, in particular
the temperature dependency of the results and the incorrect treatment
of inter-molecular forces.

The determination of potentials by the minimisation method differs
radically from that of the inverse-Boltzmann method.  The minimisation
is not based on simulation data, like inverse-Boltzmann, but on energy
minimisations at the AA scale (although the
minimisations are performed with the same MD system, strictly speaking
it is more a question of {\it molecular mechanics} than a question of
MD).

A fundamental point is that minimisations are made possible at the UA
and CG scales because strong geometric links with the {\AllAtom} scale
have been established.

The stability of the MD system used relies on a complete explanation
of the forces acting on the atoms (Chap.\ref{chapter:forcefield}) as
well as on the use of an extremely stable resolution method
($\verlet$).

\subsection*{UA Potential}
The {\UA} potentials determined by the minimisation method have
several characteristics:
\begin {itemize}
\item[$\bullet$] They have the same form as the corresponding
  {\AllAtom} potentials: the bond and valence {\UA} potentials are
  harmonic; the torsion potential is a ``triple-cosine‘’ potential;
  the inter-molecular potential is a Lennard-Jones potential.

\item[$\bullet$] The UA bond potential is identical to the AA bond
  potential.
  
\item[$\bullet$] The UA valence potential is very close to the AA
  valence potential.  It can be seen as taking into account the valence
  angle CCC of the carbons, as well as the associated CCH angles.

\item[$\bullet$] The UA torsion potential is the sum of the CCCC
  potential with the HCCH potentials sharing the same central CC bond.
    
\item[$\bullet$] The UA inter-atomic potential is a Lennard-Jones
  potential more ``excavated'' than the AA inter-atomic potential as
  it takes into account the inter-molecular forces between the
  hydrogen atoms.
\end{itemize}

Thus, to resume, the UA potential is obtained from the AA one by a
simple change of parameter values.

\subsection*{CG Potential}
Determining the {\CG} potential is considerably more difficult than
determining the {\UA} potential (although there is no torsion angle
potential in {\CG}).  The minimisation method gives potentials which
do not have the standard form of the {\AllAtom} or {\UA} potentials
and which are represented in the three figures
Fig.\ref{C4_MIN_CGBOND-curve.data},
Fig.\ref{C6_MIN_CGVALENCE-curve.data}, and Fig.\ref{figure:vdwCG}.  In
a way, the work presented here can be seen as summarised in
these three curves.

\subsection*{Reconstructions}
Reconstruction mechanisms (i.e. molecule scale changes during the
course of simulations) have been defined and implemented.  In
particular, we have considered automatic reconstructions based on the
proximity of molecules.  The reconstructions defined are not safe in
the sense that they do not systematically preserve energy.  They
should therefore be reserved for molecules with low intra-molecular
energy, being sufficiently far apart so as not to induce too large
inter-molecular energies during reconstruction.

\subsection*{Outlook}
Several possible extensions to the work presented here seem
interesting, among which the three following ones.

The first area of work concerns ``true parallelism'', using a new
version of SugarCubes (SugarCubesv5 \cite{SCv5}) in which the use of
graphical cards (GPU) to perform calculations in real parallelism
becomes possible. The use of multi-processor machines would also be of
great interest.

The second area of work would be to implement a safe reconstruction
technique, based on the minimisation of reconstructed molecules (see
Chap.\ref{chapter:reconstructions}).

The third area of work would involve extending the class of molecules
taken into account, by not limiting ourselves to alkanes. Carbon
chains, similar to alkanes, but in which one end contains oxygen atoms
(``soap'' molecule) could naturally be taken into account by
mechanisms very similar to those used for reconstructions (some
experiments have been carried out on this point).


\newpage
\section*{Context}
This text was written by Frédéric Boussinot. It describes some of the
work carried out by Bernard Monasse and Frédéric Boussinot when they
were both researchers at the CEMEF laboratory of École des Mines de
Paris. This text is the English version of \cite{FrenchSMEDM}.

The SugarCubes are the outcome of Jean-Ferdy Susini's thesis work
at Inria Sophia Antipolis, in the Mimosa team.

Thanks to Pierre Montmitonnet for his numerous comments, corrections
and suggestions concerning the French version of the document.

\bibliographystyle {plain}
{\small \bibliography {molecularDynamics}}

\end{document}